\documentclass{aa}  %

\usepackage{graphicx}
\usepackage{txfonts}
\usepackage{xcolor}
\begin{document}

   \title{A study of Andromeda to improve our knowledge on the evolution and dust production by AGB stars}


   \author{C. Gavetti\inst{1,2}, P. Ventura\inst{2}, F. Dell'Agli\inst{2}, F. La Franca\inst{1},
           E. Marini\inst{2}, M. Correnti\inst{2,3}, M. Tailo\inst{4}}

   \institute{Dipartimento di Matematica e Fisica, Università degli Studi Roma Tre, 
              via della Vasca Navale 84, 00100, Roma, Italy \and
              INAF, Observatory of Rome, Via Frascati 33, 00077 Monte Porzio Catone (RM), Italy \and
              ASI-Space Science Data Center, Via del Politecnico, I-00133, Rome, Italy \and
              Istituto Nazionale di Astrofisica– Osservatorio Astronomico di Padova, Vicolo dell'Osservatorio 5, I-35122 Padova, Italy
              }

   \date{Received September 15, 1996; accepted March 16, 1997}


 \abstract
 {}
   {We study the AGB population of the galaxy M31, based on available HST and Spitzer data,
   to characterize the individual sources in terms of mass, metallicity and formation epoch
   of the progenitors. Particular attention is dedicated to the derivation of the dust
   production rate of the stars, in the attempt of determining the global current dust production
   rate of the galaxy, divided between the silicates and the carbonaceous dust contributions.}
   {We use results from stellar evolution modelling complemented by the description of the
   dust formation process in the wind, to be used in a population synthesis approach, based on
   the star formation history and age-metallicity relationship obtained in previous investigations. 
   The comparison between the results from synthetic modelling and the data available are used
   for the characterization of AGB stars in M31.}
   {We find that the bulk of the AGB population of M31 is composed by low-mass stars of 
    different metallicity formed between 6 Gyr and 14 Gyr ago, with an additional, significant
    contribution from the progeny of $\rm 1.7-2.5~M_{\odot}$ stars formed during the secondary peak 
    in the star formation, which occurred between 1 and 2 Gyr ago. The dust production rate of the
    galaxy is mostly provided by carbon stars, whose contribution is of the order of 
    $\rm 4\times 10^{-4}~M_{\odot}$/yr, completed by silicates production from massive AGB stars,
    occurring at a rate of $\rm \sim 6\times 10^{-5}~M_{\odot}/$yr. The implications of the present
    results on the reliability of AGB modelling are also commented.
    }
   {}


   \titlerunning{The evolved stellar population of M31}
   \authorrunning{Gavetti et al.}
   \maketitle
%

\section{Introduction}
The stars of initial mass below $\rm \sim 8~M_{\odot}$ after the exhaustion of central helium 
evolve through the so called asymptotic giant branch (AGB) phase. The energy release from 
AGB stars is provided for most of the time by a H-burning shell, where CNO
nucleosynthesis takes place. Periodically, ignition of a helium-rich layer just above
the degenerate core occurs, in a series of events generally known as thermal pulses \citep{sch}.
The AGB phase, during which the stars lose the external mantle under the action of intense stellar
winds \citep{habing96}, is followed by the general contraction of the stellar structure, which starts
the evolution towards the post-AGB and PNe phases, which preceed the white dwarf
cooling.

The growing interest towards this class of stars is related to their role as pollutants of the interstellar medium and thus to the important feedback on the host environment, in relation to the gas lost via intense stellar winds \citep{martin18} and the dust formed in great quantities in their expanded circumstellar envelope.

AGB stars provide a relevant contribution to the enrichment of the interstellar medium in some chemical species, such as carbon and nitrogen \citep{ciaki20}, and are believed to play a crucial role in the determination of the chemical patterns in star forming regions of Local Group galaxies \citep{fiorenzo} and in the formation of multiple populations in globular clusters \citep{ventura01}.

Furthermore, AGB stars are efficient dust manufacturers, owing to the thermodynamic conditions of their winds, which prove extremely favourable to the condensation of gas molecules into solid grains \citep{gase85}. For this reason, they are believed to contribute significantly to dust production in nearby galaxies \citep{matsuura09, matsuura11, matsuura13, raffa14,srinivasan16, boyer17}.

Some research groups have recently implemented the 
description of dust formation in the wind of AGBs, coupled to the modelling of the
evolution of the central stars \citep{ventura12, ventura14, nanni13, nanni14}.
The results from these studies were successfully used to characterize the
evolved stellar populations of the Magellanic Clouds \citep{flavia15a, flavia15b,
nanni16, nanni19, riebel12, boyer12} and other Local Group galaxies \citep{flavia16, flavia18, flavia19}. 

Despite the steps forward accomplished in the aforementioned studies, several 
uncertainties still affect the predictive power of the results obtained in regard of the
efficiency of dust production by AGB stars: this is due to the poor knowledge of
some physical phenomena playing an important role in the evolution of AGB stars
\citep{ventura05a, ventura05b, karakas14} and to the uncertainties related to 
the dust formation process \citep{ventura14}.

To shed new light on the evolutionary and dust production properties of AGB stars,
the comparison with results from IR surveys of galaxies with resolved stellar
populations is the most efficient tool, particularly if a variety of environments,
differing in the star formation history and in the age-metallicity relation, 
is investigated. The advancement in this field is required not only for a mere
improvement in our understanding of the AGB properties, but also to develop a
new approach in the study of galaxies with resolved stellar
populations, so that the IR properties of the evolved
stellar component can be used to reconstruct the star formation history (SFH). This is
particularly important in the recently started JWST epoch, considering the increasing 
number of galaxies outside the Local Group already observed and to be targeted in the 
near future. For a significant fraction of these galaxies AGB stars will
provide the only opportunity to reconstruct their star formation history, considering that
the main sequence will be hardly sampled by the JWST observations.

In this work we focus on one of the most interesting and well-investigated galaxies:
Andromeda (M31). With a distance of 776 Kpc \citep{dalcanton12}, M31 is the spiral 
galaxy closest to the Milky Way. Thanks to its proximity , M31 is easily
reachable with HST and Spitzer, which allows a thorough investigation 
of the resolved stellar population. For these reasons, M31 represents 
a unique occasion to achieve a thorough stellar population study at solar metallicity.
A significant boost on this
side came from the Panchromatic Hubble Andromeda Treasury \citep[PHAT;][]{dalcanton12} survey, which 
imaged about $30\%$ of the star forming disk of the galaxy in different
bands, ranging from the IR to the ultraviolet. These investigations allowed a 
variety of studies, such as the derivation of the ancient \citep{williams17}
and recent \citep{lewis15} SFH, and the metallicity distribution 
\citep{gregersen15}; the latter analysis revealed that the stellar population
of M31 formed in the last 10 Gyr is dominated by a solar metallicity 
component, which is rather uncommon, if not unique, with respect to other
galaxies in the Local Group. In a recent work, Goldman et al. (2022, hereafter G22)
used available HST and Spitzer data to construct a large
sample of the AGB stars candidates ($\sim 346000$) in M31, part of which
were classified as O-rich or C-rich \citep{boyer19} according to the molecular
features detected in their spectral energy distribution (SED). An
important result obtained by G22 is that the fraction of stars classified
as ''extreme'' AGBs\footnote{G22 uses the standard criterion to define extreme AGBs the sources with $\rm ([3.6]-[4.5]) > 0.25$ mag and $\rm [4.5] < 16.4$ mag.}, those believed to provide the dominant contribution to the dust
production rate of the entire galaxy, is slightly above
$1\%$, thus significantly smaller than the 4.5\%  and 6\% found in the Large 
and Small Magellanic Clouds, respectively \citep{boyer11, riebel12}. 

The scope of the present work is to analyze the AGB population of 
M31, in order to obtain a full characterization of the sources observed,
in terms of the mass, metallicity, formation epoch of the progenitors, 
and the current evolutionary stage and dust production rate (DPR). Primary
objectives of this study are: a) the determination of the overall
DPR of M31, separated into the carbon dust and silicates contribution;
b) assessing whether the efficiency of carbon dust production
at solar metallicities is comparable with lower metallicities, as
derived from previous studies on the Magellanic Clouds and
other Local Group galaxies; c) understanding whether the small
percentage of extreme AGBs detected in M31 is related to metallicity
and/or SFH effects, or if it is an observational bias
connected to the completeness of the data.

This analysis is based on results from stellar evolution modelling
and on the description of the dust formation in the wind, used
to predict how the SED of stars of different mass (thus formed
in different epochs) and metallicity evolve during the AGB phase. This step is crucial to undertake a stellar population synthesis approach, in which the expected distribution of the stars in the various observational planes based on different filters combination is compared with the results from the observations published by G22.

The structure of the paper is as follows: the methodology and the
input adopted are described in section 2; section 3 presents the 
catalogue of AGB stars in M31 realized by G22; section 4 discusses the main uncertainties that affect the predictive power of the results; the comparison between the observations is addressed
in section 5, whereas the discussion of the results obtained is
given in section 6.

\section{The methodology}
\label{input}
The analysis presented in this work is based on the
comparison between the observed distribution of the stars belonging
to the evolved stellar population of M31 with the results obtained
by a population synthesis approach: in the latter, 
the synthetic distribution of the stars is derived on the basis of 
the time evolution of the SED of model stars of different mass and 
chemical composition.

This method requires the modelling of 
the AGB phase of the stars, which must be coupled to the description 
of the dust formation process in the wind, considering the important 
effects that the presence of dust in the circumstellar envelope of 
stars has on the shape of the SED, thus on the expected colours and 
magnitudes in the different bands.

In the following, we provide the description of the various steps 
followed in the present analysis: a) the modelling of the
evolution of the stars, from their birth, through the AGB phase,
until the start of the white dwarf cooling; b)
the schematization followed to describe the dust formation process
and to calculate the DPR during the AGB lifetime; c) the code used
to derive the AGB evolution of the SED; d) the population synthesis
approach followed to build the synthetic distribution of the stars
on the different colour-magnitude planes.

\subsection{Stellar evolution modelling}
\label{aton}
The evolutionary sequences used in this work were calculated by means
of the ATON code for stellar evolution \citep{ventura98}. The physical
input most relevant for the present investigation are the following:
\begin{itemize}

\item{{\it Convection.} The temperature gradient within regions of the star
unstable to convective currents was determined by means of the Full
Spectrum of Turbulence (FST) convective model \citep{cm91}. We also
considered Mixing Lenght Teory (MLT) modelling \citep{mlt} with variable mixing length parameter $\alpha$, to 
explore the effect of adopting convective models corresponding to different
efficiencies of the convective transport of energy. An exhaustive
discussion on how the choice of the convective model affects the description of 
the AGB phase can be found in \citet{ventura05a}.}

\item{{\it Molecular opacities.} The surface molecular opacities were calculated 
via the AESOPUS tool \citep{marigo09}, which allows considering the effects of 
the variation of the surface abundances of the CNO species. This is particularly
important to model the advanced evolutionary AGB phases of low-mass stars, when the surface
chemistry is significantly enriched in carbon \citep{vm09, vm10}.
}

\item{{\it Mass loss.} Whit regard to the red giant branch phase, 
mass loss was modelled according to the classic \citet{reimers75} prescription, 
with free parameter $\eta_R=0.4$. This regards $\rm M > 2~M_{\odot}$ stars
only, because, as discussed below, the evolution of lower mass stars
is started from the core helium burning phase.
For what concerns the AGB evolution, the determination of the mass loss rate during the 
oxygen-rich phases is based on the treatment by \citet{blo95}. To explore 
the effect of the mass-loss treatment on the description of the AGB
phase of $\rm M \geq 4~M_{\odot}$ stars we run additional simulations with
the mass loss description by Vassiliadis \& Wood (1993, VW93).
During the C-rich phases we adopted the formulae
given by the Berlin group, in \citet{wachter02} and \citet{wachter08}.}
\end{itemize}

As already highlighted in the Introduction, M31 is characterized by a variety of stars
differing in mass, chemical composition and formation epoch of the progenitors.
To describe stars of different chemistry we considered extant evolutionary 
sequences previously published by our group, of metallicity $Z=0.001$ 
\citep{ventura14}, $Z=0.004$ and $Z=0.008$ \citep{marini21},
and $Z=0.014$ \citep{ventura18}. In the low-mass domain all
the above computations, started from the pre-main sequence, were recently
extended until the start of the white dwarf cooling, as described in 
\citet{devika23}. 

For the stars undergoing the helium
flash the computations are re-started from the quiescent core helium burning phase, based on the core masses reached at the tip of the red giant branch (RGB). These stars experience a long RGB evolution, during which the surface gravities become so small that a significant fraction of the mass of their envelope is lost before they reach the tip of the red giant branch (TRGB). In the following sections, particularly in sections \ref{uncer} and
\ref{char}, and in the Fig.~\ref{fall} and \ref{fspit}, we will refer to the evolution of a given model star by mentioning the mass at the start of core helium burning, keeping in mind that it must be considered as a lower limit of the mass of the progenitor.
As the mass lost by a given star during the ascending of the RGB is uncertain, the corresponding computations from the HB were started by assuming several values of the total mass, to simulate the spread in
the RGB mass loss. Therefore, the models presented in the aforementioned investigations
were completed with additional sequences, calculated on purpose for the
present work. Further computations were added to explore the effects of
convection modelling, as described in the following sections.

\subsection{Dust formation in the wind of AGB stars}
\label{dustmod}
To model dust formation in the wind of AGB stars we followed the approach
proposed by the Heidelberg team \citep{fg01, fg02, fg06}, which was used in
previous works by our team \citep{ventura12, ventura14} and by other research
groups \citep{nanni13, nanni14}. Dust formation during a given evolutionary
phase is modelled on the basis of the following ingredients:
mass, effective temperature, luminosity, mass loss rate and surface
chemistry of the star. All these information are obtained by stellar evolution 
modelling and the ATON code, described earlier in this section. All the detailed 
equations governing the dynamic and thermodynamic stratification of stellar
winds and the growth of dust grains are listed and discussed in \citet{ventura12}.

Dust particles are assumed to form and grow in the wind, which expands isotropically from 
the photosphere of the star. In carbon-rich environments, we consider the formation of 
solid carbon and silicon carbide (SiC), whereas in the winds of oxygen-rich stars we assume 
that the dust species that form are silicates and alumina dust ($\rm Al_2O_3$); solid iron is formed 
in either cases. These are the most stable species, formed in greatest quantities
\citep{fg06}.

The dynamics of the wind is described by the momentum equation, where the acceleration 
is determined by the competition between gravity and radiation pressure acting on the 
newly formed dust grains. The coupling between grain growth and wind dynamics is given 
by the extinction coefficients, describing the effects of absorption and scattering of the radiation 
by dust particles. For the species considered here, the extinction coefficients were 
found by using the optical constants from Zubko et al. (1996) (amorphous carbon), 
\citet{peg88} (SiC), \citet{begemann94} (alumina dust), \citet{oss92} 
(silicates), and \citet{ordal} (solid iron).

The modelling of dust formation, as described above, allows the determination of: a) The size reached by the grains of the various dust species listed above; b) The 
fraction of gaseous silicon, carbon, aluminium, and iron condensed 
into dust in the circumstellar envelope of the stars (see Eqs. (20)–(23) and 
(34)–(35) in Ferrarotti \& Gail (2006)); c) The dust production rate for each 
dust species. The latter depends on the gas mass-loss rate, the surface chemical 
composition, and the fraction of gas condensed into dust 
(see Sect. 5.2 in Ferrarotti \& Gail 2006).

\subsection{The evolution of the spectral energy distribution}
\label{sedmod}
On the observational side, the most relevant effect of the presence of dust 
in the surroundings of
the stars is the reprocessing of the radiation released from the
photosphere, which is intercepted and re-processed by
dust grains, so that the whole SED is shifted to the IR part of
the spectrum. The larger the amount of dust formed, the longer
the wavelength at which the SED peaks. The SED will be also characterised
by the presence of specific emission or absorption features associated
to the formation of some dust species, such as the $11.3~\mu$m feature,
due to the formation of SiC, or the $9.7~\mu$m and $18~\mu$m features,
favoured by the presence of silicates.

To model the evolution of the SED we use the DUSTY code \citep{nenkova99},
which produces the SED of the star, based on the dust mineralogy, 
the optical depth at a specific wavelength, the effective temperature and 
the luminosity: all these information are obtained by the results from stellar
evolution + dust formation modelling, described in sections \ref{aton} and
\ref{dustmod}, applied to different evolutionary phases of the model stars
considered. The input radiation from the photosphere of carbon stars
is found by interpolation in effective temperature, surface gravity and C/O 
ratios among the COMARCS model atmospheres \citep{aringer09} of the appropriate 
metallicity. For what concerns oxygen-rich stars, we interpolated among
the NEXTGEN tables \citep{nextgen}.

\subsection{Population synthesis}
\label{popsyn}
The stellar evolution and dust formation modelling described
earlier in this section are used in a population synthesis approach, 
aimed at deriving the expected population of AGB stars nowadays 
evolving in M31, according to the star formation history given below. 
The results obtained will be tested by comparing the distributions of 
the sources by G22 and of the stars in the synthetic sample in the 
$\rm (F110W-F160W,F160W)$ and $\rm ([3.6]-[4.5],[4.5])$ colour-magnitude 
planes. The analysis in the HST plane will allow a statistical analysis, 
to characterize the individual sources in terms of the mass, chemical 
composition and formation epoch of the progenitor stars; on the other 
hand, the Spitzer plane will be used to investigate dust production by 
AGB stars evolving nowadays in M31.

We consider star formation starting from 14 Gyr ago until now, and
extract for each epoch considered a number of stars, based on the
SFH given in \citet{williams17} and, for ages below 400 Myr, 
\citet{lewis15}. We weight the number of stars extracted assuming a 
Kroupa's initial mass function \citep{kroupa01}. The metallicity adopted at each time is
chosen according to the age-metallicity relationship published in
\citet{williams17}. The number of AGB stars considered is derived on the
basis of the evolutionary time scales of the model stars of different
mass and on the duration of the AGB phase. The corresponding colours
and magnitudes are given by the results from the computations of the
synthetic SEDs, as described in section \ref{sedmod}. The foreground 
extinction of E(B-V)=0.062 \citep{schlegel98} is taken into account 
in the model . 

We note that the present investigation is applied to the disk of
M31 as a whole, for which we use the star formation history and the
age-metallicity relation given above. This choice prevents a
detailed discussion of the effects of any spatial metallicity
gradient across the galaxy, such as that investigated in detail 
by \citet{gregersen15}, and of the role of metallicity on the
formation of carbon stars, addressed in \citet{boyer19},
where 21 sub-fields of M31 were considered. On the other hand, 
this is the only possibility consistent not only with the applicability
of the description of the past history of the galaxy given by \citet{williams17},
but also of the synthetic approach itself, which demands the use of 
a large statistics, thus high number samples.

\begin{figure*}
  \centering
  \begin{minipage}{0.33\textwidth}
    \centering
    \resizebox{1.\hsize}{!}{\includegraphics[trim=0.2cm 0cm 0.6cm 1cm, clip]{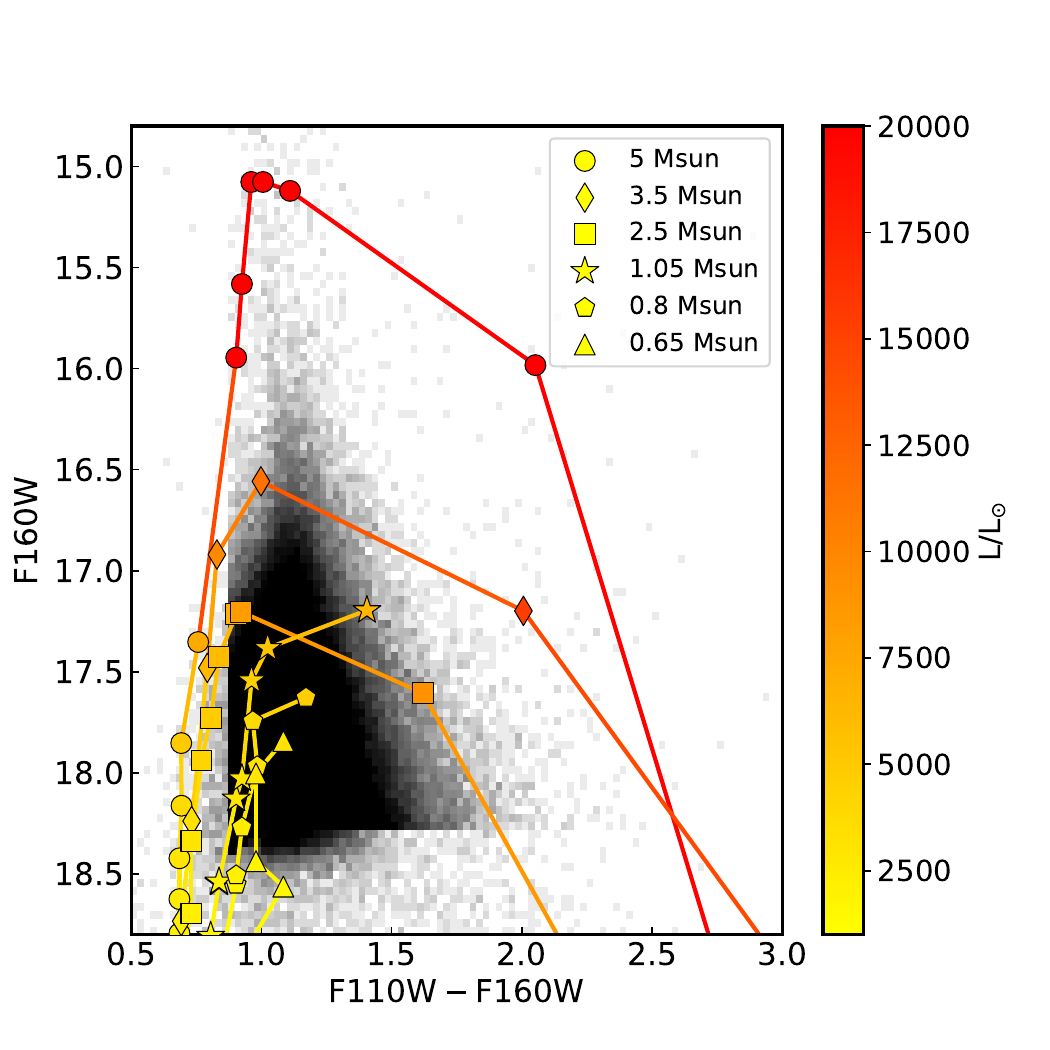}}
  \end{minipage}
  \hfill
  \begin{minipage}{0.33\textwidth}
    \centering
    \resizebox{1.\hsize}{!}{\includegraphics[trim=0.2cm 0cm 0.6cm 1cm, clip]{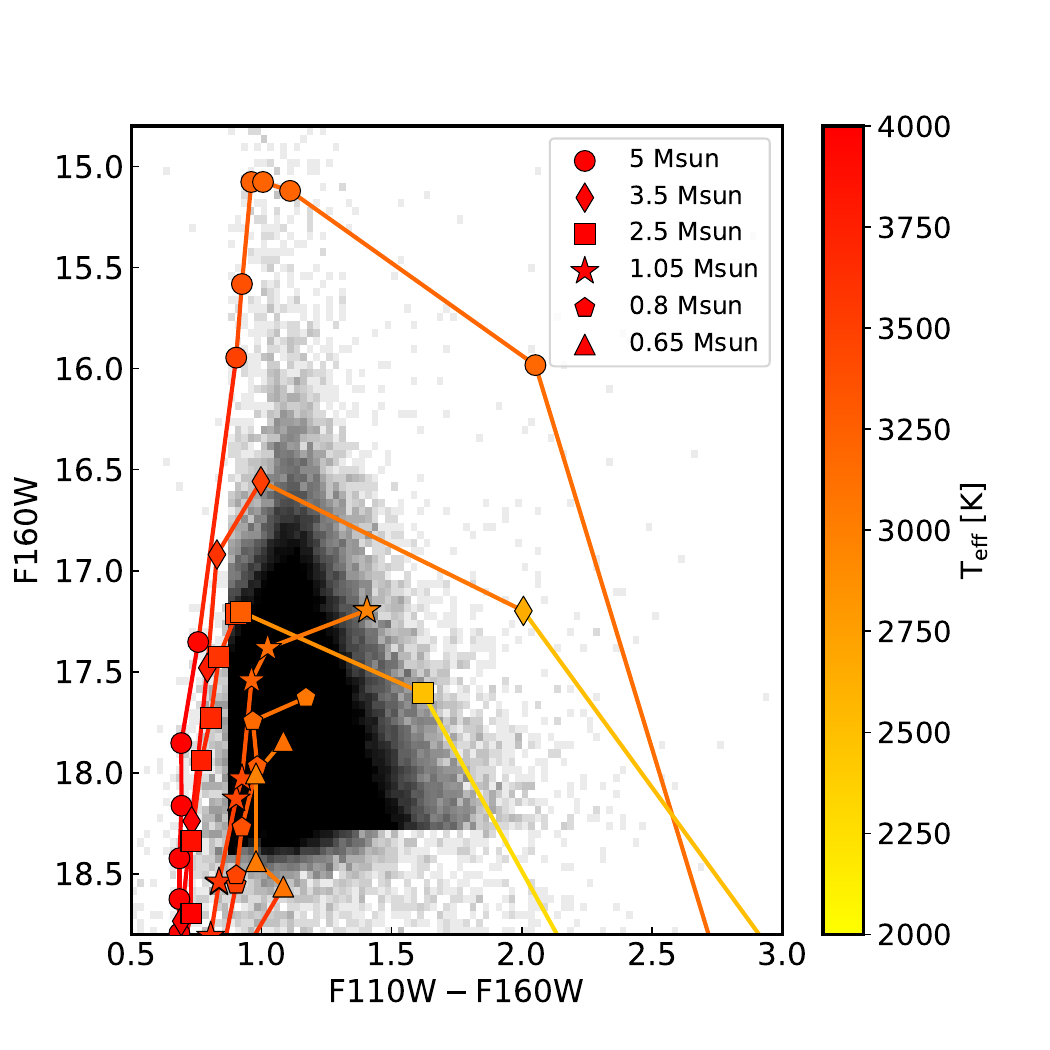}}
  \end{minipage}
  \hfill
  \begin{minipage}{0.33\textwidth}
    \centering
    \resizebox{1.\hsize}{!}{\includegraphics[trim=0.2cm 0cm 0.6cm 1cm, clip]{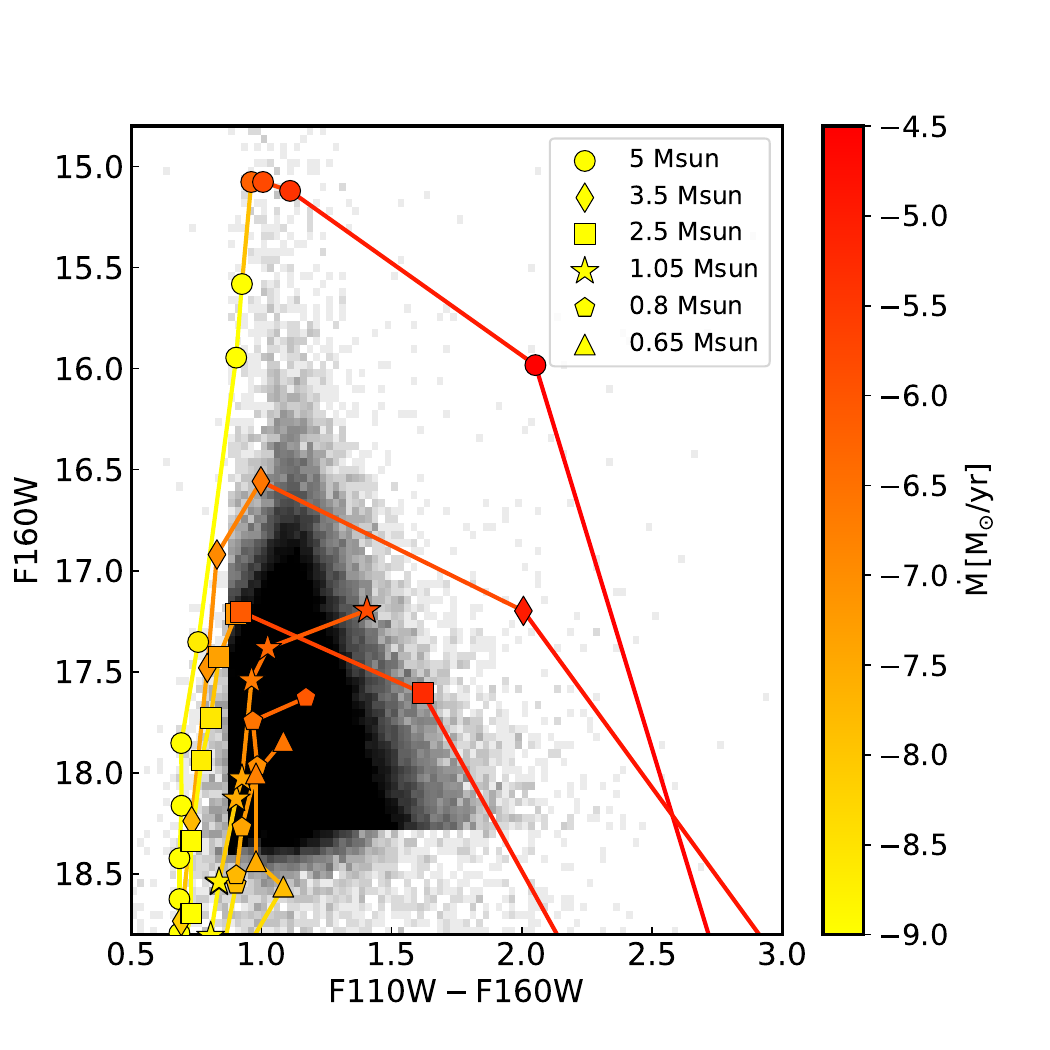}}
  \end{minipage}
  \vskip10pt
  \begin{minipage}{0.33\textwidth}
    \centering
    \resizebox{1.\hsize}{!}{\includegraphics[trim=0.2cm 0cm 0.6cm 1cm, clip]{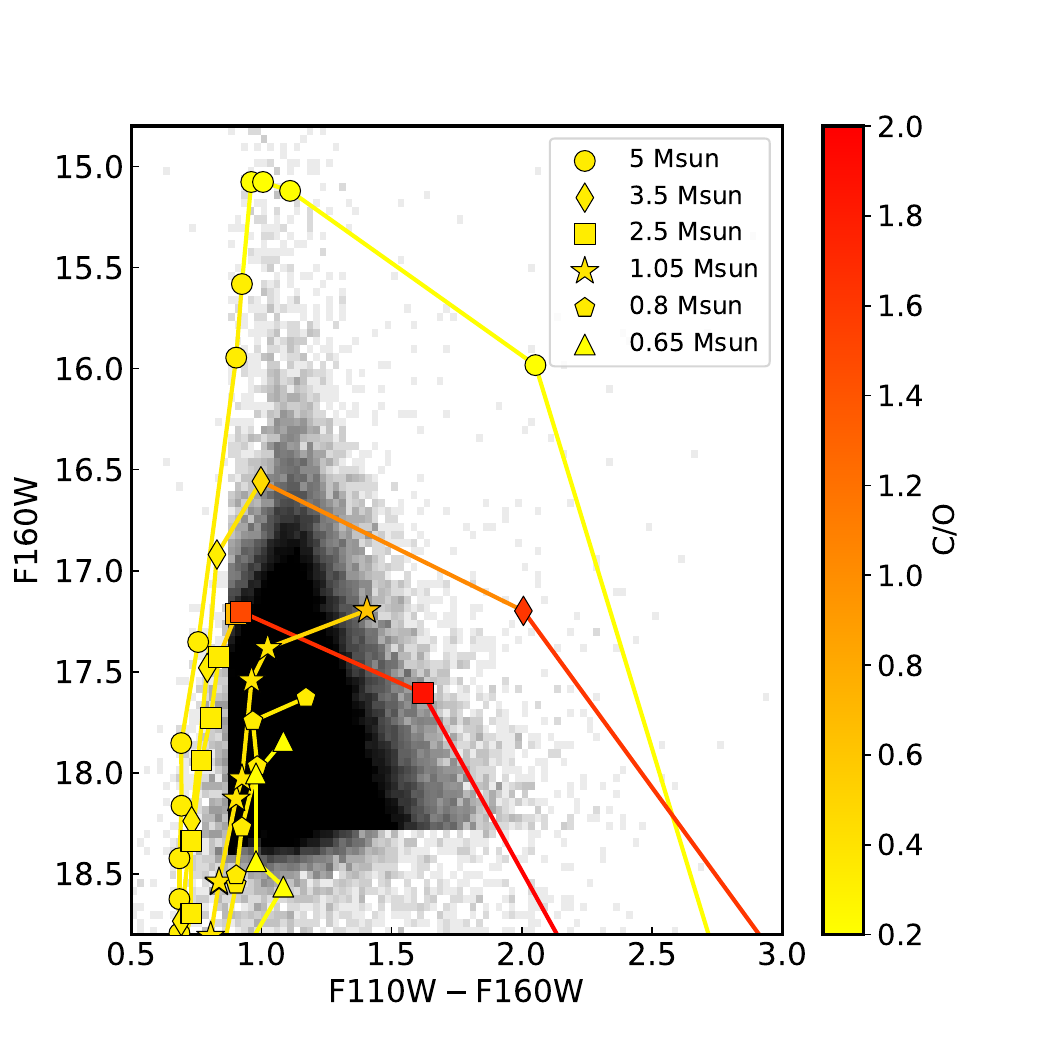}}
  \end{minipage}
  \hfill
  \begin{minipage}{0.33\textwidth}
    \centering
    \resizebox{1.\hsize}{!}{\includegraphics[trim=0.2cm 0cm 0.6cm 1cm, clip]{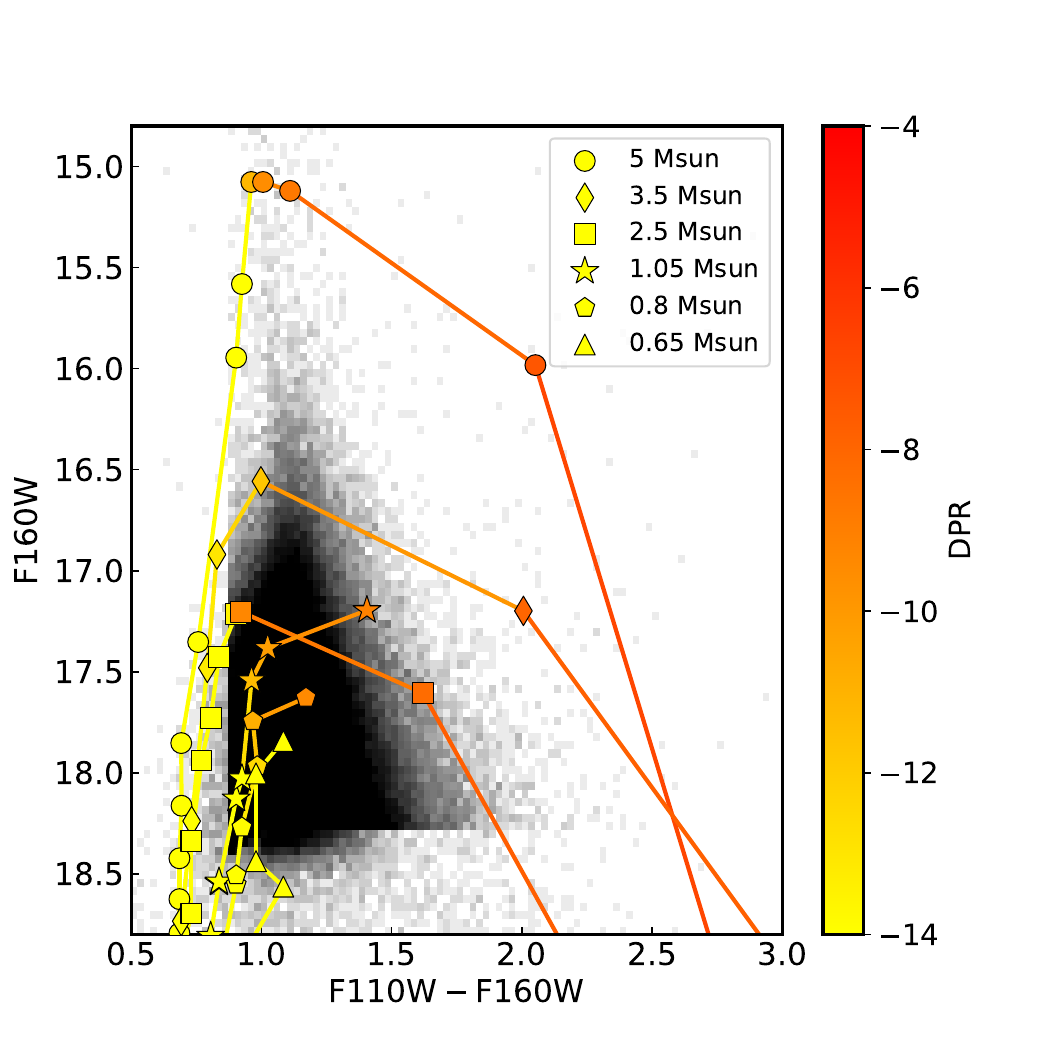}}
  \end{minipage}
  \hfill
  \begin{minipage}{0.33\textwidth}
    \centering
    \resizebox{1.\hsize}{!}{\includegraphics[trim=0.2cm 0cm 0.6cm 1cm, clip]{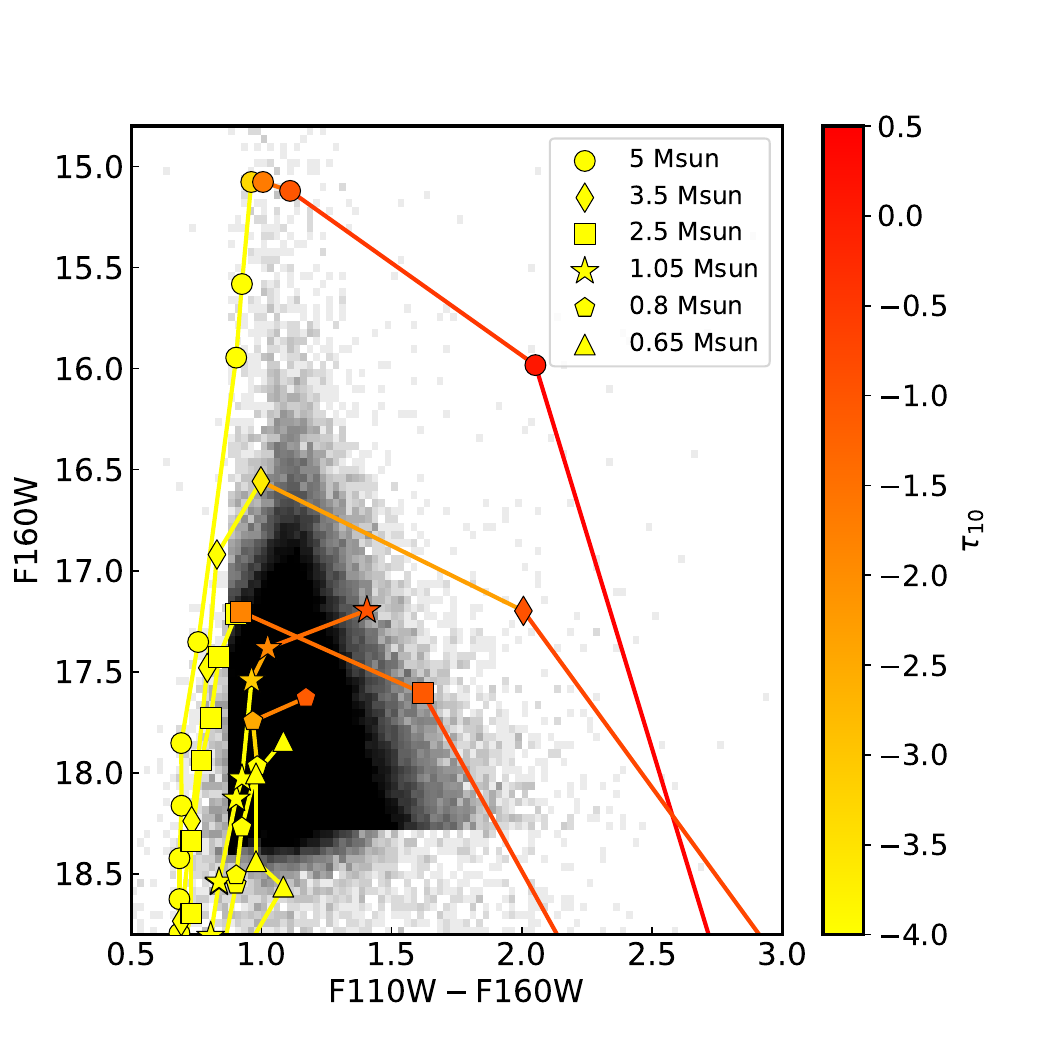}}
  \end{minipage}
  \vskip-10pt
  \caption{Data points in the catalogue of M31 AGB stars by G22 are
  shown with grey points on the colour-magnitude $\rm (F110W-F160W, F160W)$
  plane. Solid lines represent the evolutionary tracks of model stars
  of different mass and metallicity, connecting points corresponding to
  evolutionary stages taken in the middle of each inter-pulse phase.
  The different lines refer to the evolution of the 
  $\rm M=0.65~M_{\odot}$ star of metallicity $\rm Z=10^{-3}$ (triangles), 
  and to the solar metallicity model stars of mass $\rm M=0.8~M_{\odot}$
  (pentagons) $\rm M=1.05~M_{\odot}$ (stars), $\rm M=2.5~M_{\odot}$ (squares), $\rm M=3.5~M_{\odot}$ (diamonds), $\rm M=5~M_{\odot}$ (circles).   The different panels provide information on the luminosity of the   model stars (top, left), effective temperature (top, middle),
  (gas) mass loss rate (top, right), surface $\rm C/O$ ratio (bottom, left), dust production rate (bottom, middle), optical depth at $10~\mu$m
  (bottom, right).
  } 
  \label{fall}
\end{figure*}

\section{AGB star sample in M31}
\label{data}
The analysis developed in the present investigation is based on the
interpretation of the data set of M31 evolved stars published in
G22. The catalogue by G22 is based on archival HST data, matched
with the sources in the Spitzer catalogue, to obtain a collection of
near-IR photometry for the evolved stellar population of the
galaxy.

To identify AGB candidates, G22 adopted the same criteria extensively 
used by \citet{boyer19}, with the scope of excluding blue supergiants and
main sequence stars from the list of sources considered. The selection relies
on specific colour and magnitude cuts in the near-IR HST bands, according to 
the following choices:

\begin{itemize}
\item{$\rm F110W<19.28$ mag, or $\rm F160W<18.28$ mag}
\item{$\rm F110W-F160W>0.88$ mag}
\item{$\rm F814W-F160W>2.4$ mag}
\end{itemize}

To prevent the exclusion of dusty AGB stars from the sample, G22 added
stars to the catalogue that would be ruled out on the basis of the
criteria given above, in case that their Spitzer $[3.6]-[4.5]$ colours
are consistent with those expected for dusty AGB stars. More specifically,
the stars with $([3.6]-[4.5])>0.5$ mag and $[4.5] < 16.4$ mag are included in 
the catalogue, independently of their near-IR properties. This 
made 78 sources excluded on the basis of the near-IR criteria to be
added to the catalogue.

The overall catalogue of the part of the M31 disk analysed by 
G22 consists in 346623 sources with available near-IR HST photometry. 
For 56843 of these objects the Spitzer photometry is also available.

The HST data are complete down to 4 magnitudes below the tip of the
red giant branch (located at $\rm F160W \sim 18.2$ mag,  
$\rm [3.6] \sim 18.4$ mag by G22), thus can be used for statistical analysis. 
Regarding the Spitzer photometry, G22 estimate $\sim 90\%$ completeness
down to $[3.6]=15.6$ mag, and that significant incompleteness affects
the data with $[3.6]$ fainter than $\sim 18.5$ mag.

The distribution of the stars in the catalogue by G22 in the colour-magnitude
$\rm (F110W-F160W, F160W)$ plane is reported in Fig.~\ref{fall}. Overlapped to
the data points, shown with grey points, we show the evolutionary tracks
of some selected model stars of various mass and chemical composition.
We restrict the attention on the $\rm F160W < 18.8$ mag region, to
better appreciate the morphology of the evolutionary tracks of model stars
of different mass, particularly in the low mass domain. This
choice rules out the faintest sources, which, however, account for $\sim$
0.4 $\%$ only of the overall sample. As discussed in section \ref{popsyn},
these stars, surrounded by significant amounts of dust, characterised by extremely
low fluxes in the $1-1.5 ~ \mu$m spectral region, will be investigated 
in detail on the basis of their position on the $\rm ([3.6]-[4.5], [3.6]$)
plane.
The six panels of the figure give information on different physical and
chemical properties of the model stars, and on the dust production efficiency.

\section{The role of convection and mass loss}
\label{uncer}
Before entering the characterization of the evolved stellar population of M31,
we discuss the three most relevant uncertainties affecting the predictive power of
the results obtained from AGB modelling: the treatment of convection,
the mass lost by low-mass stars during the RGB phase, and the mass-loss rates
experienced by massive AGB stars descending from $\rm M \geq 4~M_{\odot}$
progenitors. Convection modelling affects the thermodynamic stratification 
of the outer regions of the star, the RGB mass loss determines the mass with 
which the stars evolve through the core helium burning and the early-AGB phases,
the mass-loss treatment reflects into the AGB lifetimes of massive AGBs.

\subsection{Convective modelling}
The thermodynamic stratification of the outermost regions of giant stars 
is characterized by the presence of convective currents, and of an overadiabaticity 
peak in the sub-photospheric 
regions \citep{salaris12}, whose height, which turns extremely sensitive to the 
convective model adopted \citep{salaris18}, has a deep impact on the effective 
temperature of the star and, as far as dust-free objects are considered, on the
colours. This does not hold when the stars are surrounded by large amounts of dust,
since in that case the radiation released from the photosphere is reprocessed
by dust particles, thus the shape of the SED is practically independent of the effective 
temperature.

\begin{figure*}
\vskip-40pt
\begin{minipage}{0.46\textwidth}
\resizebox{1.\hsize}{!}{\includegraphics{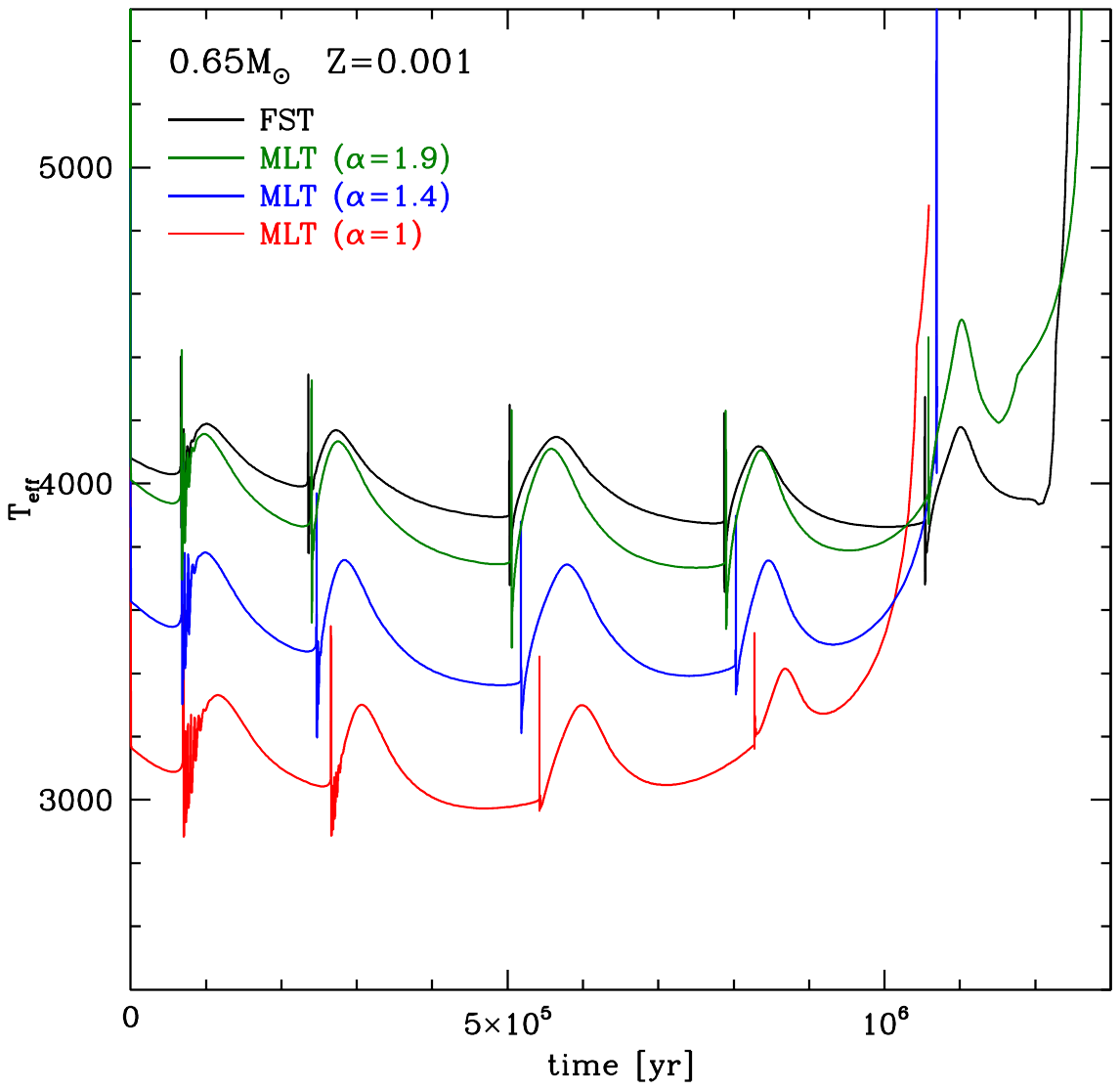}}
\end{minipage}
\begin{minipage}{0.46\textwidth}
\resizebox{1.\hsize}{!}{\includegraphics{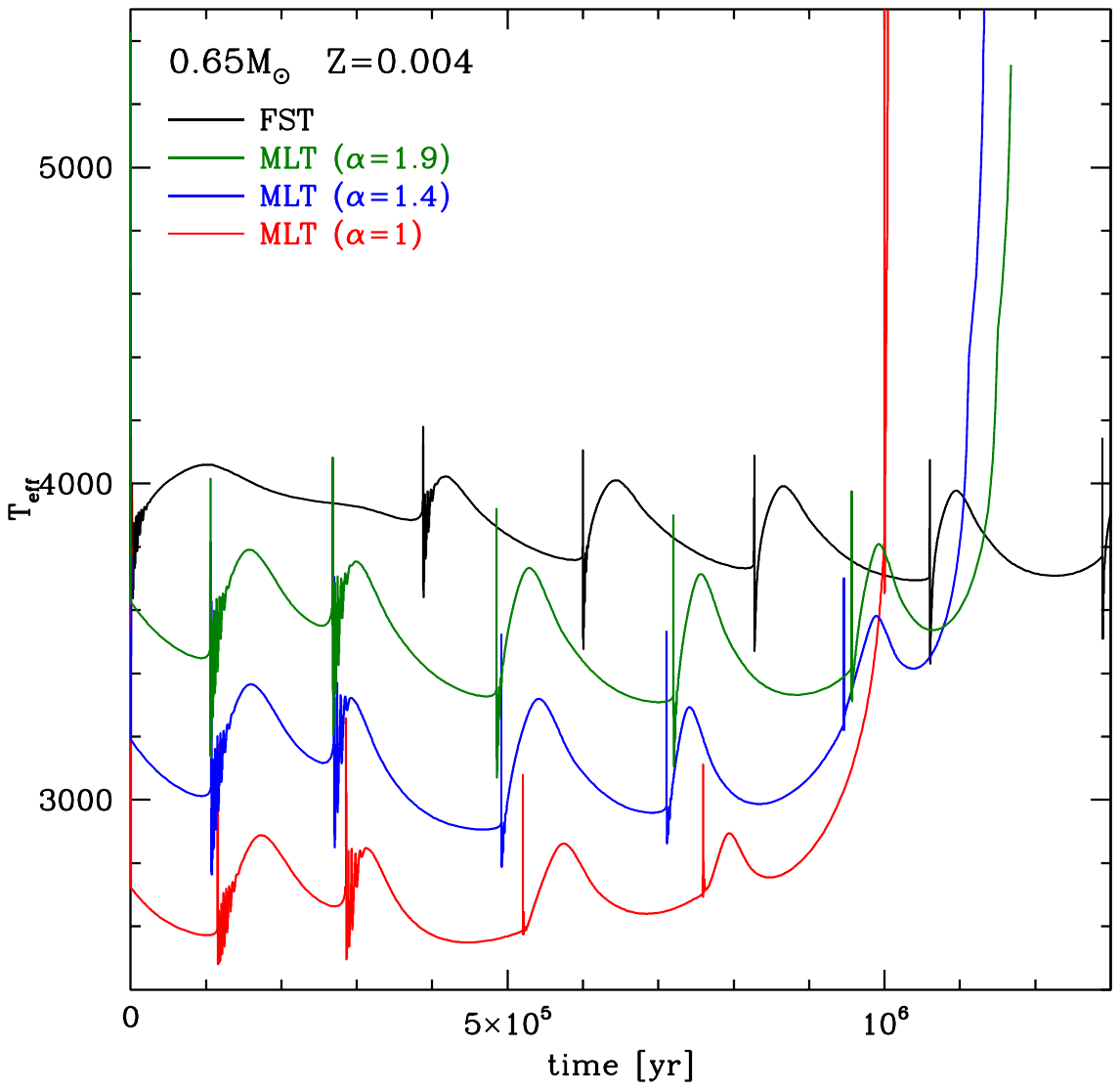}}
\end{minipage}
\vskip-90pt
\begin{minipage}{0.46\textwidth}
\resizebox{1.\hsize}{!}{\includegraphics{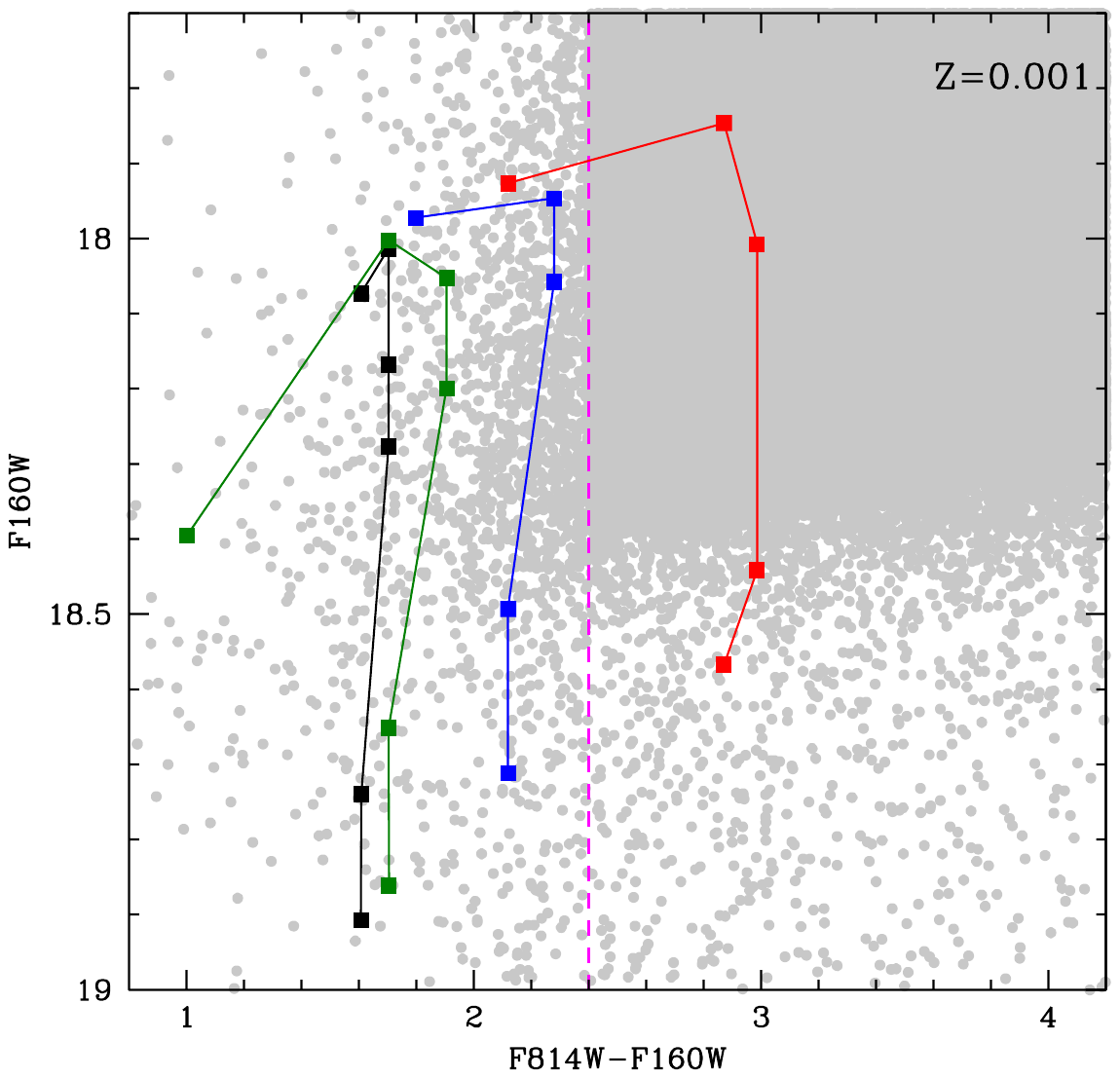}}
\end{minipage}
\begin{minipage}{0.46\textwidth}
\resizebox{1.\hsize}{!}{\includegraphics{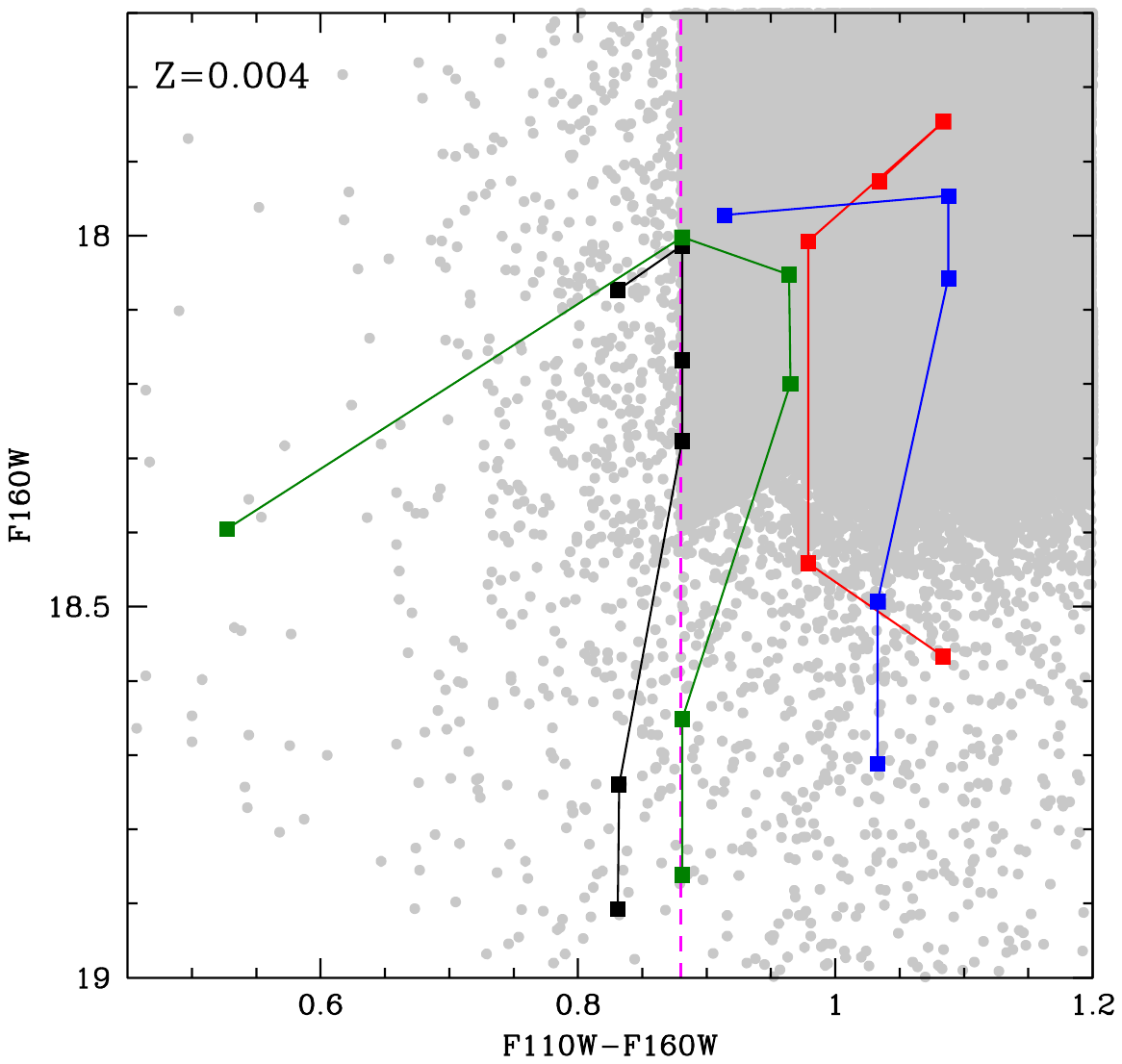}}
\end{minipage}
\vskip-60pt
\caption{Time variation of the effective temperatures of $\rm 0.65~M_{\odot}$
model stars of metallicity $\rm Z=0.001$ (top, left panel) and $\rm Z=0.004$
(top, right), calculated with different convective models. The corresponding
evolutionary tracks of the $\rm Z=0.001$ model star on the $\rm (F814W-F160W, F160W)$ and
$\rm (F110W-F160W, F160W)$ planes are shown in the bottom, left and
bottom, right panels, respectively. The points along the tracks indicate the
middle of the different inter-pulse phases. The dashed, magenta lines mark the G22 cuts.
Data points in the catalogue of M31 AGB stars by G22 are
shown with grey points
} 
\label{fconv}
\end{figure*} 

To investigate the impact of convection modelling on the results obtained
in the present work, we run simulations of model stars of various mass
and metallicity, based on different convection modelling. We mainly focus
on low-mass stars, because we will see that their progeny is the dominant 
AGB population of M31, thus the modelling of their evolution has a significant
effect on the statistical analysis of the observations. Furthermore, 
the values of the effective
temperature of these objects will affect the IR colours for the whole AGB lifetime:
indeed these stars produce little or no dust (see the results regarding the
$\rm 0.65~M_{\odot}$ and $\rm 0.8~M_{\odot}$ in the bottom, middle and right, panels
of Fig.~\ref{fall}), therefore the shape of the SED is not altered by dust reprocessing.

The top-left panel of Fig.~\ref{fconv} shows the time evolution of the
effective temperature $\rm T_{eff}$ of the metal-poor, model
star of $\rm 0.65~M_{\odot}$ (this is the mass of the star at the beginning 
of the core helium burning
phase), whose evolutionary track is represented by the line connecting the
triangles in Fig.~\ref{fall}. The results obtained with 
the FST description of convection are compared with those based on the MLT 
treatment, where different values of the free parameter $\alpha$ were
used. 

The $\rm T_{eff}$ of the different models is
approximately constant during the inter-pulse phases for
the whole AGB lifetime, with the exception of the very final evolutionary phases, when the contraction of the stellar structure favours the heating of the external regions. We note a $\sim 1000$ K difference between the two extreme cases based on the FST description ($\rm T_{eff} \sim 4000$ K) and on the MLT, $\alpha=1$ model ($\rm T_{eff} \sim 3000$ K); the $\alpha=1.4, 1.9$, MLT model stars follow an intermediate behaviour. The treatment of the convective 
instability also affects the overall duration of the AGB phase, because the cooler 
the effective temperature, the lower the surface gravity, the higher the rate at
which mass loss takes place: the time differences among the different
models are within $20\%$. 

The bottom panels of Fig.~\ref{fconv} show the evolutionary tracks of the
different $\rm 0.65~M_{\odot}$, $Z=10^{-3}$ models reported in the top, left panel on the
colour-magnitude $\rm (F814W-F160W, F160W)$ and $\rm (F110W-F160W, F160W)$ 
planes. We consider these two planes because the criteria adopted 
by G22 to select AGB stars in M31, described in section \ref{data}, are based
on cuts on the $\rm (F814W-F160W)$ and $\rm (F110W-F160W)$ colours, indicated with 
grey, vertical lines in the bottom panels of Fig.~\ref{fconv}. 

The evolutionary tracks on these planes are nearly vertical, because the effective 
temperatures are approximately constant during the AGB phase and the dust production 
rates are extremely small (see bottom, middle panel of Fig.~\ref{fall}). 
From the results shown in the top, left panel of Fig.~\ref{fall} we deduce that
the increase in the $\rm F160W$ flux is due to the rise in the luminosity, determined 
by the gradual increase in the core mass, in turn related to the nuclear
activity in the CNO burning shell. The turn to the blue of the final part of the 
evolutionary tracks indicates the start of the post-AGB phase. The uncertainty in
the determination of $\rm (F814W-F160W)$ associated to convection modelling is $\sim 1.2$ 
mag, while for $\rm (F110W-F160W)$ it is $\sim 0.2$ mag. The treatment of convection
also affects the largest $\rm F160W$ fluxes reached during the AGB phase. This is
again due to the differences in the effective temperatures: the cooler models
reaching higher near-IR fluxes for a given luminosity.

The effects of convection modelling on the determination of the effective temperature of
AGB stars changes with metallicity.
As shown in the top, right panel of Fig.~\ref{fconv}, in the $Z=4\times 10^{-3}$ 
case the $\rm T_{eff}$ difference between the FST and the MLT ($\alpha=1$) 
models is $\sim 1400$ K, which reflects into colour differences of
$\sim 2.4$ mag and $\sim 0.3$ mag for $\rm (F814W-F160W)$ and $\rm (F110W-F160W)$,
respectively.

The results obtained here will be discussed in Section 5.1.1 to derive the 
correct convection model to use as function of metallicity.

\begin{figure*}
\vskip-40pt
\begin{minipage}{0.51\textwidth}
\resizebox{1.\hsize}{!}{\includegraphics{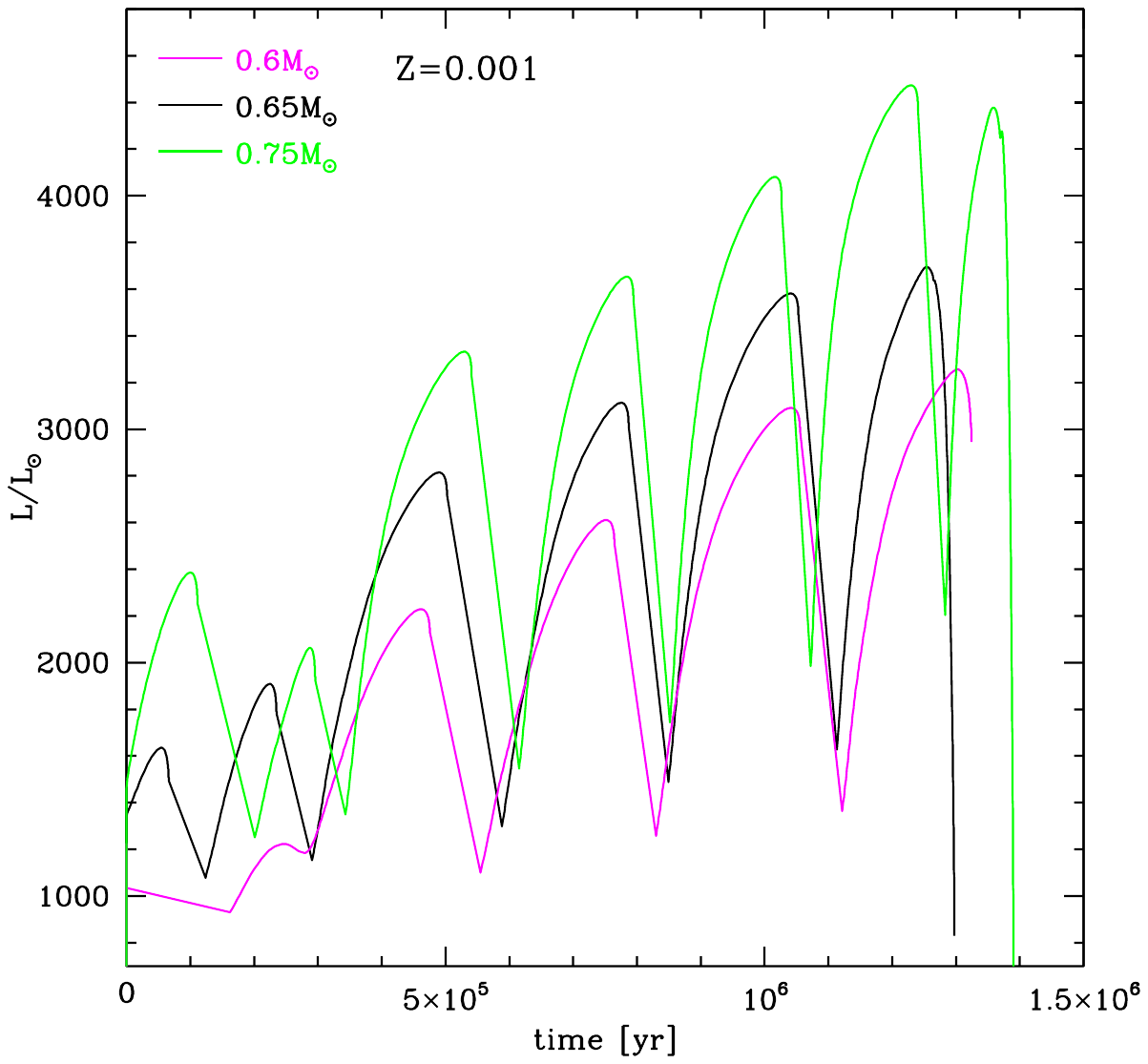}}
\end{minipage}
\begin{minipage}{0.51\textwidth}
\resizebox{1.\hsize}{!}{\includegraphics{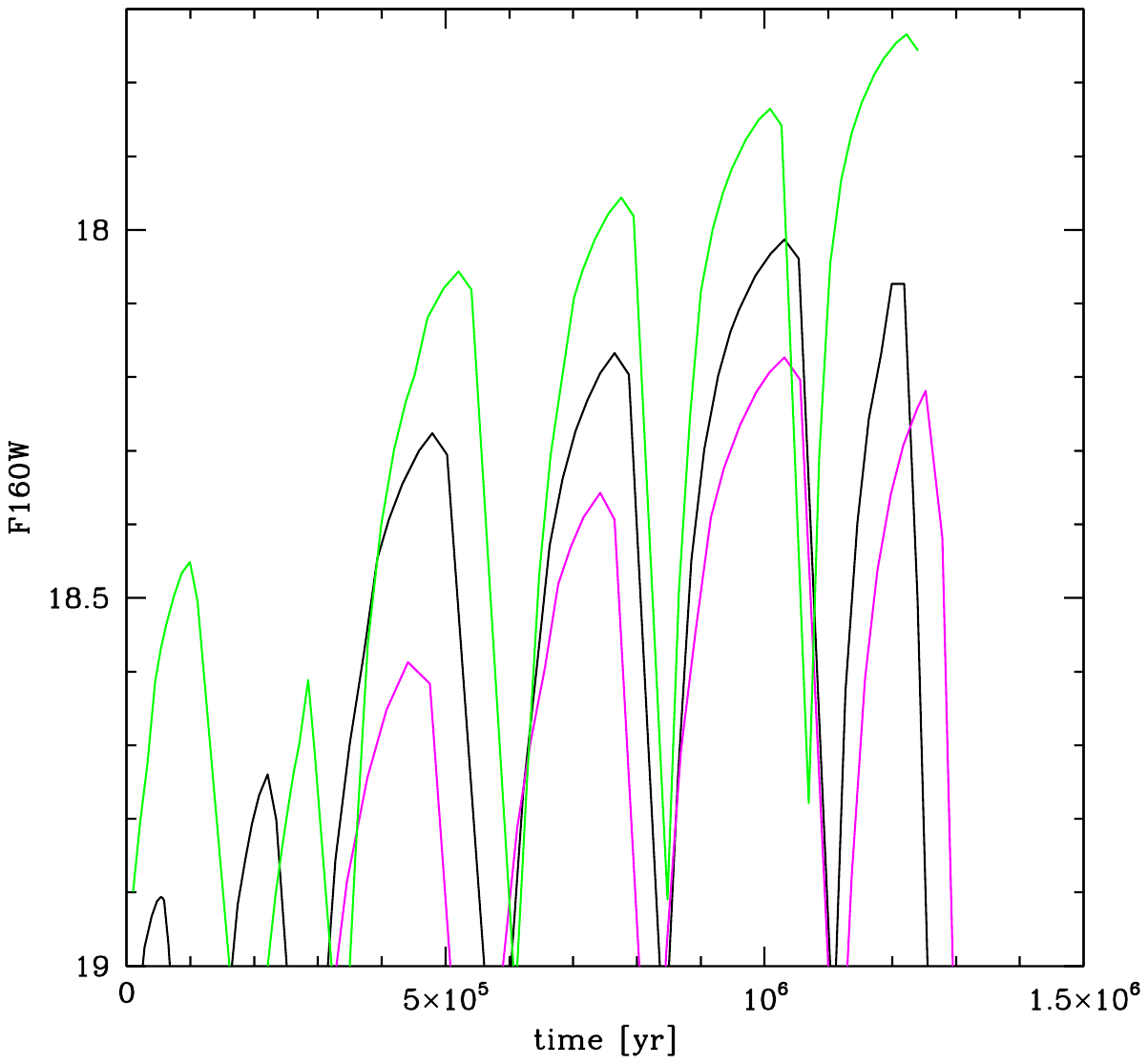}}
\end{minipage}
\vskip-60pt
\caption{Time variation of the luminosity (left panel) and of 
the $\rm F160W$ flux (right) of $Z=0.001$ model stars of initial 
mass (taken at the beginning of the core helium burning phase)
$\rm 0.6~M_{\odot}$ (magenta line), $\rm 0.65~M_{\odot}$ (black),
$\rm 0.75~M_{\odot}$ (green).} 
\label{fmlo}
\end{figure*}

\subsection{Mass loss during the RGB evolution}
\label{mlo}
Despite several recipes have been proposed to model mass loss
during the RGB evolution \citep{reimers75, catelan2000, sch05},
no firm conclusion has been reached yet on this argument,
so that the fraction of the stellar mass lost during the RGB
phase is unknown.
This issue is particularly relevant for $\rm M<1.5~M_{\odot}$ stars,
which lose a significant fraction 
of their mass during this phase. As stated in section \ref{aton}, it is 
necessary to make an assumption for the RGB mass loss, which determines 
the total mass of the star at the beginning of the core helium burning phase. 

To understand how the assumptions regarding the RGB mass loss
affect the results of the AGB evolution, we show in Fig.~\ref{fmlo}
the time variation of the luminosity (left) and of the F160W flux
(right) of model stars that start the core helium burning phase
with mass in the $\rm 0.6-0.75~M_{\odot}$ range. These can be considered
as the progeny of e.g. $\rm 0.8~M_{\odot}$ progenitors, whose mass
loss during the RGB phase was in the $\rm 0.05-0.2~M_{\odot}$ range.

The rise in the luminosity is a common feature of the AGB evolution
of the model stars considered, but the largest luminosities reached
depend on the initial mass: the $\rm 0.6~M_{\odot}$ model star reaches
$\rm \sim 3000~L_{\odot}$, whereas in the $\rm 0.75~M_{\odot}$ case we 
find $\rm \sim 4000~L_{\odot}$. The differences in the luminosity 
reflect into the expected $\rm F160W$ flux, which is about half
magnitude brighter in the $\rm 0.75~M_{\odot}$ model star in comparison
with the $\rm 0.6~M_{\odot}$ case (see right panel of Fig.~\ref{fmlo}).
We will see that these information are crucial to interpret the
observed near-IR luminosity function of the sample described in
section \ref{data}, which will offer the opportunity to discriminate
among the various possibilities regarding the RGB mass loss.

\begin{figure*}
\vskip-40pt
\begin{minipage}{0.32\textwidth}
\resizebox{1.\hsize}{!}{\includegraphics{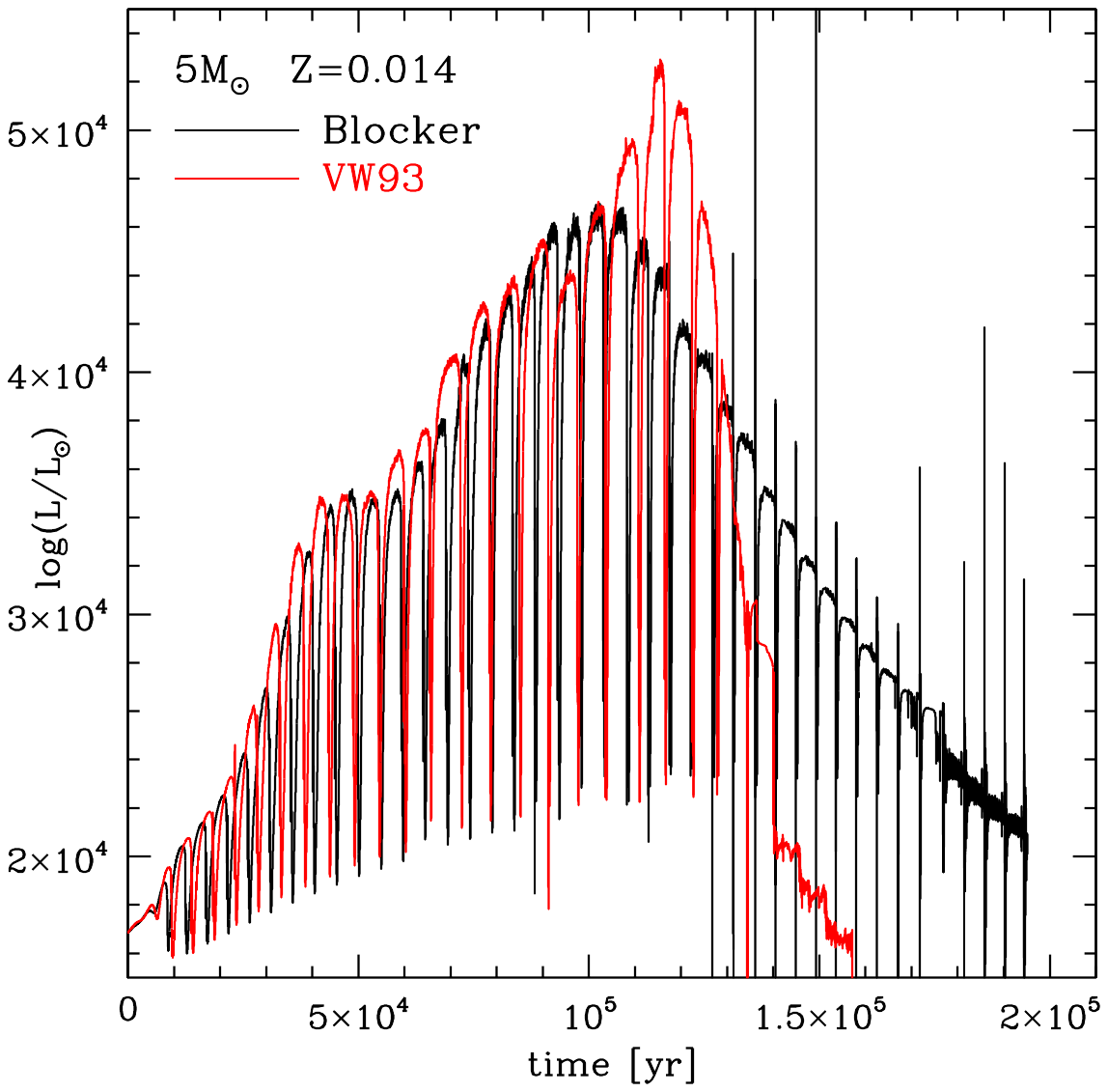}}
\end{minipage}
\begin{minipage}{0.32\textwidth}
\resizebox{1.\hsize}{!}{\includegraphics{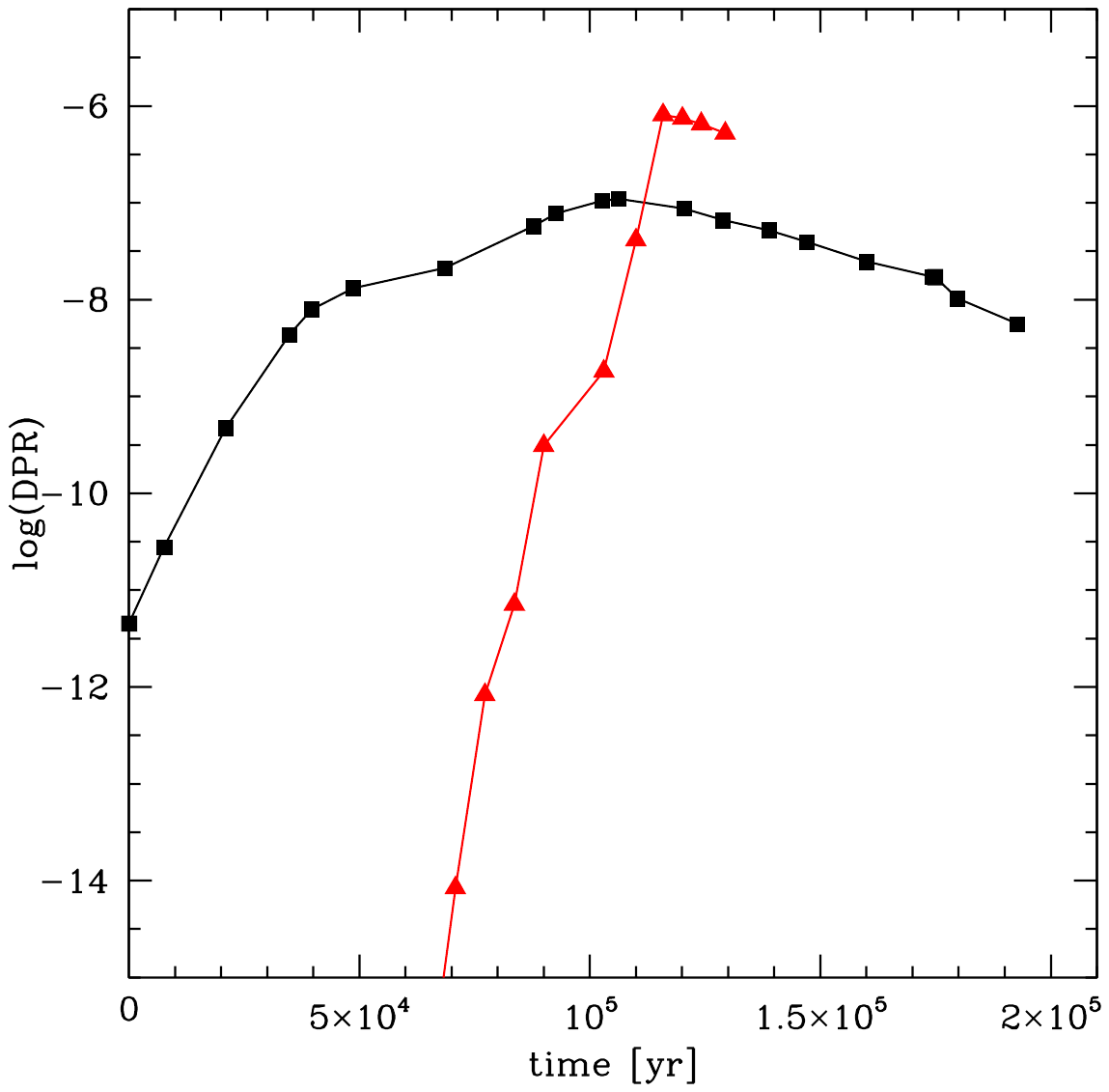}}
\end{minipage}
\begin{minipage}{0.32\textwidth}
\resizebox{1.\hsize}{!}{\includegraphics{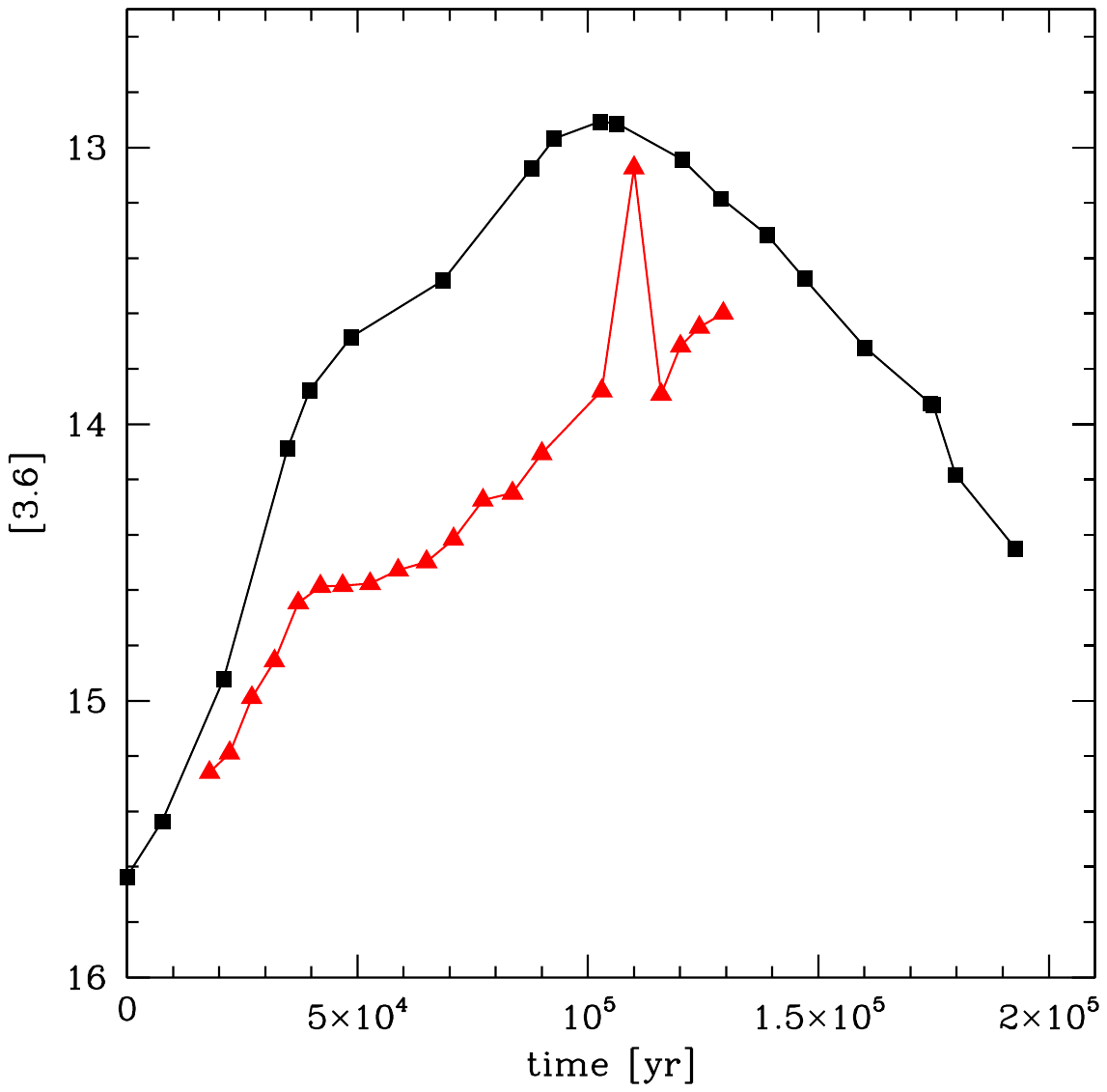}}
\end{minipage}
\vskip-40pt
\caption{AGB time variation of the luminosity (left panel),
dust production rate (middle), $[3.6]$ fluxes (right) of two model 
stars of initial mass $\rm 5~M_{\odot}$ where mass loss was described 
by means of the \citet{blo95} (black squares) and the VW93 treatments
(red triangles). 
} 
\label{fvw1}
\end{figure*}

\begin{figure*}
\vskip-40pt
\begin{minipage}{0.46\textwidth}
\resizebox{1.\hsize}{!}{\includegraphics{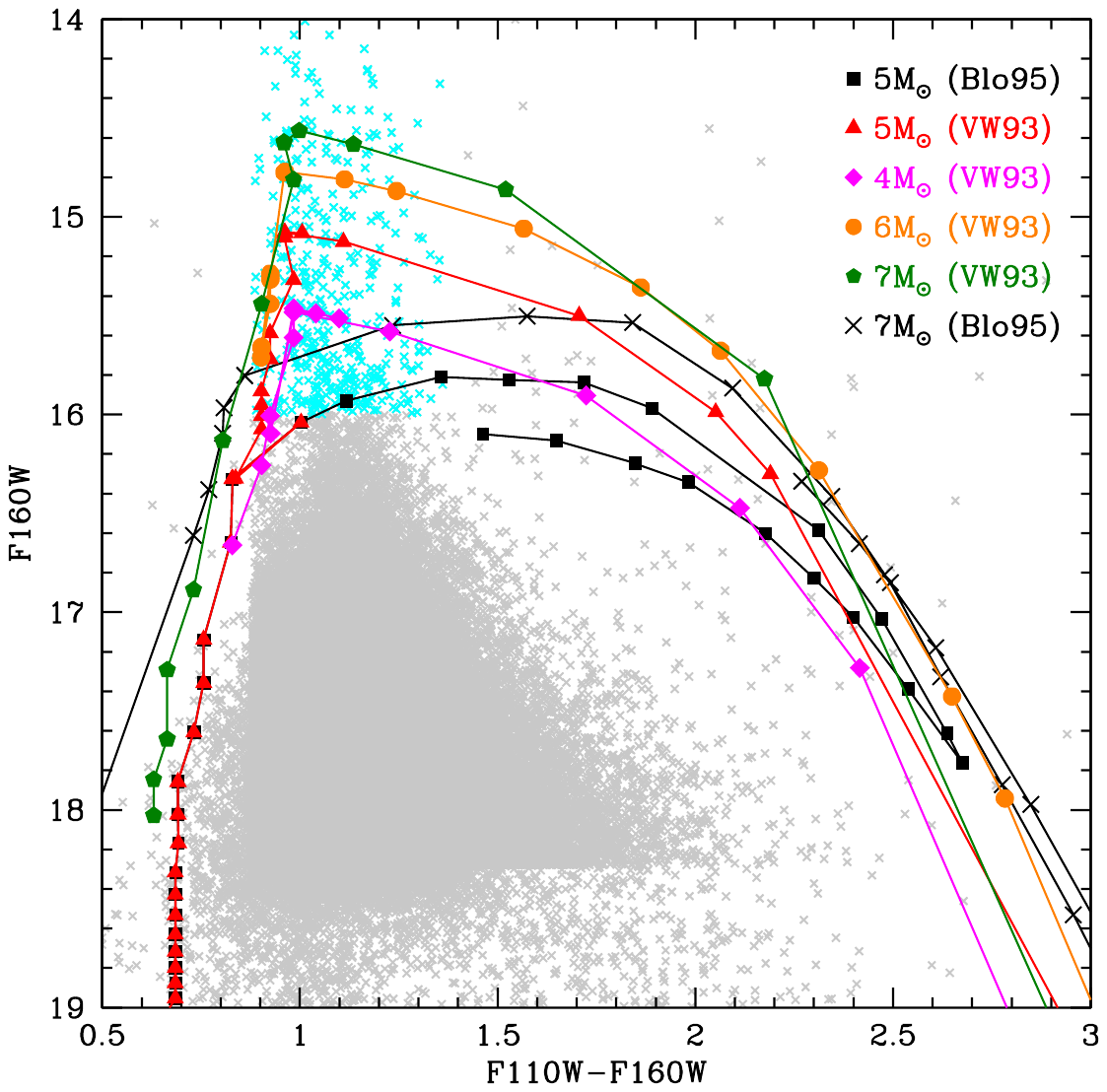}}
\end{minipage}
\begin{minipage}{0.46\textwidth}
\resizebox{1.\hsize}{!}{\includegraphics{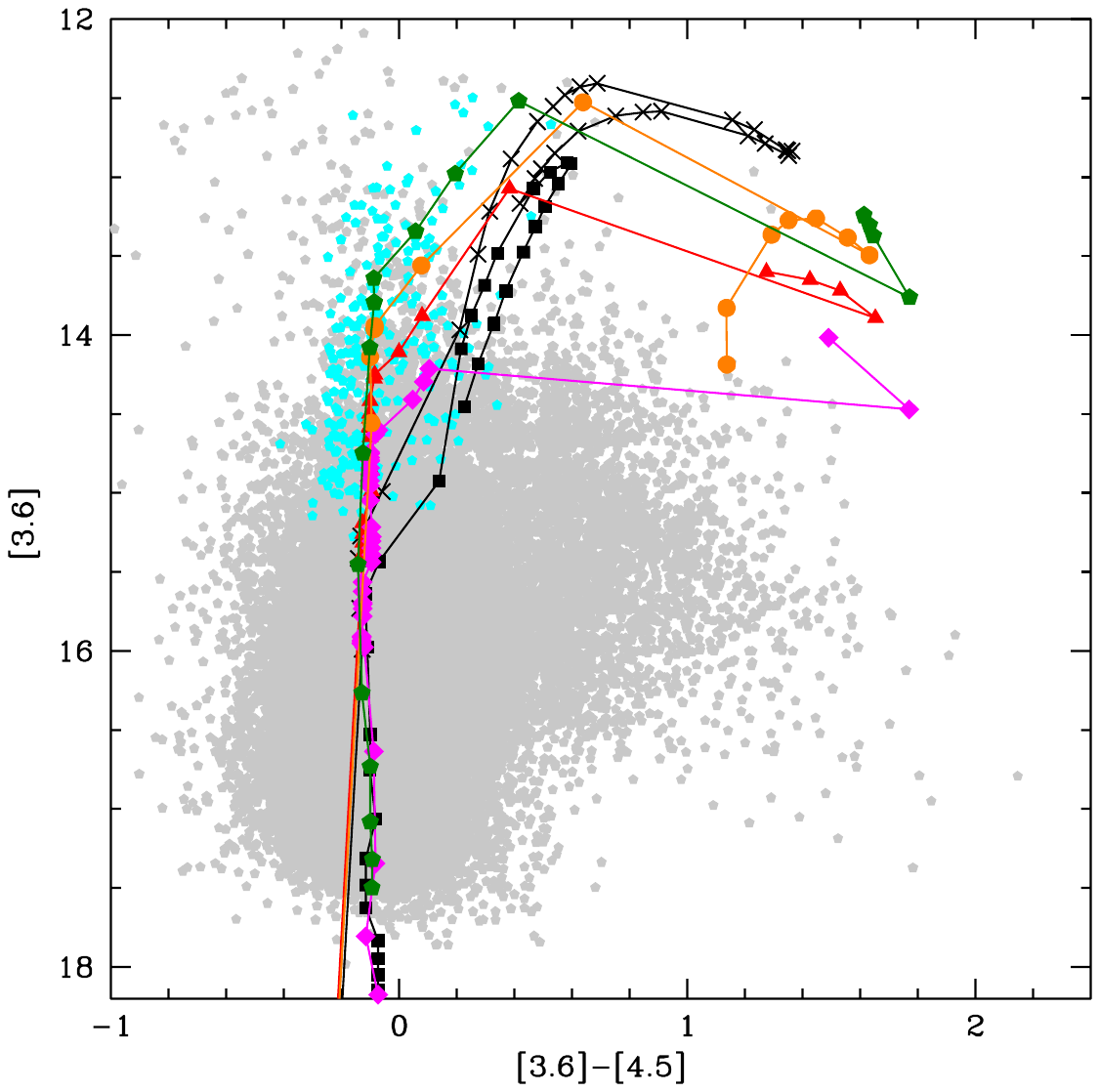}}
\end{minipage}
\vskip-60pt
\caption{Evolutionary tracks of the same $\rm 5~M_{\odot}$ model
stars reported in Fig.~\ref{fvw1} is shown with a black line 
on the $\rm (F110W-F160W, F160W)$ (left panel) and 
$([3.6]-[4.5], [3.6])$ (right panel) planes. The tracks of the corresponding $\rm 5~M_{\odot}$ model
star obtained based on the VW93 treatment of mass loss is indicated
with red triangles. Magenta diamonds, orange circles and green pentagons
refer to the evolution, respectively, of $\rm 4~M_{\odot}$, $\rm 6~M_{\odot}$
and $\rm 7~M_{\odot}$ stars, where mass loss was described following VW93. 
Black crosses refer to the evolution of a $\rm 7~M_{\odot}$ star, obtained
with the \citet{blo95} description of mass loss.
Grey points indicate the data by G22. Bright sources with 
$\rm F160W < 16$ mag are shown with cyan points.
} 
\label{fvw2}
\end{figure*}

\subsection{Mass loss rate of massive AGBs}
\label{ester}
The stars of initial mass above $\rm \sim 3.5~M_{\odot}$ are characterized 
by the ignition of proton-capture nuclear activity at the base of the
convective envelope \citep{blo91}, a phenomenon known as hot
bottom burning (HBB). 
The high luminosities associated to the ignition of HBB \citep{blo91} 
are inevitably accompanied by a considerable 
increase in mass loss rates, the treatment of mass loss has a deep
influence on the physical evolution of these stars. This is due
in particular to the strong impact of mass loss, which eventually
turns HBB off, thus triggering the decrease in the luminosity of
the star after the phase during which HBB is strongest.

The two most popular recipes adopted are those from \citet{blo95} 
and VW93. The former, based on the hydrodynamical studies by \citet{bowen}, 
predicts a tight relationship between mass loss rate and luminosity, 
with $\rm \dot M \propto L^{3.7}$. According to VW93, the luminosity scales 
with the pulsation period according to Eq.~5 in VW93, until reaching the 
super-wind phase, during which the wind is radiation-driven, so that 
$\rm \dot M=\beta L(cv_{exp})$, where $\beta$ represents the average number 
of scattering processes by dust particles experienced by each photon.

To understand the impact of the mass loss description on the evolution
of massive AGBs we show in Fig.~\ref{fvw1} the differences between the
results obtained in the modelling of a $\rm 5~M_{\odot}$ model star,
when the \citet{blo95} and the VW93 treatments are adopted. The comparison 
between the time variations of the luminosities (left panel of the figure)
shows a clar similarity during the first half of the AGB lifetime, 
as the evolution is mainly driven
by the growth of the core mass, which affects the strength of HBB. 
The luminosity of the VW93 model grows bigger than in the \citet{blo95} case, because
the VW93 mass loss rates are generally smaller, thus the envelope
consumption is slower. This situation changes during the very final
AGB phases, when the super-wind phase begins, and the mass loss rates
of the VW93 model star, of the order of $\rm 2\times 10^{-4}~M_{\odot}/$yr,
are significantly higher than in the \citet{blo95} model 
($\rm \sim 7\times 10^{-5}~M_{\odot}/$yr).

The differences in the mass loss treatment also affect the dust
production rate, as can be seen in the middle panel of 
Fig.~\ref{fvw1}. In the \citet{blo95} case the DPR is found
to be above $\rm 5\times 10^{-8}~M_{\odot}/$yr for almost the
totality of the AGB lifetime, whereas in the VW93 model 
dust production is negligible during the largest fraction of
the AGB phase, with the exception of the latest 3 inter-pulses,
when the DPR is around $\rm 10^{-6}~M_{\odot}/$yr.

The evolution of the shape of the SED is also found different in the
two cases: the on the average higher DPR found in the \citet{blo95}
model star makes the SED to be more shifted to the mid-IR, whereas
in the VW93 higher near-IR fluxes are expected. This trend can be
seen in the right panel of Fig.~\ref{fvw1}: the \citet{blo95} 
model star exhibits a larger $[3.6]$ flux (left). 

The evolutionary tracks of the two model stars presented in Fig.~\ref{fvw1} 
in the $\rm (F110W-F160W, F160W)$ and $([3.6]-[4.5], [3.6])$ planes 
are shown in Fig.~\ref{fvw2}. 
In the $\rm (F110W-F160W, F160W)$ diagram (left panel) we note that in 
the \citet{blo95} model star the track moves to the red when 
$\rm F160W \sim 16.3$ mag is reached,
as dust production favours the shift of the SED to the near-IR spectral region.
On the other hand in the VW93 case dust production is delayed (see middle panel
of Fig.~\ref{fvw1}), thus the star evolves to higher luminosities (left panel of
Fig.~\ref{fvw1}) and $\rm F160W$ fluxes, so that the track develops 
vertically until $\rm F160W \sim 15$ mag, before turning to the red. 

Moving to the $([3.6]-[4.5], [3.6])$ plane, we note that the track of
the \citet{blo95} model star is generally redder than in the VW93 case,
but for the last 3 inter-pulse phases, during which the dust production
by the VW93 model star is highest (see the last 4 triangles in the middle
panel of Fig.~\ref{fvw1}). The distributions of massive AGB stars on this plane
obtained via population synthesis in the two cases are very different: when
mass loss is modelled according to \citet{blo95}, we expect a consistent
group of stars in the 0.2 mag $\rm<([3.6]-[4.5])<0.6$ mag, 13 mag $\rm<[3.6]<15$ mag region of the
plane, whereas in the VW93 case the number of stars characterised by
the afore-mentioned $[3.6]$ and $[4.5]$ magnitudes would be
 much smaller.

\section{The characterization of the evolved stars of M31}
\label{char}
According to \citet{williams17}, the star formation history of M31 is 
characterized by an early phase of intense activity, started $\sim 14$ Gyr ago, 
extending over $\sim 6$ Gyr, with rates of star formation of the order of 
$\rm 8~M_{\odot}/yr$. It is during this phase that most of the stars of M31 formed, 
with metallicities ranging from $\rm [Fe/H]\sim -1$ to solar\footnote{
 The results reported in Fig.~8 of \citet{williams17} indicate a large
uncertainty in the estimated chemical composition of the oldest stars of M31,
the metallicity spanning the $\rm -2 <[Fe/H] < -0.5$ range. We assumed an average 
value of $\rm [Fe/H] \sim -1$ (corresponding to $Z=0.001$) to model this old population 
of the galaxy. We note that this corresponds to the low metallicity tail of the
$\rm [Fe/H]$ distribution of M31 RGB stars derived by \citet{gregersen15}.}. Based on the
evolutionary time scales of the model stars introduced in section \ref{aton}, 
we conclude that the stars formed during this period that are
nowadays evolving through the AGB descend from $\rm \sim 0.8~M_{\odot}$ stars with
$\rm [Fe/H]\sim -1$, $\rm 0.85-0.9~M_{\odot}$ stars with sub-solar chemical 
composition, and $\rm 1-1.2~M_{\odot}$ stars of solar metallicity. 

A later intense episode of star formation, characterised by star formation
rates of the order of $\rm 5~M_{\odot}/yr$, took place in M31 between 1 and
2 Gyr ago, in an environment characterised by solar or nearly solar chemical
composition. The stars formed during this second episode of star formation that are
nowadays undergoing the AGB evolution are the progeny of
$\rm 1.7-2~M_{\odot}$ stars.
Some star formation in M31 occurred also in more recent epochs \citep{lewis15}, 
which reflects into the presence of massive AGBs (initial masses $\rm 4-5~M_{\odot}$) 
in the current, evolved stellar population.

To characterize the evolved stars in M31 we use the population synthesis 
method described in section \ref{popsyn}, based on the afore-discussed SFH of 
M31 and the results from stellar evolution + dust formation modelling
(sections \ref{aton} and \ref{dustmod}). We compare the theoretical distribution of 
the stars on the $\rm (F110W-F160W,F160W)$ and $\rm ([3.6]-[4.5],[3.6])$ planes, 
as derived from population synthesis, with the dataset by G22. In building the
synthetic sample of stars we took into account the colour and magnitude cuts
described in section \ref{data}. From this analysis we aim at characterising 
the individual sources of M31, in terms of mass, formation epoch and 
chemical composition of the progenitor stars.

We will first consider the $\rm (F110W-F160W, F160W)$ diagram, to 
undertake a detailed statistical analysis, which is made possible by the 
arguments presented in section \ref{data}, regarding the completeness
of the data down to the cuts on the $\rm F110W$ and $\rm F160W$ magnitudes 
imposed by G22. This step is important to check for consistency
between the results from stellar evolution + dust formation modelling
and the observations, to draw information regarding the still poorly
known phenomena affecting the modelling of the AGB phase, with
particular attention to those described in section \ref{uncer}, and
finally to reach a thorough description of the evolved stellar population of 
M31, according to how stars of different mass and metallicity distribute 
across the $\rm (F110W-F160W, F160W)$ plane. The requirement from 
this part of the analysis is that once the constrains adopted by G22
described in section \ref{data} are considered, the numerical consistency
of the overall sample, as also the distribution of the stars across the
$\rm (F110W-F160W, F160W)$, is the same as in G22.

The second part of the analysis will be focused on the interpretation of the
distribution of M31 stars on the $\rm ([3.6]-[4.5],[3.6])$ plane obtained
with the Spitzer data. The stars in the G22 sample for which the Spitzer data are available are only a small fraction $(\sim 16.5\%)$ of the global sample, 
which prevents any statistical approach. On the other hand, this is the plane 
to be considered for the study of dust production in M31, because the SED of dusty stars peaks at wavelengths $\lambda > 3~\mu$m, whereas the $\rm F110W$ and $\rm F160W$ fluxes are low or even negligible.
This can be deduced by inspection of Fig.~\ref{fall}, where 
by looking at the tracks of the $\rm 2.5~M_{\odot}$ and $\rm 3.5~M_{\odot}$
model stars we understand that only during the first inter-pulses following the
achievement of the C-star stage the stars can be detected in $\rm F110W$ and $\rm F160W$.
Therefore, part of the stars with very low fluxes in the $1-1.6~\mu$m
region, populating the bottom side of the $\rm (F110W-F160W, F160W)$ 
diagram, are indeed the most relevant sources to study dust production, to be investigated in the $\rm ([3.6]-[4.5],[3.6])$ plane.

\subsection{Understanding the HST data}
The luminosity function (LF) of the sources in the G22 catalogue
is reported in Fig.~\ref{fhisto}: the black line indicates the number 
counts in the different $\rm F160W$ bins. Approximately half of the
stars are located into the 18 mag $\rm<F160W<18.28$ mag strip. The remaining
sources are distributed in the $\rm F160W<18$ mag range: considering
magnitude bins 0.5 mag wide, we find $37\%$ of stars with 17.5 mag $\rm<F160W<18$ mag, $12\%$ with 17 mag $\rm<F160W<17.5$ mag, $\sim 2\%$ with 16.5 mag $\rm<F160W<17$ mag, 
plus a small fraction below $1\%$, at brighter magnitudes.

The large fraction of evolved stars of M31 in the 
faintest $\rm F160W$ bins is related to the
dominant presence of low-mass stars among the AGB population, particularly
those of metal poor or sub-solar metallicity discussed at the beginning
of this section. The fractions of stars in the three fainter $\rm F160W$ bins 
found via the population synthesis approach is extremely sensitive to the
assumptions regarding convection and mass loss used to model the evolution
of low mass stars of different metallicity. In the following, we illustrate
the choices on the input macro-physics that allow to reproduce the
distribution of the stars shown in Fig.~\ref{fhisto}.

\subsubsection{The choice of the convective model}
\label{choiceconv}
In the low-metallicity domain the FST and the solar-calibrated MLT 
($\alpha=1.9$) modelling of convection are not consistent with the 
observations, as the model stars evolve at bluer $\rm (F814W-F160W)$ 
colours than the cut described in section \ref{data}: 
this is shown in the left, bottom panel of Fig.~\ref{fconv}, 
where we see that in the $\rm Z=10^{-3}$ case the MLT description with 
$\alpha=1$ is required. The $\rm Z=4\times 10^{-3}$, low mass model stars
are cooler at the surface than the lower metallicity counterparts (see top 
panels of Fig.~\ref{fconv}), thus the $\alpha=1.4$ choice also leads to 
values of $\rm (F814W-F160W)$ consistent with the cut imposed by G22.
Turning to the solar metallicity, the G22 colour criteria are 
satisfied independently of convection modelling, as far as low-mass stars
is concerned. Some differences are found in regard of the modelling of the
initial part of the AGB evolution of $\rm M \geq 1.5~M_{\odot}$ stars,
which do not meet the G22 colours cuts when the FST description
is used. This is clear in Fig.~\ref{fall}, where we 
see that the during the initial AGB phases the FST evolutionary tracks 
are bluer than the colour cut assumed by G22
(see section \ref{data}), thus they would be ruled out by the G22 selection.

The star counts in the $\rm F160W>18$ mag domain are of little help to
select the convection model most appropriate to describe solar
metallicity stars, as the
statistics is dominated by low-mass stars, with little (if any) 
contribution from $\rm M \geq 1.5~M_{\odot}$ stars. On the other hand the 
star counts at brighter $\rm F160W$ are extremely sensitive to the possible 
presence of $\rm M \geq 1.5~M_{\odot}$ stars, particularly of the progeny 
of the $\rm 1.7-2~M_{\odot}$ stars that formed during the secondary peak 
in the SFH of the galaxy.

If we rely on the MLT modelling, we find that the progeny of 
$\rm 1.7-2~M_{\odot}$ stars would populate the 17.5 mag $\rm<F160W<18$ mag
sample by G22, so that the sources in this region of the plane
would outnumber the fainter counterparts, which in fact account 
for half of the total sample. There is no combination of the
physical ingredients adopted to build the evolutionary sequences
able to repair this inconsistency, unless the stars formed during the
secondary peak in the SFH of M31 are ruled out by the cuts
adopted by G22 during the initial part of the AGB evolution: 
this is what we find when the FST modelling is used, thus we 
will rely on the FST model stars as far as the solar metallicity 
is concerned.

\subsubsection{Selecting the mass loss during the RGB evolution}
\label{choicemlo}
The choice of the amount of mass lost by the stars during the ascending
of the RGB, $\rm \delta M_{RGB}$, is a key point to set the fractions 
of the stars in the different bins. This can be understood based on
the arguments presented in Section \ref{mlo} and the results
shown in Fig.~\ref{fmlo}, where we see that the capability of low-mass
stars to attain specific luminosities or near-IR fluxes is tightly 
connected to the mass at the beginning of the core helium burning
activity, which in turn is sensitive to $\rm \delta M_{RGB}$.

\begin{figure*}
\vskip-40pt
\begin{minipage}{0.46\textwidth}
\resizebox{1.\hsize}{!}{\includegraphics{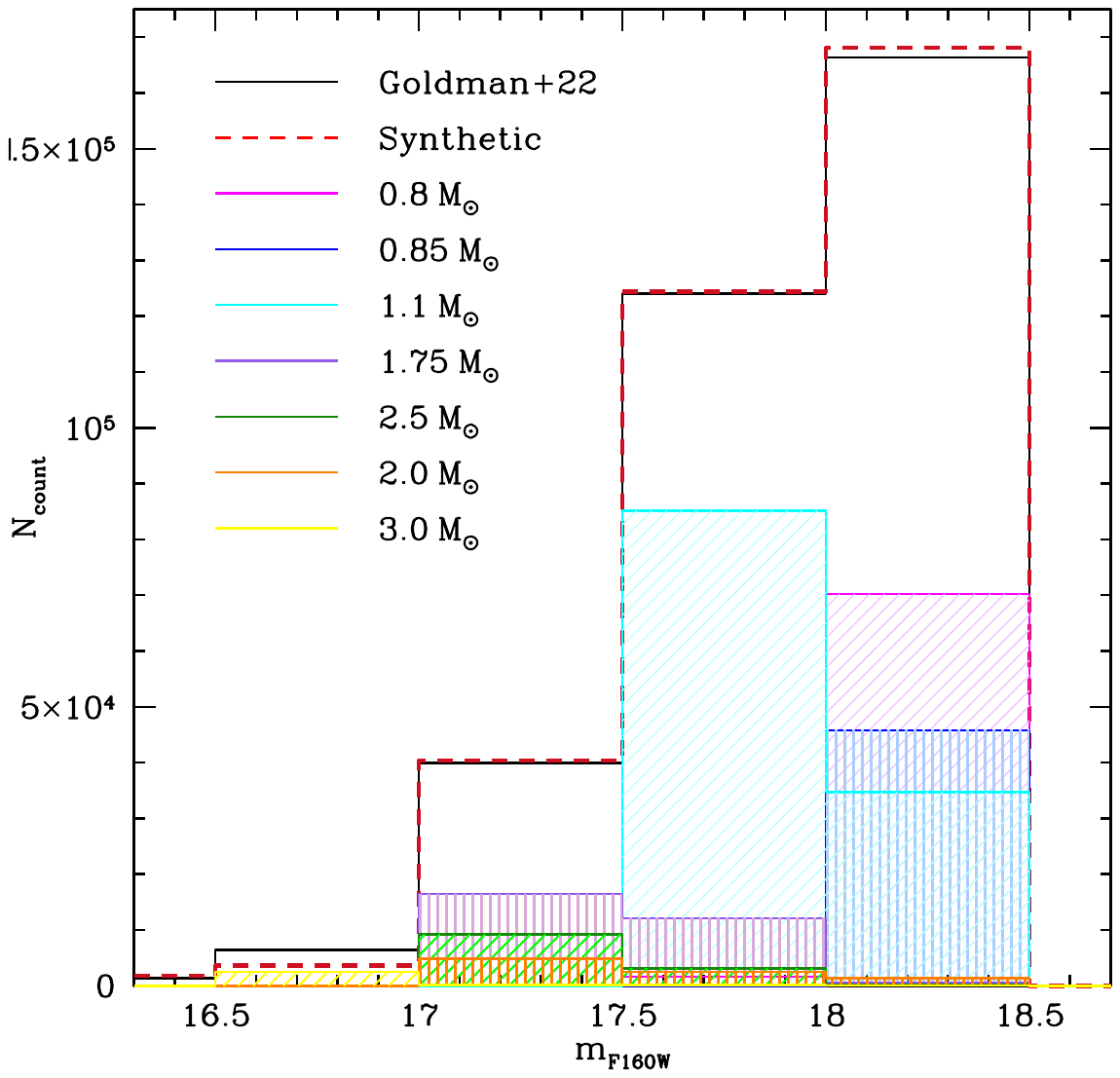}}
\end{minipage}
\begin{minipage}{0.46\textwidth}
\resizebox{1.\hsize}{!}{\includegraphics{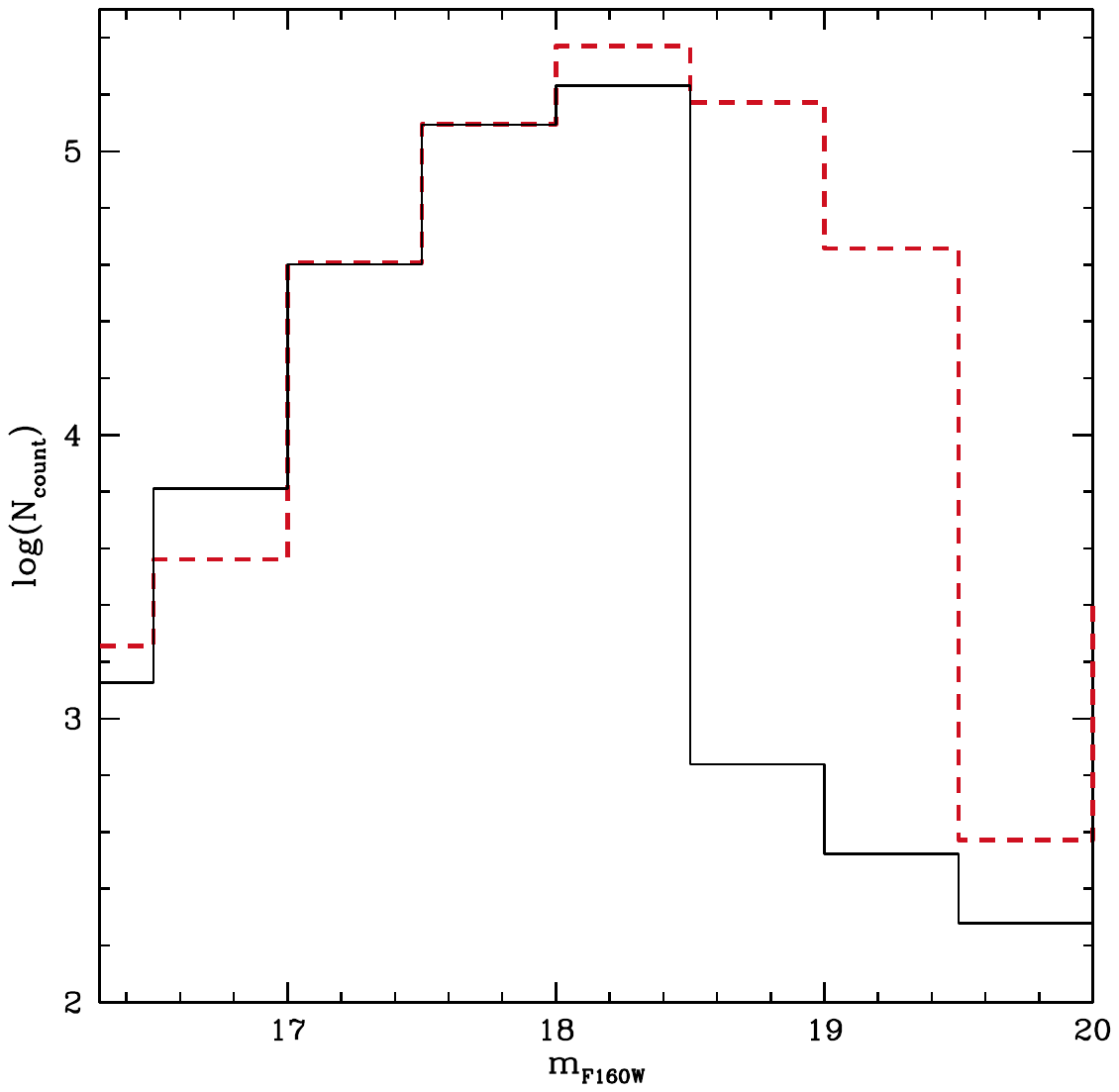}}
\end{minipage}
\vskip-60pt
\caption{Left: The $\rm F160W$ luminosity function of the stars included in the G22 sample, 
indicated with a black line, is compared with the results obtained from population
synthesis, shown with a red, dashed line. The contribution from stars of
different mass in the various $\rm F160W$ are also given. In the legend the initial
masses of the progenitor stars are indicated. Right: the same $\rm F160W$ luminosity function
reported on the left panel, in a logarithmic scale. For readability we show the
total LF only, omitting the contributions from the stars of different mass.
The large discrepancy between the dashed and the solid lines in the
$\rm F160W > 18.5$ domain is due to the cut adopted by G22 in that $\rm F160W$
range, where only the (few) sources identified as extreme AGBs were included.
} 
\label{fhisto}
\end{figure*}

The LF reported in Fig.~\ref{fhisto}, with $87\%$ of the
sources observed at $\rm F160W>17.5$ mag, and half of the sample 
at $\rm F160W>18$ mag, is reproduced if we assume that  
$\rm \delta M_{RGB}= 0.2, 0.25, 0.3~M_{\odot}$ for low-mass stars of
metallicity $\rm Z=10^{-3}$, $\rm Z=4\times 10^{-3}$ and solar,
respectively. This is the only combination allowing to reproduce
the observed percentages. Indeed numerical experiments where the
mass loss experienced during the RGB phase was artificially altered
with respect to the values given above showed the following: 
a) if the mass loss suffered either by
low-mass, $\rm Z=10^{-3}$ stars, or by their $\rm Z=4\times 10^{-3}$
counterparts was $\rm 0.05~M_{\odot}$ higher than the values 
given above, these stars would barely evolve at $\rm F160W<18.28$ mag,
so that the percentage of stars in the faintest $\rm F160W$ bin
would drop to below $30\%$; b) if for the same stars we assumed
a RGB mass loss $\rm 0.05~M_{\odot}$ lower than the values
previously mentioned, these stars would exceed the 
$\rm F160W=18$ mag threshold, and populate the 17.5 mag $\rm<F160W<18$ mag
bin, which would then become the most populated one, accounting for
$\sim 60\%$ of the entire population; c) for what regards
the low-mass, solar metallicity stars of initial mass
$\rm \sim 1.1-1.2~M_{\odot}$, the assumption of a 
$\rm \delta M_{RGB}$ value $\rm 0.05~M_{\odot}$ smaller
than the $\rm 0.3~M_{\odot}$ given above would make
these stars to evolve into the 17 mag $\rm<F160W<17.5$ mag bin,
whose numerical consistency would increase from
$12\%$ to $22\%$.

\subsection{The characterization of the different AGB populations of M31}
\label{nearir}
\begin{figure*}
\centering
\begin{minipage}{1\textwidth}
\resizebox{1.\hsize}{!}{\includegraphics{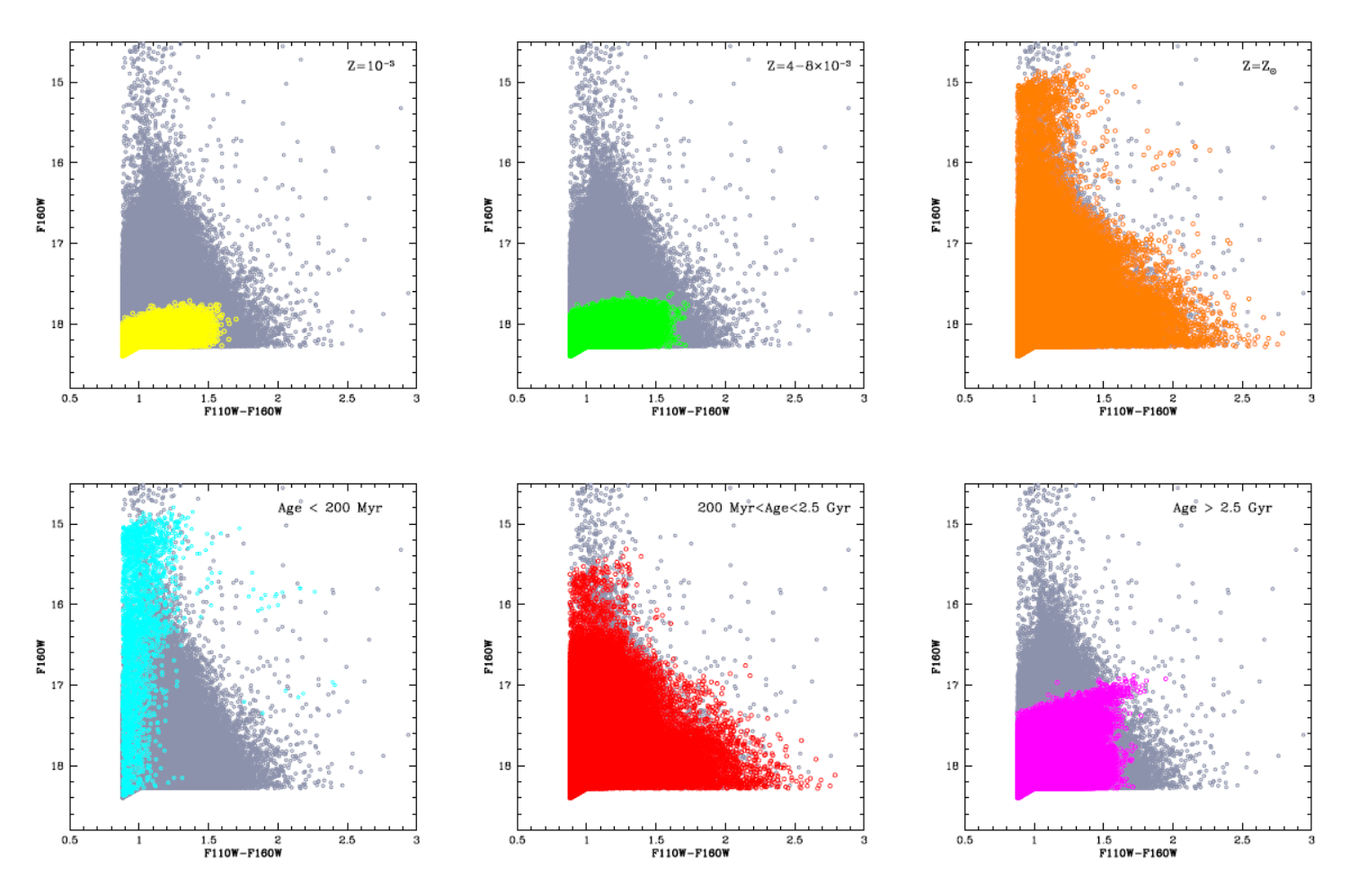}}
\end{minipage}
\begin{minipage}{1\textwidth}
\resizebox{1.\hsize}{!}{\includegraphics{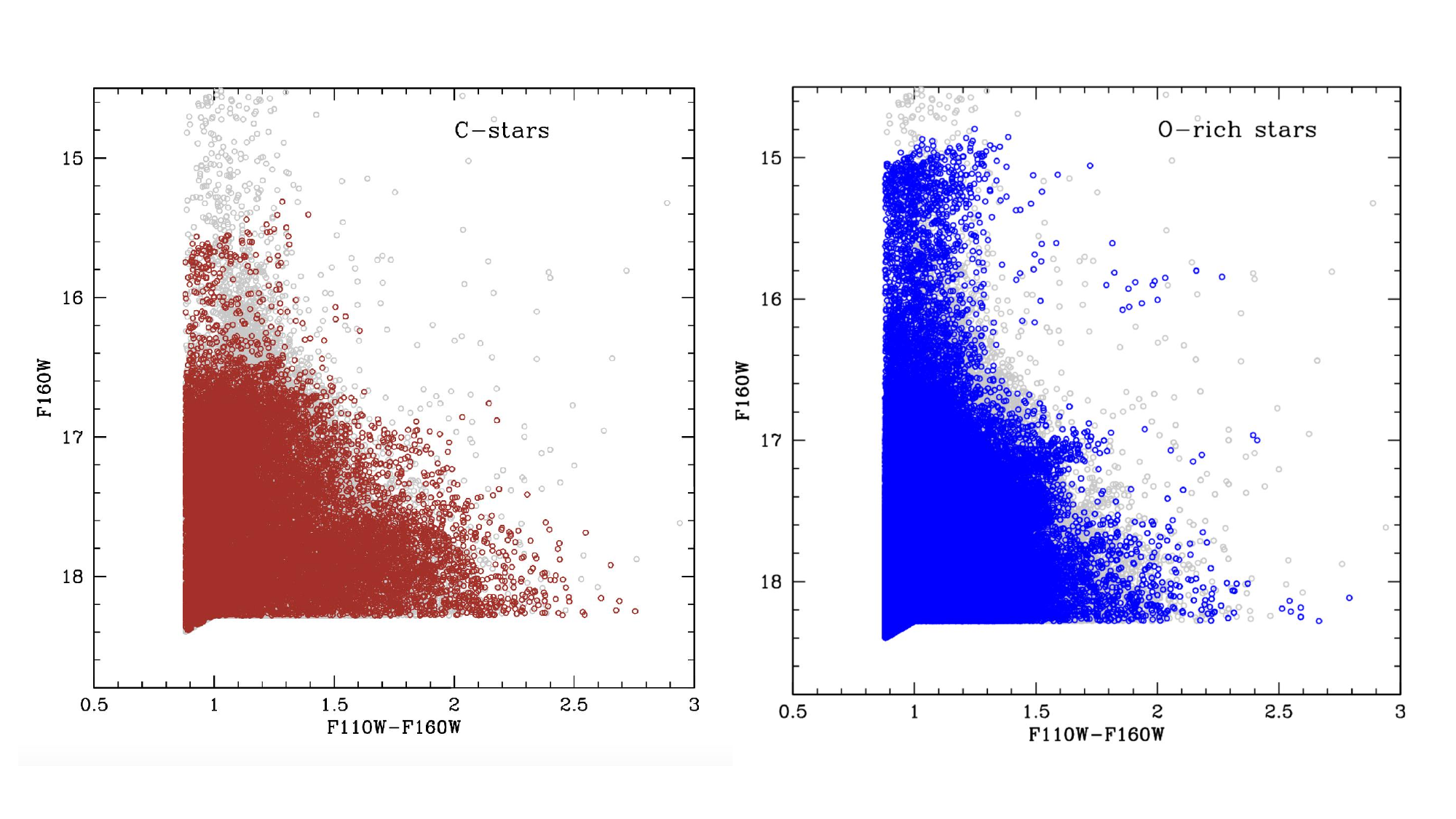}}
\end{minipage}
\vskip+20pt
\caption{AGB candidates from the G22 sample in
M31 are shown with grey points in the colour-magnitude
$\rm (F110W-F160W, F160W)$ diagram. The expected distribution of
metal-poor, sub-solar and solar metallicity stars from the synthetic modelling is indicated 
with different colour-coding in the top, left, middle and
right panels, respectively. The central panels indicate an
age classification, in which the different sources are divided
among young stars formed earlier than 200 Myr ago (left panel),
old objects formed in ancient epochs (right), and intermediate age
stars, which eventually reach the C-star stage (middle). The bottom panels show the distribution of C- and O-rich stars. }
\label{fage}
\end{figure*}

Based on the assumptions described in section \ref{choiceconv} and
\ref{choicemlo} we interpreted the near-IR observations of M31 and 
identified the stars found in the different regions of the 
$\rm (F110W-F160W, F160W)$ plane. An overview of the interpretation 
reached can be seen in Fig.~\ref{fage}: in the top, middle, and 
bottom panels the stars are divided according to the chemical composition, 
the formation epoch, and the C$/$O ratio, respectively.

In the following description we will move across the diagram, starting
from the faintest sources populating the $\rm F160W>18$ mag region.
The results of the analysis done are summarised in Figg. \ref{fhisto}, \ref{fage} 
and in Table \ref{tabnir}, where we report the comparison between the 
fraction of stars (with respect to the total sample) by G22 with that
expected from synthetic modelling, and general information on the mass and
formation epoch of the progenitors of the stars in each bin.


We are not focusing on the dust, as this issue will be discussed
more extensively later in this section, on the basis of the Spitzer 
data, as commented in the initial part of this section.

\subsubsection{The oldest stars}
\label{old}
The faintest stars in the G22 sample formed earlier than 10 Gyr ago. 
In the initial part of section \ref{char} we concluded
that these objects are mostly the progeny of $\rm \sim 0.85~M_{\odot}$ stars of 
sub-solar metallicity ($\sim 42\%$), and of stars descending from low-metallicity, 
$\rm \sim 0.8~M_{\odot}$ progenitors ($\sim 27\%$). The location of these
sources in the $\rm (F110W-F160W, F160W)$ plane is shown in the top, left and
middle, panels of Fig.~\ref{fage}, respectively.
In section \ref{choicemlo} we argued 
that these sources suffered mass loss during the RGB phase, then started core helium 
burning with masses of the order of $\rm 0.6~M_{\odot}$.

The AGB evolution of this class of stars is described in section 4.1 by \citet{ventura22}.
They experience a small number of thermal pulses (of the order of $3-4$), 
and a very few Third Dredge-Up (TDU) events. Therefore, the surface chemical 
composition is almost unchanged during the AGB lifetime, because the only modification 
takes place during the first dredge-up. As shown in
the top, right and bottom middle panels of Fig.~\ref{fall}, the
mass loss rate never exceeds 10$^{-10}$M$_{\odot}/$yr, and little dust is formed in the wind.
The evolutionary tracks of these objects move nearly vertically, thus they are 
concentrated on the bluest, lower side of the region covered by the sample stars. 
In Fig.~\ref{fall} they populate the lowest part of the plane, evolving 
at $\rm F160W>18$ mag for the entire AGB life. 

A significant fraction ($\sim 20\%$) of the sources populating the lower
part of the plane descend from $\rm \sim 1.1-1.2~M_{\odot}$ stars of solar 
metallicity, which formed around 6-9 Gyr ago. Still following the discussion 
presented in sections 
\ref{mlo} and \ref{choicemlo}, we assume that they started the post-flash evolution with masses
of the order of $\rm 0.8~M_{\odot}$.
They experience $\sim 10$ TPs before the envelope is lost, and reach luminosities 
above $\rm 4000~L_{\odot}$ during the final AGB phases. We see in bottom, middle panel 
of Fig.~\ref{fall} that even for these stars dust production is negligible.

The track of a $\rm \sim 0.8~M_{\odot}$ model star is indicated with pentagons
in Fig.~\ref{fall}. Unlike the lower mass counterparts discussed earlier,
these stars evolve at brighter magnitudes during the second part 
of their AGB life, until reaching $\rm F160W \sim 17.6$ mag; this is due in part to the
higher luminosities reached, and also to the cooler effective temperatures (up to 3200\,K), 
which further rise the flux emission in the near-IR spectral region. 

Overall, the distribution of the old AGB stellar population of M31 is the one
shown in the middle, right panel of Fig.~\ref{fage}.
$\sim 5\%$ of the stars in the $\rm F160W > 18$ mag region
of the $\rm (F110W-F160W, F160W)$ plane descend from $\rm 1.5-3~M_{\odot}$ 
progenitors, formed between 0.5 Gyr and 2.5 Gyr ago. These stars,
whose evolution is described in section 4.2 by \citet{ventura22},
became carbon stars after the accumulation of carbon in the surface regions 
(see the bottom, left panel of Fig.~\ref{fall}), under the action of 
repeated TDU events. 
As visible in Fig.~\ref{fall}, indeed the evolutionary tracks
of these stars cross the $\rm F160W > 18$ mag region of
the plane during the first inter-pulses following the achievement of
the C-star stage, when the quantity of dust in the circumstellar
envelope is still sufficiently low that they can still be observed in the 
near-IR. Therefore, the sources considered here have just reached the
C-star stage, and are characterised by surface $\rm C/O$ ratios
in the $1-1.5$ range. This is further confirmed by the 
distribution of C-stars in the bottom left panel of Fig.~\ref{fage}.

During these evolutionary phases, these carbon stars are generally
brighter than the previously discussed lower-mass counterparts
populating the same region of the plane, as their luminosities
are in the $\rm 4000~L_{\odot} < L < 13000~L_{\odot}$ range,
the brightest descending from $\rm \sim 3~M_{\odot}$
progenitors. Nevertheless, they also populate the lower side of the
plane for the following reasons: a) soon after the occurrence of the thermal 
pulse, C-stars experience a minimum in their luminosity evolution for a 
non-negligible time while the CNO shell is not fully reignited; b) the near-IR flux 
is lowered by the dust reprocessing of the radiation emitted by the photosphere, 
which favours the shift of the SED towards the mid-IR. 
Still related to reprocessing by dust particles are the
red colours of these stars, which populate the
$\rm (F110W-F160W)>1.2$ mag region.

We find a tight connection between the optical depth at
$10~\mu$m and the colour: 
$\rm \tau_{10} \sim 0.12+0.1\times [(F110W-F160W)-2]$.
The dust production rates associated with the formation of
carbon dust are below $\rm 2\times 10^{-8}~\dot M/$yr, as
shown in the bottom, middle panel of Fig.~\ref{fall}.

\subsubsection{The solar metallicity population}
\label{sol}
We discussed in section \ref{old} that among the stars present in the largest
quantities in the $\rm F160W>18$ mag region of the $\rm (F110W-F160W, F160W)$ plane, 
only those descending from solar metallicity progenitors of mass slightly higher
than solar (identified with the model star starting core helium burning with
$\rm 0.8~M_{\odot}$ in Fig.~\ref{fall}) evolve to magnitudes brighter than 
$\rm F160W=18$ mag: in the 
17.5 mag $\rm<F160W<18$ mag range they constitute the dominant population, 
accounting for almost $70\%$ of the stars. They reach these
$\rm F160W$ values during the second part of the AGB evolution, 
when the luminosity is in the $\rm 3000~L_{\odot} < L < 4500~L_{\odot}$ 
range (top, left panel of Fig.~\ref{fall}). In this magnitude bin we also 
find stars of sub-solar mass, still with solar chemistry, accounting for 
$\sim 10\%$ of the total population.

These stars evolve as oxygen-rich for the whole AGB phase. The 
dust production, limited to silicates and traces of alumina dust \citep{flavia14}, 
is extremely small, with rates $\rm \dot M_{dust} < 5\times 10^{-10}~M_\odot$/yr,
(see bottom, middle panel of Fig.~\ref{fall}).
They are concentrated on the blue side of the plane,
in the region $\rm (F110W-F160W)<1.2$ mag, as the radiation
released from the photosphere is not reprocessed by
dust particles.

The residual $20\%$ of the sources in the
17.5 mag $\rm<F160W<18$ mag magnitude range descend from
$\rm 1.5-3~M_{\odot}$ progenitors, half of which formed
during the peak in the SFH of M31 discussed in the previous sections.
These stars are known to become C-stars during the second
part of the AGB evolution, thus we expect that about
$10\%$ of the stars in this region of the plane are
C-rich, and evolve on the red side of the diagram, at 
$\rm (F110W-F160W)>1.2$ mag. The current luminosities span the 
interval $\rm 6000~L_{\odot} < L < 13000~L_{\odot}$.

Similarly to the lower luminosity C-rich counterparts
discussed in the previous point, we find that the detected
C-stars are evolving during the early phases following the
achievement of the C-star stage, before the near-IR
flux becomes negligible owing to the formation of great
quantities of carbonaceous dust. It is therefore not
surprising that the dust production rate is limited to
$\rm \dot M_{dust} < 2\times 10^{-8}~M_\odot$/yr.
The trend of the optical depth $\tau_{10}$ with $\rm (F110W-F160W)$ 
is the same as discussed above.

\subsubsection{The progeny of the burst in the SFH of M31 ($1-2$ Gyr ago)}
\label{burst}
The $\rm F160W<17.5$ mag region of the $\rm (F110W-F160W, F160W)$ plane
is populated only by the progeny of $\rm M > 1.2~M_{\odot}$ stars,
which formed not earlier
than $\sim 4$ Gyr ago. More specifically, as shown in Fig.~\ref{fhisto}, we
find that half of the population falling in the 17 mag $\rm<F160W<17.5$ mag strip 
is populated by stars of initial mass $\rm 1.7~M_{\odot} < M < 2~M_{\odot}$ 
of nearly solar metallicity, formed during the secondary peak in the SFH 
of M31 discussed at the beginning of this section, which occurred
$1-2$ Gyr ago. Other sources populating this region of the plane descend
from progenitors of mass $\rm \sim 1.5~M_{\odot}$ ($14\%$) and 
$\rm \sim 2.5~M_{\odot}$ stars ($23\%$), which formed around
2.5 Gyr and 600 Myr ago, respectively. The position of these
stars on the observational plane is shown in the middle panel of
the central line in Fig.~\ref{fage}.

As mentioned in section \ref{old}, the stars of initial mass in the 
$\rm 1.5-2.5~M_{\odot}$ range become carbon stars, owing to the 
effects of a series of TDU episodes
that rise the surface carbon.
The transition from the oxygen-rich to the C-star phase is accompanied
by the significant increase in the mass-loss rates, which favours the 
rapid loss of the external mantle, which ends the AGB evolution 
\citep{vm10, ventura22}. Indeed for the majority of the AGB life-time
they evolve as M-type objects, while the C-star phase is restricted to the 
last few TPs. This holds
in particular for $\rm M \leq 2~M_{\odot}$ stars, which evolve as C-star
only for a couple of TPs; conversely, the carbon rich phase of $\rm 2.5-3~M_{\odot}$
stars is longer, as a longer time is required before the external mantle is
lost.

On the observational side, shortly after the start of the C-star phase the 
evolutionary tracks move to the red: indeed all the stars in the 17 mag $\rm<F160W<17.5$ mag 
region of the plane with $\rm (F110W-F160W) >1.5$ mag are carbon stars.
Despite most of the stars populating this $\rm F160W$ bin
become carbon stars, we find that the fraction of C-stars barely reaches $20\%$, 
owing to the relatively short duration of the C-star phase. Furthermore, as discussed earlier in this section, after the first couple of inter-pulse phases following the achievement of the C-star stage, the 
flux in the $\rm 1-1.6~\mu$m spectral region decreases dramatically; this 
can be seen once more by looking at the evolutionary track of the 
$\rm 2.5~M_{\odot}$ model star in Fig.~\ref{fall}, where it is clear that
the formation of carbon dust is accompanied by a significant decrease in
the near-IR flux.

Therefore, the carbon stars considered here are 
caught in an evolutionary phase when the carbon excess with respect 
to oxygen is so small that the dust production rates barely exceed 
$\rm 10^{-8}~M_\odot$/yr (see bottom, middle panel of Fig.~\ref{fall}). 
The trend of the optical depth with colour is similar to those seen in 
the cases discussed before, the largest $\tau_{10}$ of carbon stars being 
of the order of 0.1 (see the bottom, right panel of Fig.~\ref{fall}).

The $\rm F160W$ magnitude range explored here is completed by a small 
fraction $(\sim 5\%)$ of stars, descending from $\rm 1.2-1.3~M_{\odot}$ 
progenitors, formed around 4 Gyr ago, which do not reach the C-star stage.  
During the final part of their AGB evolution 
they evolve at colours 1.2 mag $\rm<(F110W-F160W)<1.5$ mag, with dust production
rates $\rm \dot M_{dust} < 10^{-9}~M_\odot$/yr, and optical
depths $\tau_{10} \sim 0.1$. These sources are the counterparts of the 
reddest, low-mass oxygen rich stars, discussed in \citet{marini20}.

\subsubsection{The young population of M31}
\label{young}
The $\rm F160W=17$ mag threshold is reached only by stars of
initial mass above $\rm 3~M_{\odot}$, once their luminosity exceeds
$\rm 10^4~L_{\odot}$: this poses an upper limit of $\sim 400$ Myr to 
their age.  In particular, the 16.5 mag $\rm<F160W<17$ mag region (see the bottom, middle
panel of Fig.~\ref{fage}) is mostly populated 
by $\rm 3-3.5~M_{\odot}$ stars, formed between 200 and 400 Myr ago, with 
luminosities in the $\rm 10^4-1.6\times 10^4~L_{\odot}$ range, as seen
in the top, left panel of Fig.~\ref{fall}. 

This is the region of the plane harbouring the largest fraction ($\sim 30\%$) 
of carbon stars. The reason is that the stars in this mass domain reach the C-star 
stage when their luminosity and near-IR fluxes are within the magnitude bin 
considered, while they populate the fainter $\rm F160W$ bins during the previous
oxygen-rich phases. This can be seen by looking at the evolutionary track of the 
$\rm 3.5~M_{\odot}$ model star in Fig.~\ref{fall}. An additional reason for the
relatively large percentage of carbon stars in this part of the
plane, also clear based on the inspection of the evolutionary track in
Fig.~\ref{fall}, is that during the oxygen-rich phase these
stars evolve at colours too blue to pass the criteria described
in section \ref{data}, thus they are not included in the G22 sample.

Even in this case, as discussed for the other carbon stars with fainter 
$\rm F160W$ fluxes, the carbon accumulated in the surface regions is 
still sufficiently small to allow these sources to be detectable in the 
near-IR spectral region. Therefore the dust production rate is not expected 
to exceed $\rm 10^{-8}~\dot M/$yr, while the optical depth $\tau_{10}$ is 
below 0.05. All these information can be verified by looking at the
bottom, left panel and of the bottom panels of Fig.~\ref{fall}.

The fraction of C-stars found in this $\rm F160W$ bin is in tension with 
the results by G22, according to which C-stars account for $\sim 5\%$ of 
the stars. It is likely that this is at least partly due to the criterion 
adopted by G22 to distinguish C-stars from M-type sources, in turn based on 
the classification proposed by  \citet{boyer13,boyer17}, which relies on 
the C and M spectral models from \citet{aringer09, aringer16}. 
The application of this criterion to the present case is not obvious,
because $\rm 3-3.5~M_{\odot}$ stars, that
populate this region of the plane, attain surface $\rm C/O$ ratios
just above unity during the first inter-pulse phases after the C-star stage
is reached, while the model atmospheres by \citet{aringer09} are available
for $\rm C/O>1.05$ only. It is possible that the SED of part of the 
C-stars populating this region of the plane do not exhibit the molecular 
features associated to the presence of $\rm C_2$ and CN molecules at the 
base of the \citet{boyer17} classification. Furthermore, C-stars with a 
$\rm 1.05\leq C/O<1.08$ were probably excluded by the \citet{boyer17} 
criterion since they fall in the  K-type stars region. 

A further reason for the higher fraction of C-stars found in the present 
analysis with respect to the results obtained by G22, mentioned earlier in 
this section, could be related to the 
fact that the colours of M-type stars evolving to this $\rm F160W$ bin are 
too blue to satisfy the cuts imposed by G22, which therefore rule out
some M-type objects. Finally, this discrepancy might be connected to 
an overestimate of the efficiency of the TDU in this mass range,
which might demand a lower efficiency with respect to the lower mass
counterparts. This offers the opportunity of using these results
to draw important, independent conclusions on the efficiency of the convective
instability of the stars of mass close to the threshold for the HBB ignition, 
both in terms of the degree of overadiabaticity in the convective envelopes 
(which affects the colour) and the TDU efficiency (related to the surface carbon 
enrichment). We leave this problem open.

In the 16.5 mag $\rm<F160W<17$ mag range we also find $\sim 3\%$ of
stars descending from $\rm M\geq 4~M_{\odot}$ progenitors,
evolving at luminosities between $\rm 2\times 10^4~L_{\odot}$
and $\rm 5\times 10^4~L_{\odot}$, as indicated in the top, left
panel of Fig.~\ref{fall}, where various evolutionary stages
of the $\rm 5~M_{\odot}$ model star are indicated with open
circles. These stars evolve as oxygen-rich objects 
during the whole AGB life, as the ignition of HBB inhibits the
achievement of the C-star stage \citep{ventura22}. 
They produce silicate in their winds, so the SED is 
gradually shifted to the IR. As shown in 
Fig.~\ref{fall}, the evolutionary tracks of these stars move
to the red, crossing the range of $\rm F160W$ considered
here in the 2 mag $\rm<(F110W-F160W)<2.5$ mag region of the plane.
The rate at which dust production takes place, shown in
the bottom, middle panel of Fig.~\ref{fall}, is
$\rm 2\times 10^{-8}~M_{\odot}/$yr, and the optical depth is found to 
be correlated with the near-IR colour, according to the relationship 
$\rm \tau_{10}=1+2\times [(F110W-F160W)-2]$.

\begin{table*}
\caption{Characterization of the AGB stars of M31 populating the
different $\rm F160W$ bins in the $\rm (F110W-F160W, F160W)$ plane. \%(G22) is the percentage fraction of stars belonging to the G22 sample, \%(sint) is the percentage fraction of stars predicted by population synthesis, \%(C-stars) is the percentage fraction of carbon stars predicted by population synthesis for each F160W bins.}
\label{tabnir}      
\begin{tabular}{|c|c|c|c|c|c|}
\hline
$\rm F160W$ & $>18$ & $17.5-18$ & $17-17.5$ & $16.5-17$ & $<16.5$ \\
\hline
$\% (G22)$ & $49.0$ & $36.8$ & $11.9$ & $1.9$ & $0.4$ \\
\hline
$\rm \%(sint)$ & $49.6$ & $36.7$ & $12.0$ & $1.4$ & $0.5$ \\
\hline
Age  & 6-14 Gyr  &  600 Myr - 6 Gyr  &  600 Myr - 4 Gyr  &  200 - 300 Myr  &  $<200$ Myr  \\
\hline
Mass & $\rm 0.8-1.2~M_{\odot}$ & $\rm 1.2-3~M_{\odot}$ & $\rm 1.3-3~M_{\odot}$ 
& $\rm 3-3.5~M_{\odot}$ & $\rm \geq 4~M_{\odot}$ \\
\hline
$\rm \%(C-stars)$  &  -  &  $10\%$  &  $20\%$  &  $30\%$  &  - \\
\hline
DPR & - & $\rm <2\times 10^{-8}~M_{\odot}/$yr & $\rm <10^{-8}~M_{\odot}/$yr & $\rm <2\times 10^{-8}~M_{\odot}/$yr &  $\rm <10^{-8}~M_{\odot}/$yr \\
\hline
\end{tabular}
\end{table*}

\subsubsection{Stars undergoing HBB}
The $\rm F160W<16.5$ mag region of the $\rm (F110W-F160W, F160W)$ diagram is
mainly populated by massive AGB stars that experience HBB, and by 
$\sim 20\%$ of stars descending from $\rm \sim 3.5~M_{\odot}$
progenitors, populating the 16 mag $\rm<F160W<16.5$ mag strip, which attain these 
large near-IR fluxes during the final part of the O-rich phase, just before 
turning into C-stars. The brightest sources in the $\rm F160W<16$ mag region
are highlighted in cyan in the left panel of Fig.~\ref{fvw2}, and the same
colour coding is used to indicate their position on the $([3.6]-[4.5],[3.6])$
plane, in the right panel of the figure. The position of these sources is well 
reproduced by the evolutionary tracks of $\rm 4-7~M_{\odot}$ stars, corresponding 
to the phases when the
$\rm F160W$ flux reaches a maximum value, before dust production takes over,
the SED is shifted to the medium-IR, and the $\rm F160W$ flux decreases.

We therefore identify the bright sources in the $\rm (F110W-F160W, F160W)$ plane,
indicated with cyan points, as the progeny of $\rm M \geq 4~M_{\odot}$ stars,
formed not later than 200 Myr ago. These 
stars, whose luminosity is in the $\rm 1.5\times 10^4~L_{\odot} < L < 7\times 10^4~L_{\odot}$
range, are nowadays experiencing HBB, and are evolving through the final evolutionary
phases before the start of efficient dust formation, which prevents detection in
the near-IR. Their dust production is below $\rm 10^{-8}~M_{\odot}$/yr. The distribution
of massive AGBs in the $\rm (F110W-F160W, F160W)$ plane is reported in the middle,
left panel of Fig.~\ref{fage}: we note that they are distributed across a large
portion of the plane, as during the early AGB and the initial phases of the AGB evolution 
they evolve at $\rm F160W>16$ mag (see Fig.~\ref{fvw2}).

\begin{figure*}
\begin{minipage}{0.32\textwidth}
\resizebox{1.\hsize}{!}{\includegraphics{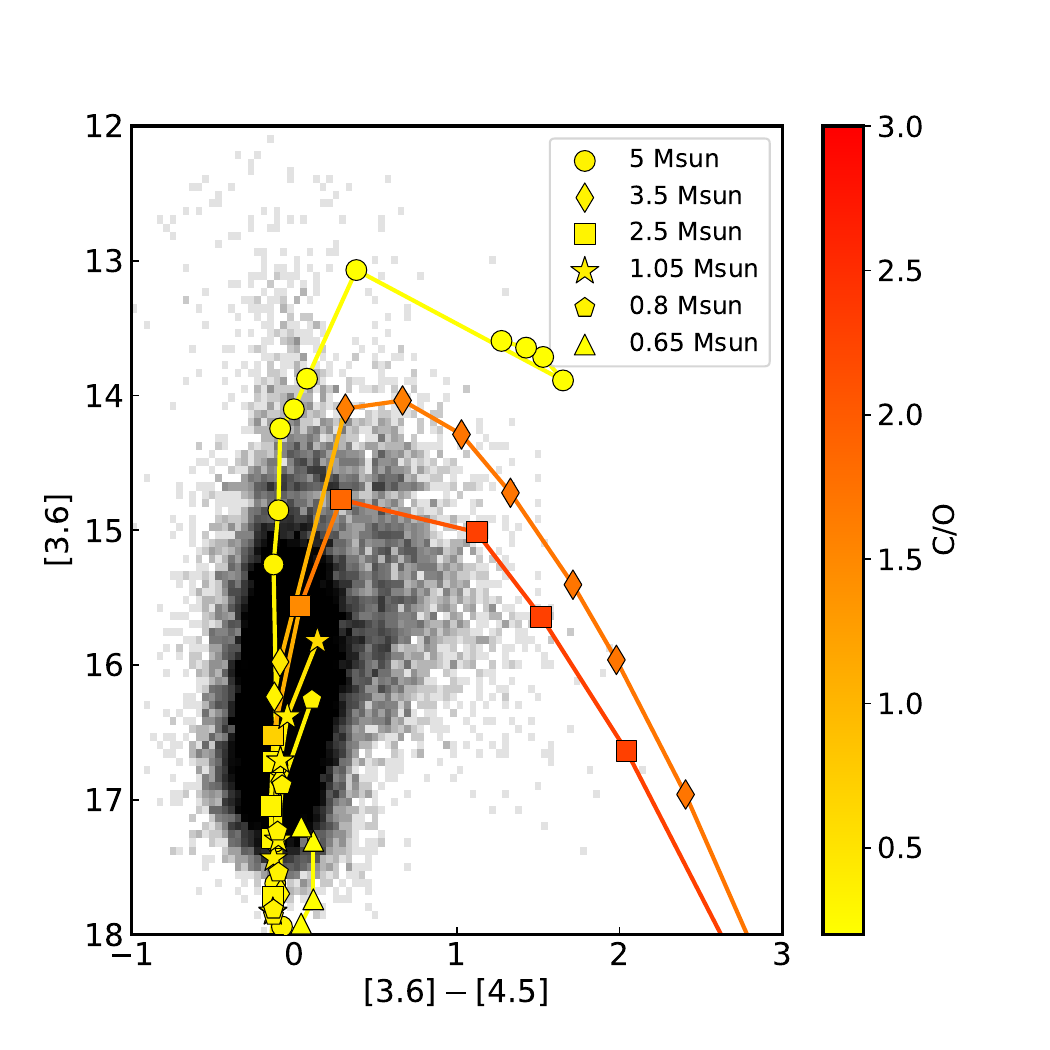}}
\end{minipage}
\begin{minipage}{0.32\textwidth}
\resizebox{1.\hsize}{!}{\includegraphics{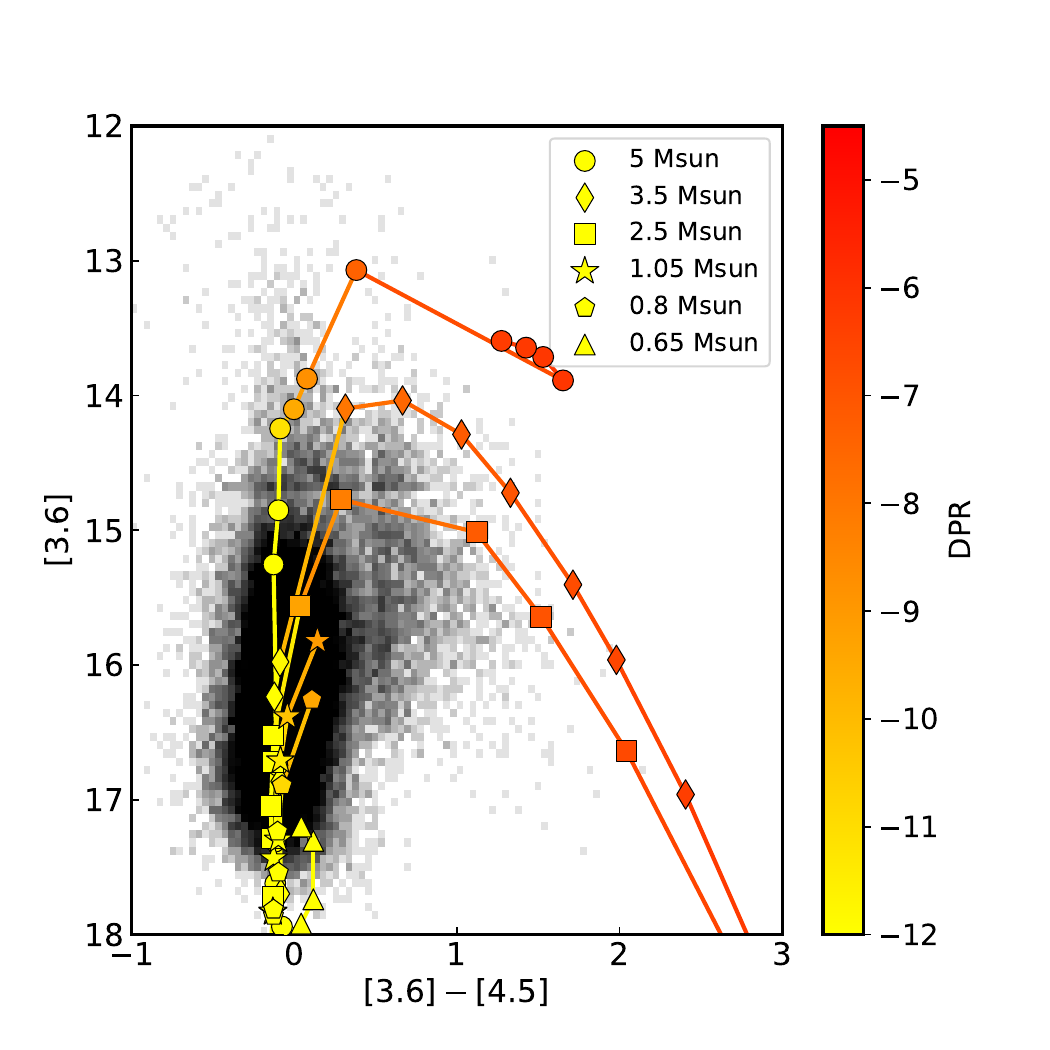}}
\end{minipage}
\begin{minipage}{0.32\textwidth}
\resizebox{1.\hsize}{!}{\includegraphics{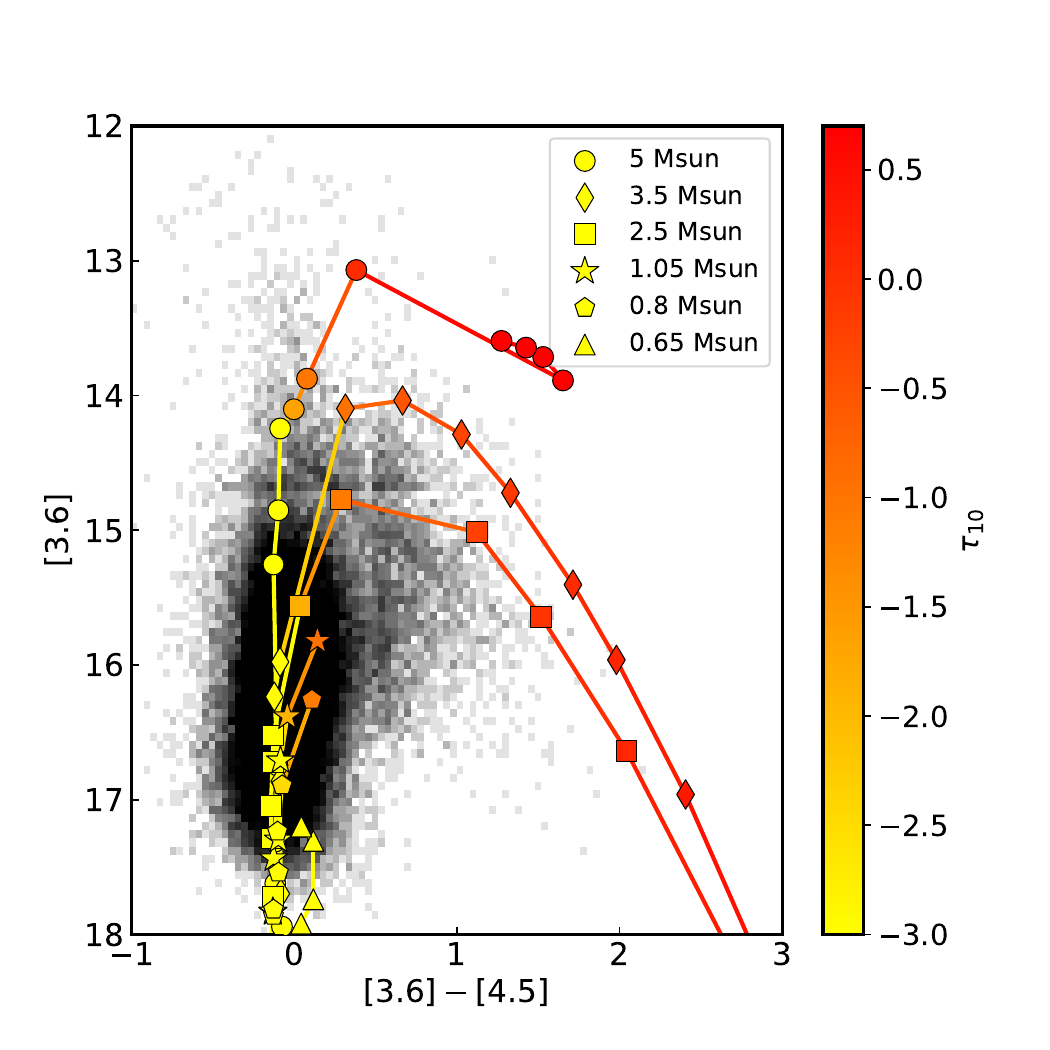}}
\end{minipage}
\caption{Data points of the M31 sources included in the catalogue
by G22 in the $\rm ([3.6]-[4.5],[3.6])$ plane
obtained with the Spitzer filters are shown with grey points. 
Solid lines represented the same evolutionary tracks of some
selected model stars reported in Fig.~\ref{fall}. The information 
regarding the C/O ratio, the dust production rate and the 
optical depth at $10~\mu$m are shown in the left, central and right panels, respectively.
} 
\label{fspit}
\end{figure*} 

\subsection{The interpretation of Spitzer data}
The distribution of the M31 sources in the catalogue described in
section \ref{data} in the colour magnitude
$([3.6]-[4.5],[3.6])$ diagram obtained with the Spitzer
filters is shown in Fig.~\ref{fspit}. Overlapped to the
data points are the evolutionary tracks of the same model
stars shown in Fig.~\ref{fall}.

A large fraction ($\sim 85\%$) of the stars populating the near-IR diagram
discussed in section \ref{nearir} is missing in this plane. We refer
to the old stars 
addressed in section \ref{old}, which, as shown in Fig.~\ref{fspit}
(see the sequence of triangles), do not evolve at $[3.6]<17.5$ mag, thus 
they would populate a region of the plane severely affected by the 
incompleteness effects discussed in section \ref{data}.
Indeed only $< 0.1\%$ of the stars are observed in the $[3.6]>17.5$ mag region, 
and no sources are detected at $[3.6]>17.7$ mag. 

The colour distribution of the data points reported in Fig.~\ref{fspit} 
shows that $\sim 92\%$ of the sample stars with available Spitzer photometry
are located into the vertical strip -0.3 mag $\rm<([3.6]-[4.5])<0.1$ mag. 
This is consistent with the theoretical expectations, as all the evolutionary
tracks shown in Fig.~\ref{fspit} run approximately 
vertically during the dust-free phases, at $([3.6]-[4.5]) \sim -0.1$ mag. 
This region of the plane is populated by stars formed in different epochs, 
descending from progenitors of various mass. The results shown in the left panel
show that the rate at which they are forming dust is extremely small, either because 
they are low-mass stars that produce little quantities of dust during their life, 
or because they are caught during the initial part of the AGB lifetime,
when dust production is still negligible.

The largest fraction of these objects ($\rm \sim 40\%$) are the progeny of solar 
metallicity stars of mass slightly above solar, populating the two faintest 
bins in the $\rm (F110W-F160W, F160W)$ plane discussed in section \ref{old} 
and \ref{sol}, whose evolutionary track can be identified with that of the 
$\rm 0.8~M_{\odot}$ model star (pentagons) in Fig.~\ref{fspit}, which
shows that these stars are located in the $[3.6]>16.1$ mag region.

Regarding the $[3.6]<16$ mag region of the plane, 
we find that approximately half of the stars descend from $\rm 1.7-2.5~M_{\odot}$ 
progenitors, formed around $1-2$ Gyr ago. The remaining half being the progeny 
of $\rm M \geq 1.5~M_{\odot}$ stars evolving through the initial part of the AGB, 
before the start of dust production.

Stars with mass in the $\rm 1.25~M_{\odot} \leq M \leq 3.5~M_{\odot}$ range, 
represented by the evolutionary tracks of the  
$\rm 2.5~M_{\odot}$ (squares) and $\rm 3.5~M_{\odot}$ (diamonds) model stars 
in Fig.~\ref{fspit}, evolve first along the vertical band discussed earlier, 
then the evolutionary tracks turn to the red, after the C-star stage is reached. 
All the sources in the $([3.6]-[4.5])>0.2$ mag, $[3.6]>14$ mag region of the plane
descend from $\rm 1.2-3.5~M_{\odot}$ progenitors of solar
metallicity, that became carbon stars. The reddest data points in
Fig.~\ref{fspit} extend to $([3.6]-[4.5]) \sim 2$ mag, where we find
stars that after experiencing several TDU events have reached surface
$\rm C/O$ in the $1.5-2$ range (see left panel of Fig.~\ref{fspit}), produce carbon dust with rates
$\rm \sim 10^{-7}~M_{\odot}/$yr (central panel of Fig.~\ref{fspit}), 
and are characterized by optical depths $\tau_{10} \sim 2$
(right panel of Fig.~\ref{fspit}).

The $[3.6]<14$ mag region of the $([3.6]-[4.5], [3.6])$ plane is
populated only by the progeny of $\rm M \geq 4~M_{\odot}$
stars, as lower mass objects do not reach these [3.6] fluxes.
As indicated by the tracks of the $\rm 4-7~M_{\odot}$ model stars
in the right panel of Fig.~\ref{fvw2}, these stars evolve 
to the red when formation of silicates begins, until reaching 
$([3.6]-[4.5]) \sim 1.5$ mag. We tentatively identify the group of stars 
in the regions centered at $([3.6]-[4.5]) \sim 0.5$ mag, $[3.6] \sim 13-13.5$ mag, and 
$([3.6]-[4.5]) \sim 1.2$ mag, $[3.6] \sim 13$ mag as stars experiencing HBB, evolving 
through the final AGB phases.

\begin{figure}
\vskip-40pt
\centering
\begin{minipage}{0.46\textwidth}
\resizebox{1.\hsize}{!}{\includegraphics{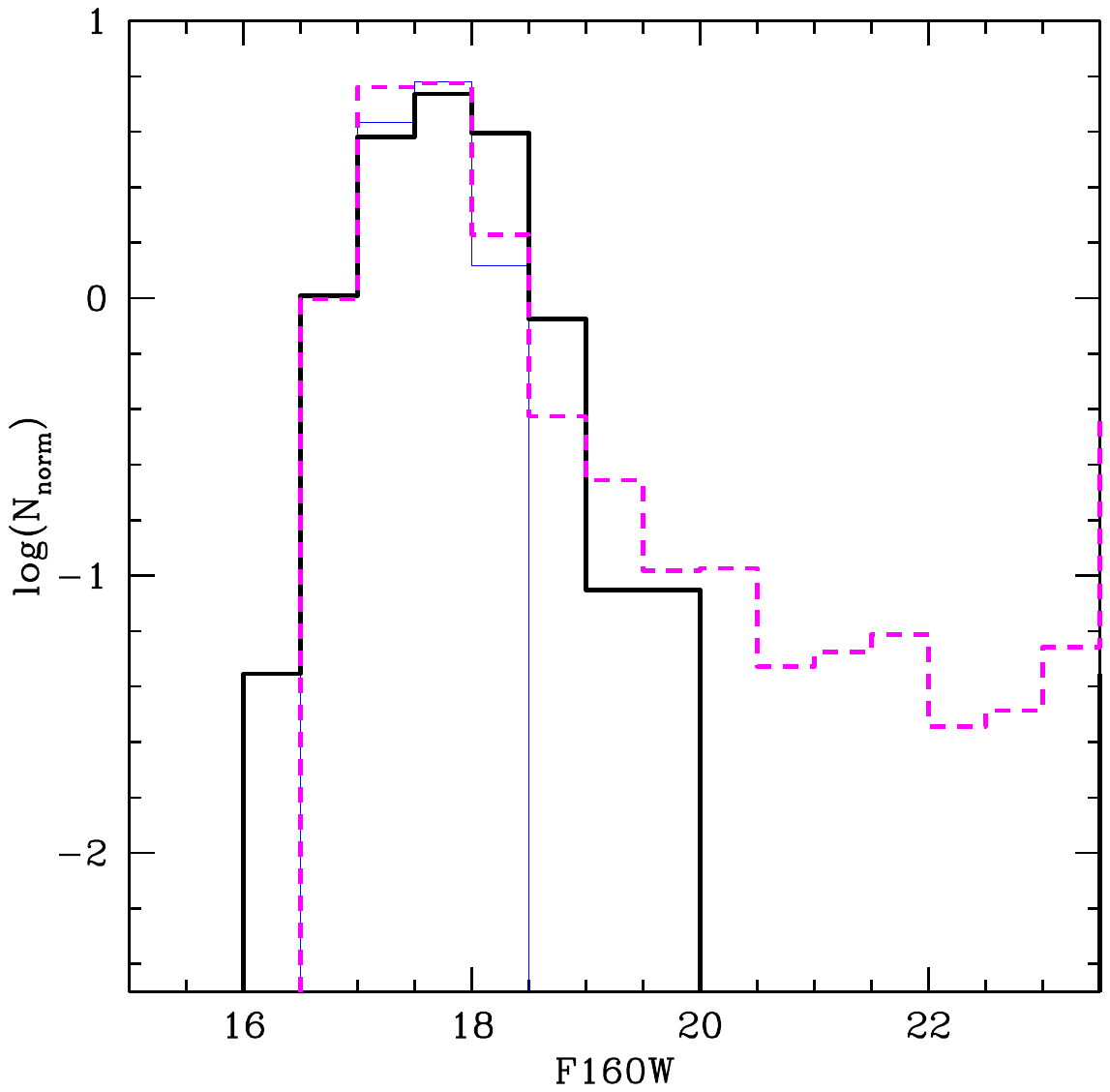}}
\end{minipage}
\vskip-80pt
\begin{minipage}{0.46\textwidth}
\resizebox{1.\hsize}{!}{\includegraphics{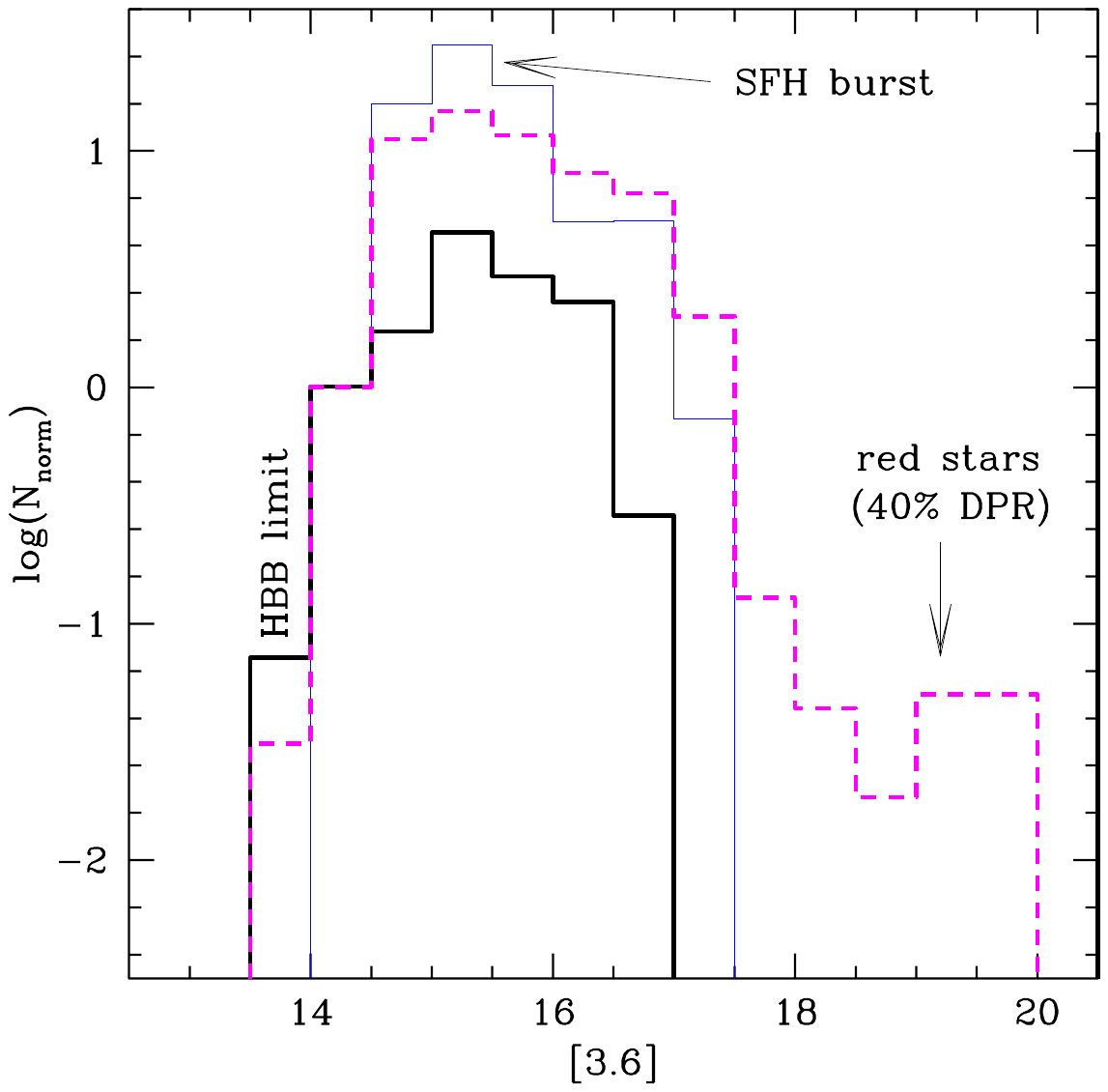}}
\end{minipage}
\vskip-60pt
\caption{The luminosity function of carbon stars in F160W (top) and [3.6] (bottom), 
normalized to 17 and 14 mag, respectively. The LF of the carbon stars sample 
as selected by \citet{boyer19} is shown in black; the LF of the carbon stars 
synthetic population produced in this work is reported in magenta, while
the synthetic LF obatined by considering the cuts proposed by G22 (see 
Section \ref{data}) is displayed in blue.}
\label{fcstar}
\end{figure}

\subsection{The luminosity function of carbon stars}
\citet{boyer19} determined the chemical type of a sub-sample of AGB stars
in M31, based on medium-band HST imaging. In fig.~\ref{fcstar} we show the LF of the 346 sources classified
as carbon stars, in the F160W (top; normalised to unity 
in the 16.5 mag $\rm<F160W<17$ mag bin) and [3.6] magnitudes (bottom; normalised to unity 
in the 14 mag $\rm<[3.6]<14.5$ mag bin). In the same figure we also report the LF
obtained via the population synthesis method described in section \ref{popsyn},
where the thin, blue and the dashed, magenta lines indicated the LF
obtained with and without the cuts discussed in section \ref{data},
respectively. In the first case we obtain a total of $\sim 37000$ carbon stars,
whereas if the colour and magnitude cuts are not applied we get 
$\sim 57000$ C-rich objects.

Understanding the near-IR LF is a tricky task, because several factors affect
its morphology. A difficulty arising in the interpretation of the LF, as
clear in the evolutionary tracks of $\rm M\geq 1.25~M_{\odot}$ model stars 
in Fig.~\ref{fspit}, is that the variation of the $[3.6]$ magnitude is not 
directly correlated with the dust surrounding the stars, in turn related to 
the amount of carbon dredge-up in the surface regions: the $[3.6]$ flux first 
increases, as the SED of the star is gradually shifted to the near-IR spectral 
region, then decreases, when the peak of the SED moves to the mid-IR 
\citep{marini21}. The evolutionary tracks of carbon stars, as visible in 
Fig.~\ref{fspit}, are characterised by the presence of a maximum
in the $[3.6]$ flux, which divides the initial phase, during which little dust
is formed in the circumstellar envelope, from the latest evolutionary phases,
when efficient dust production takes place. Each $[3.6]$ magnitude bin is therefore
crossed twice during the AGB evolution: however, following the discussion of
section \ref{burst}, where we stressed the very short evolutionary time scales 
characterizing the final part of the AGB phase, we find that 
the statistics is dominated by the least obscured stars, which account for more 
than $95\%$ of the population. These arguments hold in the $[3.6]<17.5$ mag domain
only: indeed we do not expect to find carbon stars with fainter $[3.6]$ fluxes,
if not during the very final AGB phases, when the SED peaks at
$\lambda > 5~\mu$m \citep{marini21}. On the other hand, the LF of carbon stars in F160W related to the dust production: as shown in Fig.~\ref{fall}, low-mass stars reach the brightest F160W magnitude when they enter the C-rich stage. Soon after, the increasing carbon dust production enshroud the emission at $\lambda\leq2\mu$m and the star progressively evolves towards fainter F160W magnitudes.

Other factors playing a role in the determination of the LF 
of carbon stars are the physical conditions of the stars when the C-star 
stage is reached, and the SFH of the galaxy. The former determine the dust 
production efficiency at the beginning of the C-star phase, thus the 
F160W and $[3.6]$ flux of the star at the transition from oxygen-rich to carbon-rich. 
The SFH sets the (initial) mass distribution of the carbon star population.

Inspection of Fig.~\ref{fcstar} outlines a welcome similarity between the
LFs from \citet{boyer19} and the results obtained from population synthesis; 
on the other hand the total LF, obtained when the criteria discussed in 
section ~\ref{data} are ignored, exhibits a tail in the F160W$>18.5$ mag and $[3.6]>17.5$ mag domain. 
This tail is due to the contribution of extremely red objects, with very poor flux in the 
$\rm F110W$ and $\rm F160W$ bands.

To understand the shape of the C-stars LF, and to characterize the current
carbon stars population of M31, it is important to consider which stars reach the
C-star stage, the formation epoch, and the duration of the phase during which 
they are carbon rich. Because the M31 stars potentially able to become carbon 
stars are characterized by solar metallicity, for which a minimum mass of
$\rm \sim 1.5~M_{\odot}$ is required to reach the C-star stage \citep{ventura18},
we deduce that the nowadays C-star population of M31 is made up of stars
younger than $\sim 3$ Gyr, and that all the stars formed during the initial
intense star formation activity of M31, described at the beginning of
section \ref{char}, are now evolving as oxygen rich sources. Therefore,
most of the carbon stars formed during the burst in the SFH occurred $1-2$ 
Gyr ago, when stars of mass in the $\rm 1.7-2~M_{\odot}$ range formed.
Based on these arguments, we expect that the peaks in the LF are found in 
correspondence of the F160W and $[3.6]$ fluxes characterizing the C-star evolution 
of the progeny of $\rm 1.7-2~M_{\odot}$ stars.

The LFs reported in Fig.~\ref{fcstar} exhibit a peak
in the 17 mag $\rm<F160W<18$ mag and 15 mag $\rm<[3.6]<15.5$ mag intervals, which is in fact the range spanned
by the F160W and $[3.6]$ fluxes of the $\rm 1.7-2~M_{\odot}$ stars during the first
and the second inter-pulse phases after becoming carbon stars. The faint wing of the LF in the F160W band and for $[3.6]>15.5$ mag is affected by the cut choice to remove RGB stars and incompleteness, as suggested by G22. However, we believe that this
is also due to the evolutionary properties of stars of various mass during
the C-rich phase; this can be clearly explained focusing on the [3.6] band, more sensitive to the dust emission than the F160W band. As discussed above, $\rm 1.7-2~M_{\odot}$ stars 
attain $[3.6]$ fluxes corresponding to 15.5 mag $\rm<[3.6]<16.5$ mag 
for two inter-pulses only, before dust formation is so intense that
the entire SED shifts towards the mid-IR, and the $[3.6]$ flux drops
to $\rm<[3.6] \sim 17$ mag. Therefore, these stars account for only
$30\%$ of the C-star population of M31 in the 15.5 mag $\rm<[3.6]<16.5$ mag range,
whereas the dominant contribution of $\sim 70\%$ is provided by the progeny 
of $\rm \sim 2.5~M_{\odot}$ stars. 

This behaviour is connected to the afore-discussed role played by the physical 
conditions of the star at the achievement of the C-rich phase on the carbon 
stars LF. The stars descending from $\rm 2.5~M_{\odot}$ progenitors
produce little dust during the phases immediately following the
achievement of the C-star stage, because they become C-stars after
having lost only a very small fraction of the envelope mass, which
reflects into relatively large surface gravities, effective temperatures
of the order of $3700$ K, thus physical conditions partly inhibiting the
formation of dust. This is different from the $\rm M \leq 2~M_{\odot}$ counterparts,
which become carbon stars after a significant fraction of the envelope is
lost, thus they evolve at cooler effective temperatures, of the
order of $\sim 3000$ K, and produce dust more efficiently, since the
early C-star phases. The SED of these objects is shifted to the near-IR
as soon as they become carbon stars, which explains why they are
characterized by higher $[3.6]$ fluxes during this evolutionary phase.

The significant decrease in the LF detected at $[3.6] \sim 17$ mag is related
to the fact that the minimum $[3.6]$ magnitude of $\rm 2.5~M_{\odot}$ stars  
during the inter-pulse phases is $[3.6] \sim 17$ mag, thus the $[3.6]>17$ mag bins are
populated only by carbon stars evolving through the low-luminosity phases
following each TP, which are very short in comparison to the duration of
the inter-pulse periods. This explanation does not hold for the stars responsible
for the tail in the 17.5 mag $\rm<[3.6]<20$ mag range exhibited by the 
LF (magenta line in Fig. \ref{fcstar}).
These stars are evolving
through the very final AGB evolutionary phases, during which the dust
production rate is extremely large, thus the SED peaks in the mid-IR
spectral region. 

Turning to the bright side of the LF peak, the drop in the LF in the 16 mag $\rm<F160W<17$ mag and 14 mag $\rm<[3.6]<15$ mag domains 
is because these F160W and $[3.6]$ fluxes are reached only by $\rm M>2~M_{\odot}$ stars 
during the very final AGB phases. On the contrary, the progeny of $\rm 1.7-2~M_{\odot}$ stars, which we have seen to 
provide the largest contribution to the overall carbon stars population of M31, do not reach these magnitude values.
In the LF shown in Fig.~\ref{fhisto} it is evident the severe drop at F160W$<16$ mag and $[3.6] < 14$ mag, which is the limit for the ignition of HBB (see the evolutionary tracks in Fig.~\ref{fspit}), which prevents the C-star formation.

Overall, the fraction of carbon stars found by means of the population
synthesis approach adopted here is $\sim 12\%$. This fraction includes the stars
populating the $[3.6]>17.5$ mag tail of the LF shown in Fig.~\ref{fcstar}, which
however are only a small percentage ($\sim 0.7\%$) of the total carbon stars
population of M31. The carbon stars fraction found in our analysis is higher than the quantities given by 
\citet{boyer19} for the various sub-fields considered in their analysis.
However, this is due to the different colour cuts adopted in the studies by
G22 (the same adopted in the present investigation, described in section \ref{data}),
and by \citet{boyer19}. The cuts imposed by G22 exclude a significant fraction
of stars populating the blue side of the observational planes, most of which
are oxygen-rich (see the evolutionary tracks on the blue side of the
$\rm (F110W-F160W, F160W)$ plane reported in Fig.~\ref{fall}). This situation is partly 
alleviated in the \citet{boyer19}
case, where the only colour condition adopted is $\rm (F127W-F153W)>0.1$ mag, thus a higher
number of oxygen rich stars would be recovered. To compare the $\rm C/M$ results
from the two studies we repeated the calculation by adopting the colour cut and the TRGB cut
used by \citet{boyer19} instead of the G22 conditions listed in section
\ref{data}. We found that the overall $\rm C/M$ decreases to 0.07, which is
much more consistent with the percentages found by \citet{boyer19}.

\section{Discussion}
The analysis of the previous session outlined how optical, near and mid-IR
observations are crucial for a thorough characterization of the evolved 
stellar population of galaxies. This is particularly true for structures 
characterized by a complex star formation history and metallicity distribution,
such as M31.

The interpretation of the near-IR data reported on the $\rm (F110W-F160W, F160W)$
plane allowed to draw important information, particularly on the structure and
evolution of low-mass stars: these objects, for reasons connected to the SFH 
experienced by M31 and to the shape of the IMF, which peaks at the lowest masses,
are the majority of the stellar population detected in the near-IR plane. 

The findings regarding the mass loss suffered by low-mass stars during the
ascending of the RGB are rather intriguing. The observed $\rm F160W$ LF
can only be reproduced if we assume that the average mass loss
during the RGB phase of low-mass stars is $\rm 0.2~M_{\odot}$, 
$\rm 0.25~M_{\odot}$, $\rm 0.3~M_{\odot}$ for the metallicities
$\rm Z=10^{-3}$, $\rm Z=4\times 10^{-3}$ and $\rm Z=Z_{\odot}$,
respectively. This is in agreement with several
investigations aimed at reproducing the detailed morphology of the
HBs of Galactic globular clusters \citep{caloi05, salaris16, 2419, tailo16, 
don18}, and with the recent results obtained by \citet{tailo21}, who 
calibrated the mass loss suffered by RGB stars belonging to clusters of 
different metallicity.

Recent studies based on Asterosismology (e.g. Miglio et al. 2021) suggested 
that the total mass loss experienced by low-mass RGB stars is of the order 
of $\rm 0.1~M_{\odot}$, thus smaller than those derived in the afore-mentioned 
investigations. Most of these results regard Galactic sources of higher metallicity
and younger ages than the old sources considered here. Extension of these findings 
to the metal-poor population older than $\sim 10$ Gyr would pose the issue of
partial inconsistency between results from Asterosismology and those from the
study of globular clusters. However, entering the debate regarding the RGB mass
loss is beyond the scope of the present paper, thus we leave this problem open.

The comparison between the observed and the synthetic $\rm F160W$ LF's
also provides
information on the efficiency of convection in the envelope of AGB stars,
as the latter affects the effective temperatures and colours, thus the
possibility that a given model star is consistent with the colour cuts
imposed by G22 upon selecting the sample stars to be considered.
We find that the results obtained in the metal-poor domain when the
FST or the MLT model of convection based on the solar calibration are
adopted lead to colours that are too blue to satisfy the cuts imposed
by G22, thus to a number of sources significantly smaller that that
in the G22 sample. To reconcile the results from synthetic modelling
with the observations, we find that the mixing length parameter that
is connected to the efficiency of convection is $\alpha=1.5-1.6$ in the
sub-solar case, whereas in the metal-poor domain the choice $\alpha=1$
is required. On the other hand, in regard of the solar metallicity, the
FST-based models evolve at colours consistent with the observations.
This opens the way to the possibility of calibrating the efficiency of
the convective channel to transport energy in AGB stars, something so
far restricted to main sequence and RGB stars.
An effect of the colour cut imposed on $\rm (F110W-F160W)$ described 
in section \ref{data} is that
a significant fraction of stars descending from progenitors of mass above
$\rm 2~M_{\odot}$ are excluded from the sample. Indeed during the 
oxygen-rich phase the evolutionary tracks of these stars evolve vertically, 
with colours too blue with respect to the threshold for $\rm (F110W-F160W)$ 
chosen by G22 (see Fig.~\ref{fall}). Note that this has little 
effects on the overall numerical consistency of the sample, as the majority of 
the evolved stellar population of M31 descends from low-mass progenitors. 
Furthermore, this has no effects on the determination of the dust production
rate of the galaxy, as the sources excluded are surrounded by very small amounts
of dust.

While the analysis of the distribution of the sources in the 
$\rm (F110W-F160W, F160W)$ plane allows a detailed statistical investigation
of the M31 population, we have seen that the information on the dust and on the
stars surrounded by dust can be drawn by studying the $([3.6]-[4.5], [3.6])$ diagram. 
This is particularly relevant 
to study massive AGBs experiencing HBB, which are known to be efficient silicates
manufacturers, owing to the large mass loss rates experienced \citep{ventura14, ventura22}. 
In section \ref{ester}
we discussed how the description of the evolution of this class of stars is
severely affected by the treatment of mass loss. The results from synthetic modelling
allow us discriminating among the various choices available to determine the mass
loss rates. Indeed when the \citet{blo95} recipe is adopted, the results are not
consistent with the observations, because, as discussed in section \ref{ester}, 
we should expect a number of stars in the HBB region at $([3.6]-[4.5]) \sim 0.5$ mag 
and $[3.6]<14$ mag far in excess of the observed number. 
This can be deduced by inspection of the evolutionary tracks of the
$\rm 5~M_{\odot}$ model star in the right panel of Fig.~\ref{fvw2}.
The HST results also confirm this conclusion: as shown in the left
panel of Fig.~\ref{fvw2}, and discussed in section \ref{ester}, 
the evolutionary tracks of the $\rm 4-7~M_{\odot}$ model 
stars calculated with the VW93 recipe overlap with the position of the sources 
located at $\rm F160W < 16$ mag, which otherwise remain unexplained if the \citet{blo95} 
treatment is used (see the black track on the figure). As shown in the right panel 
of Fig.~\ref{fvw2}, the same tracks are also fully consistent with the location of 
the same sources (indicated with cyan points) on the $([3.6]-[4.5], [3.6])$ plane.

The results obtained with the VW93 treatment of mass loss are far 
more consistent with the observations, which is a clear indication of the opportunity 
to use the VW93 recipe when massive AGBs of solar metallicity are modelled. 
The stars populating the $[3.6]<14$ mag region of the plane, descending from 
$\rm M \geq 4~M_{\odot}$  progenitors, are the youngest sources in 
the G22 sample, with ages below $\sim 200$ Myr. The surface chemistry of these
sources is expected to be heavily contaminated by HBB, with significant nitrogen
enhancement and isotopic carbon ratios $\rm ^{12}C/^{13}C \sim 3-4$.

We note that the results obtained here on the description of mass loss experienced by
massive AGBs of solar metallicity are consistent with the study by \citet{ester23},
who found that the SED of a few bright, long-period, oxygen-rich stars in the Galaxy 
with deep absorption silicate features can be reproduced only within the VW93
modelling of mass loss. As clearly shown in the right panel of Fig.~\ref{fvw2}, the M31 
counterparts of the stars studied by \citet{ester23} would be located into the
region of the $([3.6]-[4.5], [3.6])$ plane centered at $([3.6]-[4.5]) \sim 1.5$ mag and
$[3.6] \sim 14$ mag, where therefore we expect to find stars characterized by
intense silicates production, with rates $\rm \sim 10^{-6}~M_{\odot}/$yr,
pulsating with periods of $1500-2000$ yr.

$\rm (F110W-F160W, F160W)$ is the ideal plane for a reliable number count
of massive AGBs, because these stars can be easily identified in the $\rm F160W<16$ mag
region, as evident in the left panel of Fig.~\ref{fvw2}, where they are indicated 
with cyan points. Conversely, in the $([3.6]-[4.5], [3.6])$ diagram (see right 
panel of Fig.~\ref{fvw2}) they are partially mixed with other sources, which
complicates the statistical analysis. A rough estimate of the number of stars 
expected in the region of the $\rm (F110W-F160W, F160W)$ plane populated 
by the cyan points in Fig.~\ref{fvw2} can be found by considering that 
the time spent by $\rm 4, 5, 6~M_{\odot}$ model stars calculated with the
VW93 treatment of mass loss in the $\rm F160W<16$ mag region 
is $3\times 10^5$, $10^5$ and $3\times 10^4$ yr, respectively, and that 
star formation between 50 and 200 Myr ago took place in M31 with rates 
$\rm \sim 0.25~M_{\odot}/$yr \citep{williams17}. Based on a standard Kroupa IMF,
we estimate that the number of stars in the $\rm F160W<16$ mag
region should be around 1000, which is a factor $\sim 2$ higher than observed.
This difference might suggest that the recent star formation rate derived for M31
is overrated, or that the evolutionary time scales given above, primarily
determined by the VW93 description of mass loss, must be revised downwards.
We leave this problem open.

Overall, based on the modelling of massive AGBs that allows to better reproduce
the observations, we estimate that the dust production rate by M stars in M31 is 
$\rm \dot M_{Sil} \sim 6\times 10^{-5}~M_{\odot}/$yr. $\sim 80\%$ of the overall
silicates production is provided by massive AGBs currently experiencing HBB,
while the remaining $\sim 20\%$ is produced by $\rm M>1~M_{\odot}$ stars of
solar metallicity, during the evolutionary phases preceeding the C-star phase.

We finally consider the carbon-rich sample, which, as discussed in the previous
section, populates the red region of the $([3.6]-[4.5], [3.6])$ plane.
On this regard, a clearly discrepancy between the results based on synthetic
modelling and the observations is that in the latter the C-rich branch is
truncated at $\rm ([3.6]-[4.5]) \sim 2$ mag, whereas we see in Fig.~\ref{fspit}
that the evolutionary tracks of $\rm 1.25-3.5~M_{\odot}$ stars extend to 
$\rm ([3.6]-[4.5]) \sim 3.5$ mag. The latter extreme stars would account for the
presence of the tail in LF of carbon stars in the $[3.6]>17.5$ mag domain,
clearly visible in Fig.~\ref{fcstar} (magenta line). 
The lack of extremely red objects in M31 was also
noticed by G22, who stressed that the fraction of extreme AGBs in this galaxy,
slightly above $1.3\%$, is smaller than the values of $4.5\%$ and $6\%$ 
found in the LMC and SMC, respectively.

The scarcity of extremely red C-stars in M31, if confirmed, would indicate
that the efficiency of the dust formation process by C-rich AGBs is 
sensitive to the metallicity, thus confirming the results from \citet{jacco00} 
and \citet{jacco05}. 
This would imply that the carbon dust yields predicted by AGB + dust
formation modelling are overestimated. Unfortunately no firm conclusion
can be presently reached on this argument, because the missing C-rich
population in the $\rm ([3.6]-[4.5]) > 2$ mag region of the plane would be 
characterized by near-IR fluxes 16 mag $\rm<[3.6]<19$ mag (see the evolutionary
tracks of C-stars in Fig.\ref{fspit}), where the detection probabilities
are severely affected by incompleteness: indeed as discussed in G22, completeness 
starts to affect the distribution of M31 stars at magnitudes 
$\rm [3.6] \sim 15.2$ mag.

We are more favourable to believe that the lack in M31 of the most 
obscured C-rich stars that characterized the LMC, identified by 
\citet{flavia15a} as the progeny of $\rm 2-3~M_{\odot}$ stars, is a
simple effect of incompleteness; this would confirm the results by
\citet{sloan12} and \citet{sloan16}, that the formation of carbon dust
is independent of metallicity. 

If this is correct, we may estimate the overall carbon dust production rate 
of M31 by using the results from synthetic modelling, by summing up all the 
DPR of the individual stars, for which we find a tight correlation with the 
colour, according to the relation:
$$
\rm \log(\dot M_{car})={2\over 3}\times ([3.6]-[4.5]) -8.3
$$

We obtain $\rm \dot M_{car} \sim 3.8\times 10^{-4}~M_{\odot}/$yr.  
We note that if we exclude the contribution from the stars
populating the $\rm ([3.6]-[4.5]) > 2$ mag zone of the $\rm ([3.6]-[4.5], [3.6])$ 
plane, the overall DPR by C-stars would by underestimated by a factor $\sim 2$.

Future mid-IR JWST observations with the MIRI camera will be able to
confirm or disregard the presence of extremely red stars in M31, thus
opening the way to an accurate estimation of the overall DPR in the galaxy
and, more generally, to assess the role played by metallicity on the
dust production mechanism by carbon stars.

The present results indicate that most of the dust currently 
released by M31 AGB stars is under the form of carbon dust, with
the silicates contribution being below $\sim 20\%$. Even considering 
that the lack of the reddest C-stars would underestimate the
current carbon dust production rate by $\sim 50\%$, we would still
find that the silicates contribution barely reaches $\sim 30\%$. 
This is at odds with the suggestion given by G22 that silicates 
provide the dominant contribution to the global dust production 
rate by AGB stars in M31.
This difference is related to the fact that the estimate of the
DPR by G22 is based on the relationship between ([3.6]-[5.8]) 
and DPR proposed by \citet{martin18}, in which the authors assume 
a constant gas-to-dust ratio $\Psi=200$. Indeed, according to the 
dust production modelling described in section \ref{dustmod}, we 
find that this choice is fairly consistent with the results obtained 
for the C-star phases.
For what attains oxygen-rich stars, we
find that during the evolution of the stars descending from $\rm M \leq 1~M_{\odot}$ 
progenitors, which constitute the bulk of M-type stars of the galaxy, 
$\Psi$ barely reaches 1000. Only the stars undergoing HBB attain
values of $\Psi$ of the order of 400, during the evolutionary phases
when the HBB conditions are strongest \citep{ventura14}. These values
of the gas-to-dust ratios are in agreement with the results obtained
by \citet{goldman17} when studying a sample of bright, oxygen-rich
stars in the LMC.

In summary, we find that the contribution of the carbonaceous species to the overall
dust production rate of the galaxy is higher than that of silicates,
despite the solar chemical composition of the majority of the evolved stars in M31. 
In the stellar populations characterised by solar chemistries,
the lower mass threshold required to reach the C-star stage, of the 
order of $\rm 1.2-1.3~M_{\odot}$, is higher than in metal poor environments, 
where the lower
limit is below the solar mass \citep{ventura22}. This favours a higher relative fraction of 
M-type stars, as discussed, e.g., by \citet{boyer19}. However, this does not
reflect into a significant increase in the rate of silicates production,
since the mass loss rates experienced by M-type low-mass stars are
tipically below $\rm 10^{-6}M_{\odot}/$yr, which is not sufficient
to trigger large production rates of silicates \citep{ventura14}. On the other hand, 
turning to younger objects, we find a much higher contribution from 
the dust released by massive AGBs experiencing HBB, in comparison with more
metal poor environments \citep{ventura18}: however, the
fraction of these stars over the entire AGB population of M31
is below $1\%$, far too small to counterbalance the dust produced by
$\rm 1.2-3.5~M_{\odot}$ stars, during the C-rich phase.

Further observational confirmations are required before these results
can be definitively validated. For instance, these predictions could be 
tested by means of the criterion proposed by
\citet{boyer17} to distinguish M-type sources from C-stars, 
based on the position of the individual sources on the $\rm (F127M-F139M, 
F139M-F153M)$ plane, where the two classes of objects populate
distinct regions. Therefore, the future availability of these data for
the sources investigated here will allow a further validation of
the present interpretation.

\section{Conclusions}
We use a population synthesis approach based on stellar evolution and dust formation modelling to interpret HST and Spitzer data of the stars in the galaxy M31, in order to
characterize the AGB population and to clarify the reasons for the lack of the
extremely low fraction of red objects detected in this galaxy, in comparison to
other environments, such as the Magellanic Clouds.

The detailed comparison between the distribution of the data set in the observational
colour-magnitude planes built with the near-IR HST and Spitzer magnitudes
and the expectations from synthetic modelling allowed the identification of the 
stars populating the different regions of the planes and the derivation of a general
view of the AGB population of M31.

The analysis of HST data shows that $\sim 70\%$ of the AGB population descend from
$\rm 0.8-1.2~M_{\odot}$ progenitors of various metallicity, formed between 6 Gyr and 14 Gyr ago, during the initial phase of intense star formation activity experienced by M31. A significant fraction ($\sim 15\%$) of the sources classified as AGBs descend from $\rm 1.7-2.5~M_{\odot}$ stars of solar metallicity, formed during the secondary peak in the SFH of M31, which occurred between 1 Gyr and 2 Gyr ago. A small fraction (below $1\%$) of the AGB sample are identified as the progeny of $\rm M\geq 4~M_{\odot}$ stars, formed during the last 200 Myr, currently experiencing HBB. The statistical study of HST data offers a notable opportunity to improve our
understanding of some still poorly known aspects related to the AGB evolution, which regard in particular the efficiency of the convective transport of energy in the envelope of AGB stars of different metallicity, the mass loss suffered by low-mass stars during the ascending of the RGB, and the mass loss rates experienced by massive AGB stars during the HBB phase.

The study of the distribution of the stars in the $([3.6]-[4.5], [3.6])$ plane proves crucial to address the dust formation issue, considering the extremely low near-IR emission of
dusty stars. On the basis of the morphology of the evolutionary tracks of stars of
different mass during the AGB evolution, it is possible to select the regions within this observational plane that are populated by stars of different mass with little or no dust in their surroundings, and those harbouring the stars providing the most relevant contribution to the current dust budget, namely the progeny of $\rm M\geq 4~M_{\odot}$ stars, nowadays experiencing 
HBB, which produce large quantities of silicates, and carbon stars, descending from 
$\rm 1.2-3.5~M_{\odot}$ progenitors, surrounded by carbonaceous dust. 
$80\%$ of the overall dust production rate in M31 is provided by carbon stars, which contribute
with $\rm 4\times 10^{-4}~M_{\odot}$/yr, while the formation 
of silicates takes place with a rate of $\rm 6\times 10^{-5}~M_{\odot}$/yr. 

The latter estimates are based on the assumption that the lack of carbon stars 
populating the $([3.6]-[4.5])>2$ mag region of the colour-magnitude plane of M31, and more
generally the well known smaller fraction of extreme AGBs in the galaxy with respect to 
the Magellanic Clouds, is a mere effect of incompleteness of the M31 data-set in the 
$[3.6]$ band. Should the lack of extremely red carbon stars be real and not related
to the completeness issue, the dust production rate by carbon stars would be
approximately half the value indicated above. JWST mid-IR observations are urgently
required to clarify this still open point.

\begin{acknowledgements}

\end{acknowledgements}

%

\begin{thebibliography}{}

\bibitem[Abel et al.(2008)]{abel08} Abel, N.~P., van Hoof, P.~A.~M., Shaw, G., et al.\ 2008, \apj, 686, 1125

\bibitem[Aller \& Czyzak(1983)]{aller83} Aller, L.~H. \& Czyzak, S.~J.\ 1983, \apjs, 51, 211. 

\bibitem[Aringer et al.(2009)]{aringer09} Aringer, B., Girardi, L., Nowotny, W., et al.\ 2009, \aap, 503, 913

\bibitem[\protect\citeauthoryear{Aringer et al.}{2016}]{aringer16} Aringer B., Girardi L., Nowotny W., Marigo P., Bressan A., 2016, MNRAS, 457, 3611.

\bibitem[Begemann et al.(1994)]{begemann94} Begemann, B., Dorschner, J., Henning, T., et al.\ 1994, \apjl, 423, L71

\bibitem[Bl\"ocker \& Sch\"onberner(1991)]{blo91} Bl\"ocker, T. \& Sch\"onberner, D.\ 1991, \aap, 244, L43

\bibitem[Bl\"ocker(1995)]{blo95} Bl\"ocker, T.\ 1995, \aap, 297, 727

\bibitem[Bowen(1988)]{bowen} Bowen, G.~H.\ 1988, \apj, 329, 299

\bibitem[\protect\citeauthoryear{Boyer et al.}{2011}]{boyer11} Boyer M.~L., Srinivasan S., van Loon J.~T., McDonald I., Meixner M., Zaritsky D., Gordon K.~D., et al., 2011, AJ, 142, 103. 

\bibitem[Boyer et al.(2012)]{boyer12} Boyer, M.~L., Srinivasan, S., Riebel, D., et al.\ 2012, \apj, 748, 40

\bibitem[\protect\citeauthoryear{Boyer et al.}{2013}]{boyer13} Boyer M.~L., Girardi L., Marigo P., Williams B.~F., Aringer B., Nowotny W., Rosenfield P., et al., 2013, ApJ, 774, 83. 

\bibitem[Boyer et al.(2017)]{boyer17} Boyer, M.~L., McQuinn, K.~B.~W., Groenewegen, M.~A.~T., et al.\ 2017, \apj, 851, 152

\bibitem[Boyer et al.(2019)]{boyer19} Boyer, M.~L., Williams, B.~F., Aringer, B., et al.\ 2019, \apj, 879, 109

\bibitem[Caloi \& D'Antona(2005)]{caloi05} Caloi, V. \& D'Antona, F.\ 2005, \aap, 435, 987

\bibitem[Canuto \& Mazzitelli(1991)]{cm91} Canuto V.~M.~C., Mazzitelli I., 1991,
      ApJ, 370, 295

\bibitem[Catelan(2000)]{catelan2000} Catelan, M.\ 2000, \apj, 531, 826

\bibitem[Dalcanton et al.(2012)]{dalcanton12} Dalcanton, J.~J., Williams, B.~F., Lang, D., et al.\ 2012, \apjs, 200, 18

\bibitem[Dell'Agli et al.(2014)]{flavia14} Dell'Agli, F., Garc{\'\i}a-Hern{\'a}ndez, D.~A., Rossi, C., et al.\ 2014, \mnras, 441, 1115.

\bibitem[Dell'Agli et al.(2015a)]{flavia15a} Dell'Agli, F., Ventura, P., Schneider, R., et al.\ 2015a, \mnras, 447, 2992

\bibitem[Dell'Agli et al.(2015b)]{flavia15b} Dell'Agli, F., Garc{\'\i}a-Hern{\'a}ndez, D.~A., Ventura, P., et al.\ 2015b, \mnras, 454, 4235

\bibitem[Dell'Agli et al.(2016)]{flavia16} Dell'Agli, F., Di Criscienzo, M., Boyer, M.~L., et al.\ 2016, \mnras, 460, 4230

\bibitem[Dell'Agli et al.(2018)]{flavia18} Dell'Agli, F., Di Criscienzo, M., Ventura, P., et al.\ 2018, \mnras, 479, 5035

\bibitem[Dell'Agli et al.(2019)]{flavia19} Dell'Agli, F., Di Criscienzo, M., Garc{\'\i}a-Hern{\'a}ndez, D.~A., et al.\ 2019, \mnras, 482, 4733

\bibitem[Di Criscienzo et al.(2015)]{2419} Di Criscienzo, M., Tailo, M., Milone, A.~P., et al.\ 2015, \mnras, 446, 1469

\bibitem[Ferrarotti \& Gail(2001)]{fg01} Ferrarotti, A.~S. \& Gail, H.-P.\ 2001, \aap, 371, 133

\bibitem[Ferrarotti \& Gail(2002)]{fg02} Ferrarotti, A.~S. \& Gail, H.-P.\ 2002, \aap, 382, 256

\bibitem[Ferrarotti \& Gail(2006)]{fg06} Ferrarotti, A.~S. \& Gail, H.-P.\ 2006, \aap, 447, 553

\bibitem[Gail \& Sedlmayr(1985)]{gase85} Gail, H.-P. \& Sedlmayr, E.\ 1985, \aap, 148, 183

\bibitem[Goldman et al.(2017)]{goldman17} Goldman, S.~R., van Loon, J.~T., Zijlstra, A.~A., et al.\ 2017, \mnras, 465, 403

\bibitem[Goldman et al.(2022)]{goldman22} Goldman, S.~R., Boyer, M.~L., Dalcanton, J., et al.\ 2022, \apjs, 259, 41, G22

\bibitem[Gregersen et al.(2015)]{gregersen15} Gregersen, D., Seth, A.~C., Williams, B.~F., et al.\ 2015, \aj, 150, 189

\bibitem[Groenewegen \& Sloan(2018)]{martin18} Groenewegen, M.~A.~T. \& Sloan, G.~C.\ 2018, \aap, 609, A114

\bibitem[\protect\citeauthoryear{Habing}{1996}]{habing96} Habing H.~J., 1996, A\&ARv, 7, 97.

\bibitem[\protect\citeauthoryear{Harris \& Zaritsky}{2009}]{harris09} Harris J., Zaritsky D., 2009, AJ, 138, 1243.

\bibitem[Hauschildt et al.(1999)]{nextgen} Hauschildt, P.~H., Allard, F., Ferguson, J., et al.\ 1999, \apj, 525, 871

\bibitem[H{\"o}fner \& Olofsson(2018)]{hofner18} H{\"o}fner, S. \& Olofsson, H.\ 2018, \aapr, 26, 1

\bibitem[Iben(1974)]{iben74} Iben, I. Jr. 1974, ARA\&A, 12, 215
 
\bibitem[Kamath et al.(2023)]{devika23} Kamath, D., Dell'Agli, F., Ventura, P., et al.\ 2023, \mnras, 519, 2169

\bibitem[Karakas \& Lattanzio(2014)]{karakas14} Karakas A.~I., Lattanzio J.~C.\ 2014, PASA, 31, e030
 
\bibitem[Kobayashi et al.(2020)]{ciaki20} Kobayashi, C., Karakas, A.~I., \& Lugaro, M.\ 2020, \apj, 900, 179

\bibitem[\protect\citeauthoryear{Kroupa}{2001}]{kroupa01} Kroupa P., 2001, MNRAS, 322, 231. 

\bibitem[Laor \& Draine(1993)]{laor93} Laor, A. \& Draine, B.~T.\ 1993, \apj, 402, 441. 

\bibitem[Lewis et al.(2015)]{lewis15} Lewis, A.~R., Dolphin, A.~E., Dalcanton, J.~J., et al.\ 2015, \apj, 805, 183

\bibitem[Marigo(2002)]{marigo02} Marigo, P.\ 2002, \aap, 387, 507

\bibitem[Marigo \& Aringer(2009)]{marigo09} Marigo, P. \& Aringer, B.\ 2009, \aap, 508, 1539

\bibitem[Marini et al.(2020)]{marini20} Marini, E., Dell'Agli, F., Di Criscienzo, M., et al.\ 2020, \mnras, 493, 2996

\bibitem[Marini et al.(2021)]{marini21} Marini, E., Dell'Agli, F., Groenewegen, M.~A.~T., et al.\ 2021, \aap, 647, A69
   
\bibitem[\protect\citeauthoryear{Marini et al.}{2023}]{ester23} Marini E., Dell'Agli F., Kamath D., Ventura P., Mattsson L., Marchetti T., Garc{\'\i}a-Hern{\'a}ndez D.~A., et al., 2023, A\&A, 670, A97

\bibitem[Matsuura et al.(2009)]{matsuura09} Matsuura, M., Barlow, M.~J., Zijlstra, A.~A., et al.\ 2009, \mnras, 396, 918
 
\bibitem[Matsuura(2011)]{matsuura11} Matsuura, M.\ 2011, Why Galaxies Care about AGB Stars II: Shining Examples and Common Inhabitants, 445, 531

\bibitem[Matsuura et al.(2013)]{matsuura13} Matsuura, M., Woods, P.~M., \& Owen, P.~J.\ 2013, \mnras, 429, 2527.

\bibitem[Miglio et al.(2021)]{miglio21} Miglio, A., Chiappini, C., Mackereth, J.~T., 
et al.\ 2021, \aap, 645, A85

\bibitem[Nanni et al.(2013)]{nanni13} Nanni A., Bressan A., Marigo P., et al.\ 2013, 
 \mnras, 434, 2390
   
\bibitem[Nanni et al.(2014)]{nanni14} Nanni A., Bressan A., Marigo P., et al.\ 2014, 
 \mnras, 438, 2328
 
\bibitem[Nanni et al.(2016)]{nanni16} Nanni, A., Marigo, P., Groenewegen, M.~A.~T., 
 et al.\ 2016, \mnras, 462, 1215

\bibitem[Nanni et al.(2019)]{nanni19} Nanni, A., Groenewegen, M.~A.~T., Aringer, B., 
 et al.\ 2019, \mnras, 487, 502
   
\bibitem[Nenkova et al.(1999)]{nenkova99} Nenkova, M., Ivezic, Z., \& Elitzur, M.\ 1999, 
Thermal Emission Spectroscopy and Analysis of Dust, Disks, and Regoliths, 20

\bibitem[Ordal et al.(1988)]{ordal} Ordal, M.~A., Bell, R.~J., Alexander, R.~W., et al.\ 1988, \ao, 27, 1203

\bibitem[Ossenkopf et al.(1992)]{oss92} Ossenkopf, V., Henning, T., \& Mathis, J.~S.\ 1992, \aap, 261, 567

\bibitem[Pegourie(1988)]{peg88} Pegourie, B.\ 1988, \aap, 194, 335

\bibitem[Reimers(1975)]{reimers75} Reimers, D.\ 1975, Memoires of the Societe Royale des Sciences de Liege, 8, 369

\bibitem[Riebel et al.(2012)]{riebel12} Riebel, D., Srinivasan, S., Sargent, B., et al.\ 2012, \apj, 753, 71

\bibitem[Rouleau \& Martin(1991)]{roleau91} Rouleau, F. \& Martin, P.~G.\ 1991, \apj, 377, 526.

\bibitem[Salaris(2012)]{salaris12} Salaris, M.\ 2012, Red Giants as Probes of the Structure and Evolution of the Milky Way, 26, 45

\bibitem[Salaris et al.(2016)]{salaris16} Salaris, M., Cassisi, S., \& Pietrinferni, A.\ 2016, \aap, 590, A64

\bibitem[Salaris et al.(2018)]{salaris18} Salaris, M., Cassisi, S., Schiavon, R.~P., et al.\ 2018, \aap, 612, A68

\bibitem[Schneider et al.(2014)]{raffa14} Schneider R., Valiante R., Ventura P., et al.\ 2014, MNRAS, 442, 1440 

\bibitem[\protect\citeauthoryear{Schlegel et al.}{1998}]{schlegel98} Schlegel D.~J., Finkbeiner D.~P., Davis M., 1998, ApJ, 500, 525. 

\bibitem[\protect\citeauthoryear{Schneider \& Maiolino}{2023}]{raffa23} Schneider R., Maiolino R., 2023, arXiv, arXiv:2310.00053

\bibitem[Schr{\"o}der \& Cuntz(2005)]{sch05} Schr{\"o}der, K.-P. \& Cuntz, M.\ 2005, \apjl, 630, L73

\bibitem[Schwarzschild \& H{\"a}rm(1965)]{sch} Schwarzschild, M. \& H{\"a}rm, R.\ 1965, \apj, 142, 855

\bibitem[Sloan et al.(2012)]{sloan12} Sloan, G.~C., Matsuura, M., Lagadec, E., et al.\ 2012, \apj, 752, 140

\bibitem[Sloan et al.(2014)]{sloan14} Sloan, G.~C., Lagadec, E., Zijlstra, A.~A., et al.\ 2014, \apj, 791, 28

\bibitem[Sloan et al.(2016)]{sloan16} Sloan, G.~C., Kraemer, K.~E., McDonald, I., et al.\ 2016, \apj, 826, 44

\bibitem[Srinivasan et al.(2016)]{srinivasan16} Srinivasan, S., Boyer, M.~L., Kemper, F., et al.\ 2016, \mnras, 457, 2814

\bibitem[Tailo et al.(2016)]{tailo16} Tailo, M., Di Criscienzo, M., D'Antona, F., et al.\ 2016, \mnras, 457, 4525

\bibitem[Tailo et al.(2021)]{tailo21} Tailo, M., Milone, A.~P., Lagioia, E.~P., et al.\ 2021, \mnras, 503, 694

\bibitem[VandenBerg \& Denissenkov(2018)]{don18} VandenBerg, D.~A. \& Denissenkov, P.~A.\ 2018, \apj, 862, 72

\bibitem[van Loon(2000)]{jacco00} van Loon, J. T.\ 2000, \aap, 354, 125

\bibitem[van Loon et al.(2005)]{jacco05} van Loon, J. T., Marshall, J. R., Zijlstra, A. A.\ 2005, \aap, 442, 597

\bibitem[Vassiliadis \& Wood(1993)]{vw93} Vassiliadis, E. \& Wood, P.~R.\ 1993, \apj, 413, 641, VW93

\bibitem[Ventura et al.(1998)]{ventura98} Ventura, P., Zeppieri, A., Mazzitelli, I.,   D'Antona, F., 1998, A\&A, 334, 953

\bibitem[Ventura et al.(2001)]{ventura01} Ventura, P., D'Antona, F., Mazzitelli, I., et al.\ 2001, \apjl, 550, L65.

\bibitem[Ventura \& D'Antona(2005a)]{ventura05a} Ventura P., D'Antona F.\ 2005a, 
\aap, 431, 279

\bibitem[Ventura \& D'Antona(2005b)]{ventura05b} Ventura, P. \& D'Antona, F.\ 2005b, \aap, 439, 1075

\bibitem[Ventura \& Marigo(2009)]{vm09} Ventura, P. \& Marigo, P.\ 2009, \mnras, 399, L54

\bibitem[Ventura \& Marigo(2010)]{vm10} Ventura, P. \& Marigo, P.\ 2010, \mnras, 408, 2476

\bibitem[Ventura et al.(2012)]{ventura12} Ventura P., Di Criscienzo M., Schneider R., 
et al.\ 2012, \mnras, 420, 1442

\bibitem[Ventura et al.(2013)]{ventura13} Ventura, P., Di Criscienzo, M., Carini, R., et al.\ 2013, \mnras, 431, 3642

\bibitem[Ventura et al.(2014)]{ventura14} Ventura P., Dell'Agli F., Schneider R., et al.\ 2014, \mnras, 439, 977

\bibitem[Ventura et al.(2018)]{ventura18} Ventura, P., Karakas, A., Dell'Agli, F., et  al.\ 2018, \mnras, 475, 2282. 

\bibitem[Ventura et al.(2022)]{ventura22} Ventura, P., Dell'Agli, F., Tailo, M., et al.\ 2022, Universe, 8, 45.
 
\bibitem[Vincenzo et al.(2016)]{fiorenzo} Vincenzo, F., Belfiore, F., Maiolino, R., et al.\ 2016, \mnras, 458, 3466

\bibitem[Vitense(1953)]{mlt} Vitense, E.\ 1953, \zap, 32, 135

\bibitem[Wachter et al.(2002)]{wachter02} Wachter, A., Schr{\"o}der, K.-P., Winters, J.~M., 
et al.\ 2002, \aap, 384, 452

\bibitem[Wachter et al.(2008)]{wachter08} Wachter, A., Winters, J.~M., Schr{\"o}der, K.-P., et al.\ 2008, \aap, 486, 497

\bibitem[Williams et al.(2017)]{williams17} Williams, B.~F., Dolphin, A.~E., Dalcanton, J.~J., et al.\ 2017, \apj, 846, 145

\end{thebibliography}
%

\end{document}